\documentclass[11pt, a4paper, onecolumn]{google}

\usepackage{longtable}
\usepackage{multirow}
\usepackage[square,numbers,sort&compress]{natbib}
\usepackage{amssymb}
\usepackage{float}
\usepackage{makecell}

\usepackage{xcolor}

\definecolor{linkblue}{RGB}{0,0,238}

\title{Beyond Explainable AI (XAI): An Overdue Paradigm Shift and Post-XAI Research Directions}

\renewcommand{\today}{}

\author[1]{Saleh Afroogh*}
\author[2]{Syed Ishtiaque Ahmed}
\author[3]{Petra Ahrweiler}
\author[4]{David Alvarez-Melis}
\author[5]{Mansur Maturidi Arief}
\author[6]{Emilia Barakova}
\author[7]{Falco J. Bargagli-Stoffi}
\author[8]{Erdem Biyik}
\author[9]{Hanjie Chen}
\author[7]{Xiang 'Anthony' Chen}
\author[10]{Robert Alan Clements}
\author[11]{Keeley Crockett}
\author[12]{Amit Dhurandhar}
\author[13]{Fethiye Irmak Dogan}
\author[14]{Mollie Dollinger}
\author[15]{Motahhare Eslami}
\author[16,17]{Aldo A Faisal}
\author[1]{Arya Farahi}
\author[18]{Melanie F. Pradier}
\author[7]{Saadia Gabriel}
\author[19]{Diego Garcia-Olano}
\author[20]{Marzyeh Ghassemi}
\author[21]{Shaona Ghosh}
\author[13]{Hatice Gunes}
\author[22]{Ehsan Hajiramezanali}
\author[23]{Stefan Haufe}
\author[24]{Biwei Huang}
\author[8]{Angel Hwang}
\author[5]{Md Tauhidul Islam}
\author[1]{Junfeng Jiao}
\author[25]{Amir-Hossein Karimi}
\author[4]{Saber Kazeminasab}
\author[2]{Anastasia Kuzminykh}
\author[4]{William La Cava}
\author[26]{Brian Y. Lim}
\author[27]{Xiaofeng Liu}
\author[28,29]{Mohammad R. K. Mofrad}
\author[30]{Alicia Parrish}
\author[31]{Maria Perez-Ortiz}
\author[5]{Shriti Raj}
\author[8]{Swabha Swayamdipta}
\author[28]{Salmonn Talebi}
\author[12]{Kush R. Varshney}
\author[18]{Mihaela Vorvoreanu}
\author[24]{Lily Weng}
\author[32]{Alice Xiang}
\author[1]{Yiming Xu}
\author[15]{Ding Zhao}
\author[8]{Jieyu Zhao\vspace{0.5cm}}

\correspondingauthor{saleh.afroogh@utexas.edu}%\\[1cm]

\affil[1]{\small\textcolor{linkblue}{University of Texas at Austin}}
\affil[2]{\small\textcolor{linkblue}{University of Toronto}}
\affil[3]{\small\textcolor{linkblue}{JG University of Mainz}}
\affil[4]{\small\textcolor{linkblue}{Harvard University}}
\affil[5]{\small\textcolor{linkblue}{Stanford University}}
\affil[6]{\small\textcolor{linkblue}{Eindhoven University of Technology}}
\affil[7]{\small\textcolor{linkblue}{University of California, Los Angeles}}
\affil[8]{\small\textcolor{linkblue}{University of Southern California}}
\affil[9]{\small\textcolor{linkblue}{Rice University}}
\affil[10]{\small\textcolor{linkblue}{University of San Francisco}}
\affil[11]{\small\textcolor{linkblue}{Manchester Metropolitan University}}
\affil[12]{\small\textcolor{linkblue}{IBM Research}}
\affil[13]{\small\textcolor{linkblue}{University of Cambridge}}
\affil[14]{\small\textcolor{linkblue}{Curtin University}}
\affil[15]{\small\textcolor{linkblue}{Carnegie Mellon University}}
\affil[16]{\small\textcolor{linkblue}{Imperial College London}}
\affil[17]{\small\textcolor{linkblue}{Universität Bayreuth}}
\affil[18]{\small\textcolor{linkblue}{Microsoft Research}}
\affil[19]{\small\textcolor{linkblue}{Meta}}
\affil[20]{\small\textcolor{linkblue}{Massachusetts Institute of Technology}}
\affil[21]{\small\textcolor{linkblue}{Nvidia}}
\affil[22]{\small\textcolor{linkblue}{Genentech}}
\affil[23]{\small\textcolor{linkblue}{Technical University Berlin}}
\affil[24]{\small\textcolor{linkblue}{University of California San Diego}}
\affil[25]{\small\textcolor{linkblue}{University of Waterloo}}
\affil[26]{\small\textcolor{linkblue}{National University of Singapore}}
\affil[27]{\small\textcolor{linkblue}{Yale University}}
\affil[28]{\small\textcolor{linkblue}{University of California, Berkeley}}
\affil[29]{\small\textcolor{linkblue}{Lawrence Berkeley National Laboratory}}

\affil[30]{\small\textcolor{linkblue}{Google DeepMind}}
\affil[31]{\small\textcolor{linkblue}{University College London}}

\affil[32]{\small\textcolor{linkblue}{Sony AI}}

\begin{abstract}
This study provides a cross-disciplinary examination of Explainable Artificial Intelligence (XAI) approaches -- focusing on deep neural networks (DNNs) and large language models (LLMs) -- and identifies empirical and conceptual limitations in current XAI. We discuss critical symptoms that stem from deeper root causes (i.e., two paradoxes, two conceptual confusions, and five false assumptions). These fundamental problems within the current XAI research field reveal three insights: experimentally, XAI exhibits significant flaws; conceptually, it is paradoxical; and pragmatically, further attempts to reform the paradoxical XAI might exacerbate its confusion -- demanding fundamental shifts and new research directions. 

To move beyond XAI's limitations, we propose a four-pronged synthesized paradigm shift toward reliable and certified AI development. These four components include: verification-focused Interactive AI (IAI) to establish scientific community protocols for certifying AI system performance rather than attempting post-hoc explanations, AI Epistemology for rigorous scientific foundations, User-Sensible AI to create context-aware systems tailored to specific user communities, and Model-Centered Interpretability for faithful technical analysis -- together offering comprehensive post-XAI research directions.\\
\hfill
\end{abstract}

\begin{document}

\maketitle

\section{Introduction}
As AI systems increasingly influence consequential decisions affecting human lives, the need to understand their internal mechanisms has evolved from a technical preference to a societal imperative \cite{1}. With the development of LLMs, VLMs, and foundation models, transparent decision-making has become even more important, as the trained datasets are less controlled and, in some cases, more biased \cite{2,3}. The black-box challenge facing modern XAI (see Table~\ref{tab:1} for definitions) is, paradoxically, a consequence of AI’s remarkable success. While early logic-based symbolic AI systems \cite{4} were interpretable (see Table~\ref{tab:1}), but limited, today's successful predictive and pattern recognition-based ML models often achieve unprecedented performance by using opaque internal representations that frequently disregard interpretability for effectiveness \cite{5,6}. This has motivated a diverse range of stakeholders—from governmental regulators and public sector agencies to domain scientists, healthcare providers, and corporate marketing teams—to actively invest in methods that could illuminate the internal mechanisms of opaque AI systems. 

Although the idea had been around for a while, explainable AI (XAI) has solidified as a formal research enterprise through DARPA’s 2016 program \cite{7}, which sought a systematic study of AI’s ``black-box'' problem by funding post-hoc techniques like LIME \cite{8} and Grad-CAM \cite{9}. While interpretability research (i.e., research on making models less of a black-box) predates this initiative—from rule-based systems  \cite{10, 11, 12}, and case-based neural architectures \cite{13} to methods for explaining individual classifications \cite{14} and critiques of linear model interpretability \cite{15}—the DARPA initiative propelled post-hoc efforts into the mainstream under the banner of XAI \cite{16,17,18} (whereas inherently interpretable architectures became somewhat of an afterthought instead of a primary goal). Simultaneously, the EU’s GDPR, adopted in 2016, enshrined the ``right to explanation'' for algorithmic decisions \cite{19}, making XAI a compliance necessity that shaped both corporate practices and academic research globally \cite{20}. 

After nearly a decade, the legacy of the XAI program is increasingly questioned, with critics arguing that challenges have outweighed achievements. It is described as being ``in trouble'' \cite{6}, and some scholars suggest it should be “stopped” for high-stakes decisions \cite{21} or has no role in the future of human-centric AI approaches \cite{22}; others view it as myth \cite{23} or consider it already “dead” \cite{24}. A study by \citet{25} demonstrates that most XAIs provide shallow and inadequate explanations. Through a systematic analysis of 34 XAI systems published during 2019-2021, selected from an initial pool of 165 articles using strict criteria requiring actual machine-generated explanations, they classify ``explanations'' of AI systems into seven categories: (1) basic surface features, (2) success instances, (3) failure instances, (4) AI reasoning, (5) diagnosis of failures, (6) exploration, and (7) interactive adaptation - this final category contributes the most to sense-making and understanding the workings of AI systems. Their empirical evaluation shows that current XAI tends to produce technical systems at lower cognitive support levels or non-interactive ones. Thus, the majority of XAI systems at present fail to support the sense-making necessary for cultivating trust among AI users. \citet{25}'s findings reveal that interactive adaptation (category 7) contributes most significantly to sense-making and user understanding yet represents the most underutilized approach in current XAI systems. This implies that the core issue in XAI is not the pursuit of explainability tools themselves—which are useful for tasks like debugging—but rather the field's ambition to explain black-boxes. This ambition leads to an overemphasis on post-hoc explanations instead of a focus on verifying the AI's output in practice or building inherently interpretable models. (More on this in Section~\ref{subsec:interactive-verification}.)

Moreover, the field of XAI suffers from both fundamental disagreements about its scope and pervasive terminological ambiguity.  \citet{6} highlight the disagreements on the field's scope  \cite[see, for instance,][]{26,27,28,29,30}. Further, there is no consistency in the definitions of fundamental terms that form the foundation of XAI, such as ``explainability'' and ``interpretability,'' leading to terminological inconsistency. While these two terms are sometimes used interchangeably \cite{23}, some authors propose distinctions \cite{21, RudinEtAlSurvey2022,31,32,33,34} or use hybrid phrases such as "interpretable explanations" \cite{35} or "explainable interpretations,"\cite{36} which further exacerbates the ambiguity within the field. Beyond definitional problems, the XAI literature reveals a lack of agreement on fundamental questions: who XAI is for, what it should achieve, and how its success should be measured. Some projects target end-users seeking trust \cite{37}, others focus on developers needing debugging tools \cite{38}, while still others aim to satisfy regulators requiring accountability \cite{39}. The result is a field where researchers pursue incompatible goals, using inconsistent terminology, all under the same umbrella term of "XAI." Consequently, we are confronted with a meta-problem: the endeavor to open AI's black-boxes has itself created another: the black-box of XAI per se. 

\begin{figure}[!tb]
\centering
\includegraphics[width=0.98\textwidth]{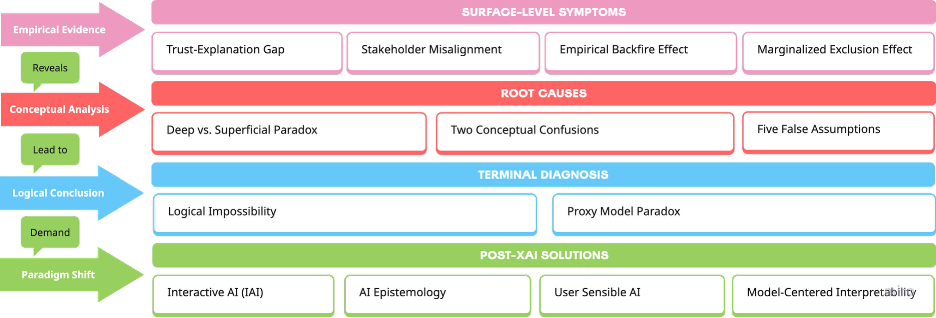}
\caption{From Symptoms to Solution: XAI’s Systematic Diagnosis Framework}
\label{fig:framework}
\end{figure}

This meta-problem, along with the drivers to continue working despite these criticisms, has left the field in a quandary or impasse. This present landscape of XAI motivates us to re-examine the landscape by examining the fundamental grounds of the XAI field and systematically diagnosing its symptoms and root causes that have stalled the field’s progress (see Figure~\ref{fig:framework}). We examine 13 critical issues (four surface-level symptoms and nine root causes: two paradoxes, two conceptual confusions, and five false assumptions), demonstrating that while the existing XAI field has provided valuable foundations, these constraints indicate the field has reached a pivotal moment requiring fundamental reconceptualization and new research trajectories.

This position paper draws on insights from a diverse team of contributors, including leading high-tech companies, prominent university research groups, and cross-disciplinary perspectives—spanning computer science (e.g., AI, XAI, HCI, NLP, formal methods, AI alignment), statistics (e.g., data science, biostatistics and causal inference), cognitive science, psychology, and philosophy. We structure this paper as follows: Section~\ref{subsec:2} defines the problem’s scope, core concepts, and stakes for science and applications. Section~\ref{subsec:3} examines XAI's fundamental issues, including empirical symptoms like the trust-explanation gap and cognitive overload, along with deeper root causes such as the deep-vs-superficial paradox, conceptual confusions, and false assumptions. Section~\ref{subsec:4} discusses a four-pronged alternative direction: Interactive AI as a paradigm shift toward scientific community verification and certification protocols, AI Epistemology for establishing rigorous scientific foundations, User-Sensible AI for adaptive and context-aware systems that create expert-mediated pathways to serve diverse communities, and Model-Centered Interpretability for legitimate technical analysis without false explanatory claims about real-world phenomena. Section~\ref{subsec:5} concludes.

\subsection{Working Definitions for Analytical Coherence}

The XAI field suffers from terminological inconsistency, where definitions are not universally agreed upon and are used largely interchangeably in the literature \cite{6}. This conceptual ambiguity enables some XAI apologists to remain strategically flexible when confronting critical evaluations. In this section, we do not aim to resolve this issue by proposing universal definitions or achieving terminological consensus within this research community. Instead, for the purpose of this paper, we provide a glossary of provisional working definitions (see Table~\ref{tab:1}) that enables consistent analysis throughout this paper without claiming to resolve broader field disputes.

{\fontsize{10}{10.9}\selectfont
\renewcommand{\arraystretch}{1.5} % Adjust row height
\begin{longtable}{p{0.20\linewidth} p{0.75\linewidth}}
\caption{Glossary of Working Definitions for Analytical Coherence. \label{tab:1}}\\
\toprule
Term & Definition \\
\midrule
\endfirsthead
\toprule
Term & Definition \\
\midrule
\endhead
\midrule
\multicolumn{2}{r}{Continued on next page} \\
\midrule
\endfoot
\bottomrule
\endlastfoot
Black-Box Model & A (temporarily or persistently) non-interpretable AI model whose internal decision-making process is opaque and complex for stakeholders. Opacity is subject-dependent, and stakeholders include AI developers, domain experts, or non-expert end-users. Therefore, its opacity includes the following types: (1) the formal and computational interpretations are opaque even for AI developers, (2) the underlying scientific intuitions (if any) are unknown for domain experts and scientists, and (3) there is no commonsense understanding for non-expert end-users. (Otherwise, the model would be a glass box for AI developers, domain experts, and non-expert end-users, respectively.) While black-box model inputs and outputs are observable, the functional and scientific mapping between them is unknown or excessively complex (if any). Consequently, the truth of its outputs is inherently uncertain, requiring additional methods to properly evaluate and manage them. The XAI research field has emerged, for the most part, as a response to this challenge. \\
End-user & Includes both non-expert end-users and (non-AI) domain experts/scientists who interact with AI systems at different levels of technical expertise. \\
Explainable AI & A recent subfield that responds to the challenge of black-box models by using a second approximate model or variable importance measure to explain black-box models' behavior. It often includes two types of post-hoc approaches: architectural -- a second model (known as a proxy or surrogate model) is created to explain the first black-box model; semantic -- the second model consists of approximate values and variables that map the relationship between input and output of the first black-box model. These values and variables are supposed to be either scientifically relevant to build trust among domain experts, or intuitively and narratively meaningful to help non-expert end-users understand the system.  \newline If the second model's mechanisms, values, and variables are clearly the same as or (scientifically) relevant to those of the original model—so that the second model is completely faithful to the original—then the first model is no longer a black-box and requires no XAI approach; it is simply an interpretable model. Therefore, the values and variables of the proxy model must differ meaningfully from those of the first black-box model. In other words, the second model is either unfaithful, or its faithfulness is unknown. \\
Interpretable AI & A (immediate or achieved) transparent ``glass-box'' model whose internal decision-making process is directly understandable for AI developers and domain experts. Its values, variables, and mechanisms adhere to both formal methods and essential scientific knowledge of the domain. These models provide their own explanations, which are faithful to what the model actually computes. \\
Explanation Attempt vs Interpretation Attempt & An explanation attempt aims to make a model's decisions (not necessarily the internal dynamics) understandable to end-users to build trust. This is often a post-hoc process, using a second approximate model to explain the black-box's behavior. However, these explanations are often unfaithful to the model's actual computations and lack detail. \newline In contrast, an interpretation attempt focuses on understanding how a model truly works from the inside out. The goal is to make models interpretable both computationally for AI developers and scientifically for domain experts. \\
Scientific Explanation & Justifications grounded in causal relationships, domain knowledge, or experimental and empirical evidence (including DN models) that connect predictions to real-world theories, making intuitive sense to domain experts. \\
\vspace{-10pt}\makecell[l]{Causal \\ Truth-conducive\\  Explanation} & \vspace{-15pt}Justifications grounded in attempts to seek complete and faithful causal understanding and conducive to discovering causal truth through scientific methodology (excluding pseudoscience, superficial views, magic, etc.) and domain knowledge, yet might be incomplete or unsuccessful in many cases, therefore relying on pragmatic verification and effectiveness testing. \\
``Explanation'' (in XAI) & ``Explanations'' often come from a separate, simpler, model that aims to replicate most of the behavior of a black-box, or it comes from variable importance estimates. The term generally refers to understanding how a model works, not how the world works. Semantically, there is a vast disagreement, spanning rigorous scientific explanations and the superficial, chain-of-thought, or pseudo-scientific, and narratives people use to make sense of everyday life. \newline In this paper, we use this term with quotation marks to highlight the vague usage of ``explanations'' in the XAI literature, indicating the definitional disagreement surrounding the term and the potential failure of these purported explanations to meet the standards of genuine scientific explanation. \\
Self-explanation (by LLMs) & Explanations that large language models (LLMs) produce themselves about their own predictions or outputs, providing justification for their own behavior. \\
Mechanistic Interpretability & A model-centered explanation attempt  to identify how individual components (neurons, layers, attention heads) work together to implement algorithms and generate predictions. They provide a detailed, low-level understanding of the model's internal logic, which may itself be complex. Key Attribute: Uncovering the ``how'' at the most fundamental level of the model's architecture. The goal is to provide formal explanations of the model. \\
Completeness (of an Explanation) & The extent to which an explanation (second model) provides sufficient detail about the variables and values used by the original black-box model to understand what it is actually doing (not just vague importance metrics). Many explanations are so incomplete that they look like summaries, make no sense, or cannot convey which variables the model uses or how it processes them in its reasoning. \\
Transparency (of a Model) & The property of an AI model where its internal logic is fully visible and comprehensible. A transparent model allows an observer to trace the entire decision pathway for any given input. It is synonymous with a "glass box" or "white box" model. This is the defining feature of Interpretable AI. Two kinds exist: mechanistic interpretations transparency vs. causal transparency for domain scientists. \\
Accuracy (of a Model) & A performance metric that quantifies how often a model's predictions are correct. It is a statistical measure of predictive performance without any regard for how or why the predictions are made. It is a measure of predictive performance, not of transparency or trustworthiness. \\
Faithfulness (of a Model) & The degree to which an explanation (i.e. the proposed second model) accurately represents what the original model actually computes. Explanations cannot always have perfect fidelity and completeness for black box models - if they did, they would equal the original model and it would thus not be a black box. \\
Verification (of AI predictions) & Systematic protocols conducted by qualified domain experts to test, validate, and confirm AI system predictions against established knowledge and real-world outcomes, providing objective performance assessment without requiring explanatory claims about system internals. \\
Trust (in AI) & Psychological confidence in a system based on understanding, familiarity, or perceived competence, often developed through repeated positive interactions. \\
Reliance (on AI) & Practical dependence on a system based on consistent performance and verifiable outputs, independent of understanding internal mechanisms. \\

\end{longtable}
}

\section{Problem Setting and Scope}
\label{subsec:2}
\textbf{\textit{This paper focuses on post-hoc explanations for complex systems.}}
Its scope is not general AI transparency or a broad conception of XAI that encompasses all AI methods. It is also not about ante-hoc explanations or intrinsically interpretable models. Furthermore, the goal of this study is not to dismiss all XAI model analysis as meaningless or to ignore the legitimate secondary benefits of XAI tools. Our focus is on the narrow view of XAI, which specifically addresses machine learning model explanations (particularly in complex systems and deep neural networks (DNNs) such as large language models (LLMs)—that are deep neural networks trained on massive text datasets) rather than the broader conception that encompasses all AI methods and human-centered perspectives \cite{40,41,42,43}. Moreover, we address post-hoc “explanations” for complex and opaque systems, which aim to inform and calibrate users’ trust and enable meaningful recourse—allowing users to contest incorrect decisions. 

\textit{\textbf{We adopt the DARPA framework: user-oriented trust is the primary objective.}} 
Acknowledging various approaches to XAI, including alternative conceptions \cite{41,44} and the computational versus human-centered explanation dichotomy identified by \citet{45}, this paper adopts the user-oriented definition established by the DARPA XAI Program \cite{17} This program represents a collaborative community initiative that explicitly identifies trust as its central objective. They aim to develop AI systems that ensure transparent and understandable decision-making processes for users \cite{46,47}, thereby promoting human-AI trust. Thus, we address post-hoc ``explanations'' for users’ trust (as the primary target)— not  secondary potential benefits of XAI techniques for model performance \cite{48,49} or their positive side effects, such as enhanced debugging capabilities \cite{50,51}. Nevertheless, in Section~\ref{subsec:4.4}, we discuss some implications of our critical analysis for these secondary benefits. 

The primary goal of XAI, as per DARPA's framework, is to calibrate end-users' trust, including regulators, domain experts, business stakeholders, compliance officers, and especially non-experts, through clarifying an AI's decision-making process, which can take two forms: ``explaining'' the \textit{how} (the model's internal mechanics) or the \textit{why} (the data features and heuristics that drove the outcome)\cite{52}.\footnote{This analysis focuses on benign end-users, as adversarial or malicious users may actually find XAI beneficial for discovering new exploits. Also, throughout this paper, our use of the term “user” refers specifically to “end-users” (whether a domain expert or a non-expert) rather than AI developers.} In complex AI systems (such as DNNs and LLMs), decisions are often made through multi-layered processes; thus, end-users often cannot possibly interpret the underlying logic \cite{53}. Many researchers contend that understanding of such inner workings is key to building users' trust, and crucial to such understanding is explainability. This argument appears to make XAI relevant, practical, and necessary for building a trustworthy, responsible AI, as emphasized in many studies and guidelines \cite{54,55,56,57}. However, as we will demonstrate, this approach is fundamentally flawed—research shows that such explanations can potentially impair rather than improve human judgment \cite{58}. Systematic verification and certification by domain expert communities provide more reliable pathways to trustworthy AI deployment.

\textbf{\textit{Perfect accuracy eliminates the need for explanation.}}
Every ML system processes input to produce output, but the critical factor is the interpretability of the processes and the relevant decision-making. When simple, inherently understandable models perform adequately, they could be preferred, as they are computationally accurate and transparent for scientists and causally intelligible for domain experts. If a system is 100\% accurate in every scenario, then no explanations are needed in practice – it always works. Thus, we consider complex black-box models that are not 100\% accurate in this manuscript. Note that ``explanations'' are also not 100\% accurate, which means they explain the underlying model incorrectly, and we will not necessarily know when the explanations (and thus the predictions) are incorrect \cite{21}. Further, explanation models tend to be less accurate near the ML model’s decision boundaries, but these are exactly the ``non-obvious'' cases – the cases where explanations would be needed, if they were needed at all \cite{RudinEtAlSurvey2022}. 

We point out that when complex black-box models like deep neural networks are used for some applications, a significant interpretation effort (see Table~\ref{tab:1}) may be required; if AI scientists can computationally understand the model but domain experts cannot, the model requires validation through potential causal or experimental explanations. If explanation efforts fail entirely, it is pragmatically essential to stop using misleading post-hoc explanations. Epistemologically, we may acknowledge these as either false explanations or open research questions—and refrain from deploying the model unsafely.

\textbf{\textit{The choice is not between explanation (in XAI) and opacity (of black-boxes). }}
Our analysis reveals that the pursuit of end-user explanations creates a false dichotomy—the mistaken belief that we must choose between either explaining AI systems to users or accepting completely unaccountable black-boxes. To move beyond this binary choice, we propose distinguishing between a pragmatic level—which addresses the practical need to verify AI recommendations—and an epistemic level—which addresses the scientific goal of understanding the internal mechanisms of AI models in generating valid knowledge. We can consider black-box systems, as they are, at the pragmatic level while establishing rigorous verification protocols through scientific community interaction \cite{vashistha2025i, vashistha2024u}. This approach recognizes that users benefit from certified AI outputs through established professional trust relationships rather than (in some or all cases) requiring direct comprehension of AI internals. The medical field demonstrates this strategy effectively, with acetaminophen serving as a prime example—safely utilized for several decades even though its mechanism remains incompletely understood, proven effective through comprehensive clinical testing instead of full mechanistic comprehension \cite{59}. This choice does not preclude ongoing efforts to understand the internal dynamics of black-boxes at the epistemic level, as we will discuss in Section~\ref{subsec:4.4}.

\textbf{\textit{As a scientific field, XAI is not pseudoscience.}} That is, [premise \#1:] if we consider XAI as a scientific field, it cannot be constituted by non-scientific or pseudoscientific narrations. 
Also, [premise \#2:] any scientific field is grounded in (not irrelevant to) facts and real-world phenomena. If you agree with these two premises, you have already agreed that, as a scientific field of research, XAI cannot legitimately produce explanations that are pseudoscientific or detached from empirical reality. XAI's explanations should, by definition, be scientific and empirically grounded, not abstract or world-independent. It is supposed to provide genuine scientific explanations to support AI models’ predictions and recommendations in different domains like medical diagnosis, risk analysis, drug repurposing, etc.—all involving real-world phenomena where real-world relevance and empirical accuracy are paramount. To achieve this, XAI should (when attempting to solve scientific problems) offer truth-conducive and fact-sensitive explanations such as the classical Deductive-Nomological Model—a model that seeks to identify causal relationships between what is being explained (the explanandum) and the explanation itself (the explanans) \cite{60}. It does this by providing logically consistent and empirically supported accounts. This conception encompasses the rigorous scientific forms of explanation—a priori, a posteriori, logical, mathematical, and empirical—that follow established methodological frameworks including induction, deduction, and controlled experimentation. This focus on genuine scientific explanation also helps counter the hype and confusion about XAI’s ability to explain scientific recommendations and decisions for real-world problems in fields like medicine, climate science, finance, agriculture, etc. —an issue we elaborate on in Section~\ref{subsec:3.2.3}  We also distinguish scientific explanation from casual information. While some argue that XAI can be anecdotal and there is no need to be causally rigorous or maintain scientific rigor, this ignores contextual standards. The casual explanations adequate for personal preferences (like "Red gets my vote!") are dangerously insufficient when applied to high-stakes scientific discovery problems, e.g., the discovery of genes that cause outcomes. The demand for scientific XAI isn't about imposing formalism everywhere but about ensuring explanations meet the necessary standards in consequential contexts to prevent hype and terminological ambiguity. 

\subsection{The stakes: why this matters}

While explanations help clinicians better trust and engage with AI \cite{panigutti2022understanding}, they can increase overreliance on AI, even when it is wrong \cite{jacobs2021machine}. However, the utility of explanations is increasingly questioned. Empirically, the stakes extend beyond misleading explanations to the counterproductive effects of transparency itself. Evidence demonstrates that providing explanations can impair users' ability to detect AI errors \cite{58}, suggesting that verification-based approaches offer more reliable pathways to trustworthy AI deployment than explanation-based strategies. A healthcare study of 457 clinicians diagnosing acute respiratory failure found that biased AI with explanations decreased diagnostic accuracy by 9.1 percentage points, with explanations failing to help clinicians recognize the bias \cite{63}. AI systems, despite appearing neutral, inherently lack neutrality \cite{64}—their explanations can normalize biased decisions without acknowledging the bias, creating a dangerous illusion of objectivity. Theoretically, misleading explanations may increase user trust in the short term. However, repeated over-reliance on incorrect outputs eventually reduces trust \cite{65} and leads to stagnation in AI development by suppressing motivation for seeking genuine (complete and faithful) explanations \cite{66}. Pragmatically, in high-stakes cases like healthcare and finance, reliance on XAI for attaining transparency and trust can have serious repercussions, as it may fail to deliver genuine understanding, leading to injustices, discrimination, and health issues \cite{53}. 

\section{Fundamental Limitations of XAI}
\label{subsec:3}

\begin{figure}[!tb]
    \centering
    \includegraphics[width=1\linewidth]{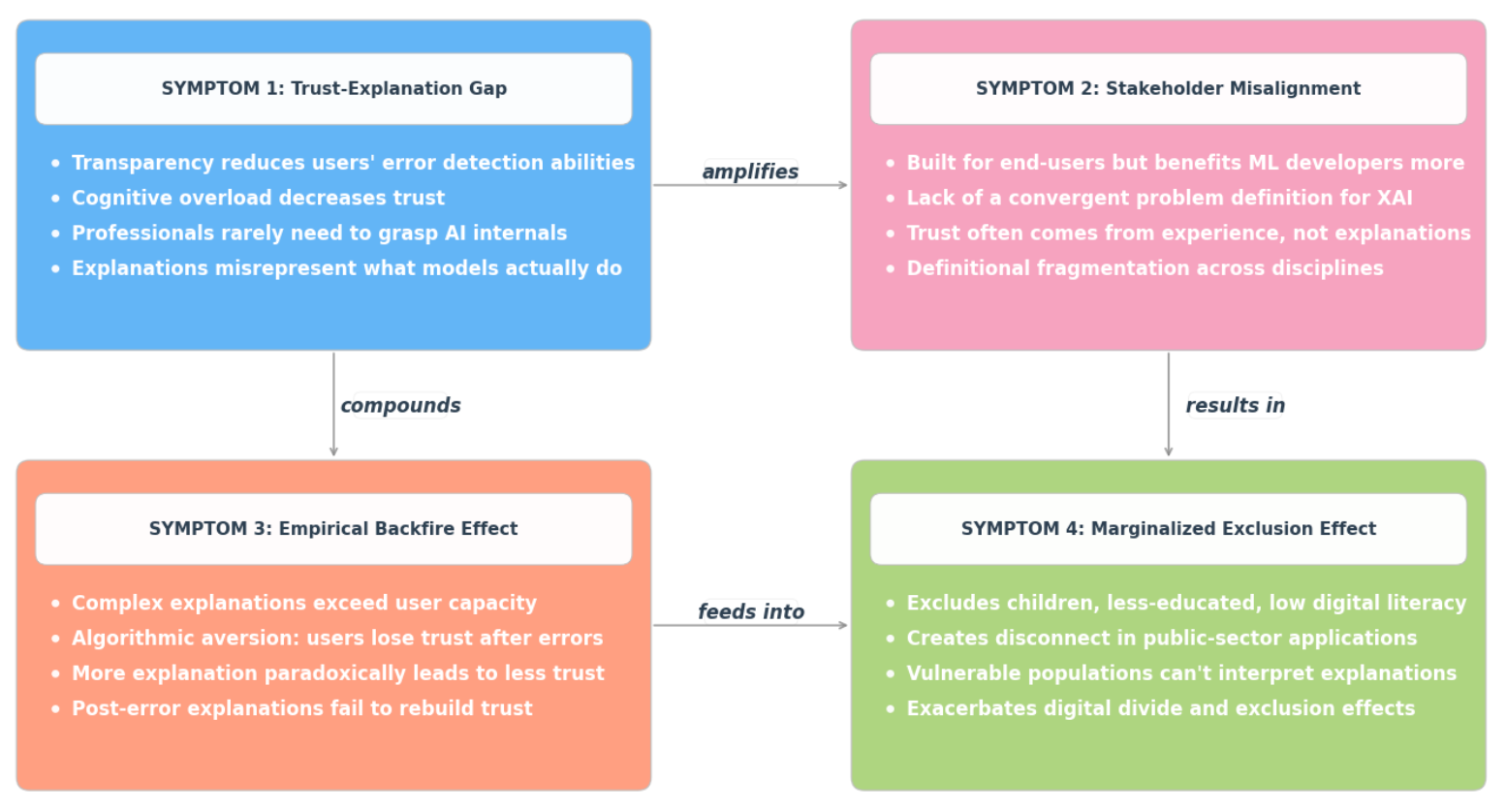}
    \caption{XAI Surface-Level Symptoms: Empirical Flaws Analysis}
    \label{fig:surface_level}
\end{figure}

In this section, we examine XAI's inherent limitations through a three-tiered diagnostic analysis: from observable symptoms to underlying causes. We first classify XAI’s empirical flaws into four symptom categories -- the Empirical Trust-Explanation Gap, Stakeholder Misalignment, the Empirical Backfire Effect, and the Marginalized Exclusion Effect -- and we will review some reported flaws in real-world XAI implementations (Section~\ref{subsec:3.1} and Figure~\ref{fig:surface_level}). Secondly, we conducted an etiological analysis—progressing beyond implementation issues to logical, methodological, and semantic examination—to identify the fundamental root causes. The aforementioned symptoms indicate deeper root causes in XAI: a paradox between accuracy and understandability, two conceptual confusions (explanatory levels and causation), and five false assumptions underlying the XAI field (Section~\ref{subsec:3.2}). Finally, we demonstrate the second paradox of XAI (Section~\ref{subsec:3.3}). Thus, XAI's limitations do not just pertain to implementation; rather, they demand a paradigmatic shift in our approach to AI transparency and trust. 

\subsection{Surface-Level Symptoms: Empirical Evidence}
\label{subsec:3.1}

Empirical evidence reveals four critical symptoms that challenge XAI's fundamental assumptions, suggesting current approaches fail to achieve their intended objectives across diverse user populations.

\textbf{\textit{Symptom 1: Empirical Evidence Contradicts XAI’s Effectiveness in Fostering and calibrating trust.}} The underlying assumption of XAI is that transparency yields greater trust; however, empirical research demonstrates that this relationship is complex. Studies of large language models show that explanations increase user reliance on both correct and incorrect responses, demonstrating that transparency does not automatically foster appropriate trust calibration \cite{67}. Furthermore, understanding a system's inner workings is not always necessary for user trust \cite{6}. The role of explanations is context-dependent: when failures align with user expectations, explanations become less critical for maintaining trust, but when expectations are low or failures are unexpected, explanations play a more substantial role in repairing trust \cite{70}. As some studies suggest, understanding AI algorithms is not even necessary for many professionals \cite{72}. Furthermore, for complex AI systems like LLMs, explanations are often unfeasible or misleading \cite{73}. Although these models may yield seemingly coherent “explanations”, they often misrepresent the underlying processes, which increases distrust in AI. Similarly, Turpin et al. \cite{74} argue that Chain-of-Thought (CoT) “explanations” often fail to capture the actual reasoning process behind AI decisions. 

The model faithfulness problem \cite{75,76}, namely the gap between explanations and what actually happens within the model, weakens AI systems’ trustworthiness in practical applications, as frequently observed in unreliable Chain-of-Thought reasoning. This faithfulness issue extends to feature attribution methods, where the standard definition—marking input features whose omission degrades model performance—is often considered a gold-standard substitute for "explanation correctness" but suffers from fundamental limitations. Specifically, XAI methods often highlight suppressor variables that are statistically unrelated to the prediction target but necessary for optimal model performance, providing confusing rather than informative explanations to end-users \cite{73}. This empirical evidence undermines the role of understanding model internals in developing user trust. However, explainability tools may still influence trust indirectly—when ML experts (correctly or incorrectly) find explanations useful for justifying model predictions, they gain confidence in releasing such models, which can indirectly impact non-expert end-user trust. 

% \begin{figure}[!tb]
% \centering
% \includegraphics[width=1.0\textwidth]{assets/F012.png}
% \caption{XAI Surface-Level Symptoms: Empirical Flaws Analysis}
% \label{fig:surface_level}
% \end{figure}

\textit{\textbf{Symptom 2: XAI Better Serves Those Who Were Not Its Primary Targets.}} According to DARPA, the target audience of XAI is end-users \cite{18}; however, empirical research shows that XAI methods have been largely beneficial to ML developers for technical debugging  \cite{18,51,77,78}. In fact, trust as a psychological state is often fostered through interaction and experience \cite{79}, rather than through technical understanding of the system’s internal workings. This is particularly evident in vulnerable populations, less-educated individuals, professionals from other fields, and those with low digital literacy \cite{80,81}. 

While end-user’s trust may not be the sole objective of XAI—as explanations for AI developers, deployers and auditors are also important—the fundamental issue remains that none of these groups are currently served well by existing XAI approaches. Despite the vast body of XAI methods created to bridge model complexity and explainability, a widely-accepted concrete problem that XAI should solve remains undefined, resulting in methods that lack theoretical and empirical evidence for explanation correctness \cite{73}. 

\textbf{\textit{Symptom 3: When Explanations Backfire.}} As research on algorithmic aversion suggests, users tend to lose trust in AI systems after observing even minor errors, regardless of how well those errors are explained \cite{82}. It is also shown that explaining AI mistakes after the fact doesn't rebuild user trust in some cases \cite{83}. Transparent explanations of AI systems' internal processing can create cognitive load problems that exceed human cognitive capacity, ultimately preventing users from developing trust \cite{5,84,85}; it may even be counterproductive \cite{68}, potentially misleading users and increasing distrust \cite{69,71}. Moreover, recent research on chatbot apologies reveals cases where more explanations lead to less trust. In bias scenarios, explanatory apologies were perceived as ``justifying the chatbot's sexist behavior rather than taking responsibility,'' with participants interpreting responses as ``deflecting blame onto external factors, such as the training data'' rather than accepting accountability. Providing detailed ``explanations'' alerts users to system limitations, diminishing their trust \cite{86}, which can be viewed as a failure (for increasing trust) or a success (for understanding when not to trust the system \cite{RudinEtAlSurvey2022}). 

\textbf{\textit{Symptom 4: XAI Fails Those Who Need It Most.}} “Different user types require different types of explanations” \cite{5}. XAI struggles to make AI decisions understandable to non-specialists, such as children, less-educated individuals, and those with low digital literacy \cite{81,87,88,89}. 
For many of these groups, trust may be fostered either by relying on domain experts or by practical experience and interaction that helps users build their own mental models of the system. Moreover, XAI’s main approaches suffer from biases that can propagate and amplify sociodemographic disparities. Explanation methods exhibit significant disparities in fidelity and accuracy across sensitive attributes like race and gender \cite{94}.  Rather than functioning as corrective tools that surface potential biases, XAI explanations often mask undesirable behaviors—failing to identify social biases behind clearly biased decisions \cite{95, deck2024critical}—and thereby reinforcing rather than mitigating harm to vulnerable populations. In some cases, explanations can make models seem to depend on sensitive variables like race when, in fact, they depend on less-sensitive variables (like age) due to the correlation between sensitive and less-sensitive variables \cite{RudinWaCo2020}.

\subsection{The Root Causes: Deep-Superficial Paradox, Conceptual Confusions and False Assumptions}
\label{subsec:3.2}

The failures cited above are symptomatic of deeper theoretical limitations in XAI. The root problem consists of three interconnected categories: an incompatibility between the faithfulness of the explanations and their comprehensibility, two conceptual confusions about explanatory levels and causation, and five false assumptions about trust and knowledge formation. These are not implementation limitations but structural issues, rendering XAI conceptually incoherent.

\subsubsection{The Deep-Superficial Paradox}
A central challenge in XAI stems from a paradox between two types of explanations for complex models — deep technical explanations and superficial simplified ones \footnote{The distinction between "deep, technical explanations" and "superficial, simplified ones" may require clarification. Superficial explanations may include simple bar charts showing feature importance or simplified explanations using proxy linear regression models with relevant features that align with end-users' knowledge but are not faithful to the inner mechanism of the original model. However, "deep" explanations (if any) refer to technical explanations of the original model that are both genuinely scientific and yet not inherently interpretable. (If a model is inherently interpretable—like a faithful, well-specified linear regression with relevant meaningful features—it requires no additional XAI efforts, as the model's decision process is transparent by design. That being said, while seemingly similar trade-offs exist in the literature \cite{46}, this approach differs and stems from logical issues that will be discussed in Section~\ref{subsec:3.3}.)
}. On the one hand, deep explanations aim to accurately capture the complex and intricate workings of AI’s underlying black-box algorithms. These explanations provide a more faithful representation of the system’s decision-making processes, offering insight into how data is handled, examined, and applied to generate predictions or decisions. However, the complexity of these explanations often results in cognitive overload for users. This often makes deep explanations accurate but practically inaccessible to most users.

\begin{figure}[!ht]
    \centering
    \includegraphics[width=0.8\linewidth]{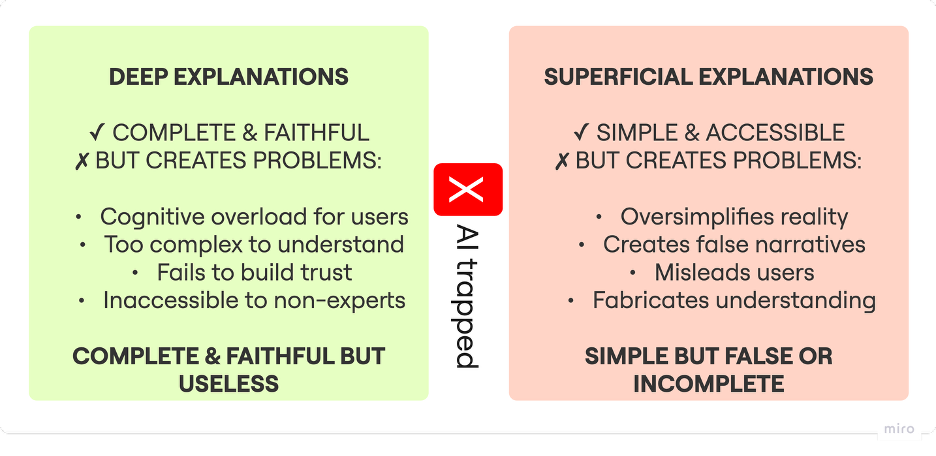}
    \caption{The XAI Paradox: Trapped in a Dilemma}
    \label{fig:paradox}
\end{figure}

On the other hand, some superficial “explanations” simplify AI processes to make them more comprehensible and user-friendly. While this approach improves accessibility, it comes at the cost of completeness or faithfulness - as attempts to make them understandable result in simplifications of their processes, which inevitably entail superficial “explanations” that cannot do justice to the intricate underlying processes. The simplification often leads to incomplete (too little information) or even fabricated (unfaithful) narratives that do not truly represent the AI’s decision-making process. Instead, users are given “explanations” that might be easy to understand but fail to reflect the real complexity of the system’s internal mechanisms. As a result, these “explanations” create a false sense of understanding, where users believe they grasp the system’s processes when, in reality, they are being misled by oversimplified accounts (see Figure~\ref{fig:paradox}).

This paradox in XAI exposes a fundamental tension: deep explanations are more faithful and complete but not understandable, while superficial “explanations” are understandable but not accurate. Therefore, XAI is stuck in a situation where it cannot simultaneously achieve both goals, failing to deliver on its promise to provide explanations that are both truthful and useful.

\subsubsection{Level Confusion of “Explanation” in XAI: From Black-Box to Behavioral Box}
In explaining an AI system, just as in explaining human intelligence, a distinction between three levels helps to avert some confusion: biological/algorithmic, psychological, and behavioral. The effectiveness of explanations depends on the target audience: developers benefit from algorithmic insights, domain experts require behavioral explanations aligned with professional standards, while end-users need actionable outcomes. Current XAI approaches have been focused on algorithmic and psychological “explanations.” After demonstrating that the first two levels fail to achieve understanding and foster trust, we propose moving beyond the algorithmic black-box and psychological pseudo-explanations to a "behavioral box" approach, which focuses on the third level, enabling us to interact with the system, verify its outputs, and gain a practical understanding.  

At the \textbf{biological/algorithmic} level, both human and AI intelligence are often too complex to be sufficiently explainable and rendered transparent, as they involve processes that are inherently opaque and act as black-boxes \cite{96}. For humans, the biological neural networks comprising human brains and the emergence of consciousness remain largely mysterious and incompletely understood \cite{97}, yet trust in the human minds’ outputs (e.g., decisions or beliefs) remains contingent and not impossible -- or at least there is more trust in humans than in AI models in many cases. Similarly, AI’s deep learning models are inscrutable at the algorithmic level. As \citet{98} note, trying to explain AI’s decision-making at this level leads to oversimplifications and misunderstandings. In fact, understanding this level is not necessary for user trust, as attested by the phenomenon of algorithmic aversion. Research indicates that acceptance levels vary depending on some pragmatic features such as task difficulty \cite{99}, domain specificity \cite{100} dimensions of decisions \cite{6,101}, and users’ ability to modify algorithmic recommendations \cite{102}, rather than on users' comprehension of each detailed calculation involved in algorithmic mechanisms. While valuable for developers and researchers, algorithmic-level explanations are sometimes inappropriate for end-users. Thus, ``explaining'' the internal mechanism of AI is not often necessary to build trust —even if such faithful explanations were possible. Nevertheless, we affirm the importance of fundamental research in biology and algorithms, a topic we explore in the sections on AI Epistemology (\ref{subsec:4.2}) and model-centered interpretability (\ref{subsec:4.4}).

While AI systems and human cognition operate through different processes at the “\textbf{psychological}” level, using a simplified common-sense approach can be useful for our goal in this study, even though it has limitations. Under this simplified framework, human intelligence can be understood by examining emotions and intentions that psychological experiments can reveal \cite{103,104}; Meanwhile, AI systems—especially LLMs—appear to replicate or simulate these pseudo-psychological "traits" by generating text and processing information \cite{105,106}. In domains like lie-detection or comprehension modeling, the complexity of mental states can make "explanation" models useful for ML engineers \cite{107,108}, with research detecting deception across contexts—from fake product reviews \cite{109} to political misinformation \cite{110} and user responses to false content \cite{111}. 

However, the risk of explanatory falsity at the shallow level for end-users is even greater than at the deep level. A recent study has demonstrated that "explanations” can lead to unwarranted trust or confidence in the model. Additionally, there exists ``the concern of illusory understanding, with which one subjectively over-estimates the understanding they gain from XAI'' \cite{112}. Even more concerning, studies show that trust levels in explanations were comparable, regardless of whether the explanations were genuine or were placebos containing no useful information \cite{113}. 

Finally, at the \textbf{behavioral} level, human actions and AI outputs are both observable and measurable \cite{114,116}. This represents a move from the traditional ``black-box'' problem to a ``behavioral box'' solution: in this approach, contrary to current post-hoc XAI methods, transparency is achieved not by exposing hidden mechanisms, but through consistently observable and verifiable outputs. As we will elaborate in Section~\ref{subsec:interactive-verification}, rather than requiring user understanding, this approach establishes certification pathways through qualified domain experts. It also aligns with the recognition that ``users' goal with XAI is not an understanding defined in a vacuum, but an actionable understanding that is sufficient to serve the objective that they seek explanations for'' \cite{112}. A pragmatic solution does not always necessitate explainability of black-box models,  nor does it necessarily always require trust in AI. Instead, we need can verify the output before relying on it. To establish reliable AI deployment, there is not necessarily a need for end-user explanations of psychological or algorithmic processes. Instead, systematic verification of consistent performance by qualified experts can provide a foundation for some trustworthy AI systems. Users can benefit from these verified outputs through established professional trust relationships, just as patients rely on medical expertise without understanding pharmacological mechanisms. Keeping with the human analogy, to trust an individual, we don’t need to know their biological/neural processes or their deep psychological states. What is more, work in human-computer interaction indicates that effective teamwork between humans and machines arises when operators develop a ``clear understanding of the automation’s functioning'' \cite{117}. This highlights the significance of context-sensitive interaction and expert-mediated verification protocols (rather than explanatory) strategies for trust or reliance building. This is evidenced by the fact that studies on user adoption of automation rarely distinguish between black-box and transparent systems \cite{118,119}. These findings imply that inner processes of AI systems may sometimes not be particularly relevant to user acceptance. Furthermore, research by \citet{69} shows that user decision-making with AI can improve when they concentrate on observable and actionable outputs instead of the underlying AI mechanisms.

In the case of humans, confusion of these explanatory levels leads to confabulations, such as explaining one’s choice of a seat in terms of comfort or lighting, while it is, in fact, due to, say, hand dominance \cite{122}. This is an example of divergence between motivating and explanatory reasons \cite{122,123,124}. In  LLMs and AI systems, confusion of these levels can result in hallucinations. In AI systems, hallucinations result from training and evaluation systems that reward confident guesses over expressions of uncertainty \cite{125}, leading to an LLM/AI system explaining its choice of an answer to a question in terms of a specific kind of reasoning, while it is indeed grounded in training data and algorithms. For humans, there is a correction mechanism through scientific community interactions and feedback, whereas such feedback is often absent in the case of LLM/AI. Such hallucinations may appear coherent and yet are disconnected from reality. This is even worse in the context of XAI, where there is a mismatch between the level (often the algorithmic one) where “explanations” are produced and the level (often psychological or behavioral) where they are interpreted. Although most XAI techniques focus on the algorithmic level (offering “explanations” in terms of algorithmic features that impact the system’s decisions), these outputs are often articulated in descriptive terms, involving psychological states and processes (e.g., belief and reasoning articulation in LLMs) to the system. This invites users to project human-style rationality onto statistical operations. Worse still, because these “explanations” are not grounded in scientific principles or logical consistency, they can be internally plausible but factually false—hallucinated rationales that offer reassurance while masking the true causes of output. Empirical evidence confirms this problem: in a human study of news claim verification, 25\% of GPT-4 generated explanations provided incorrect rationales for correct news reliability labels \cite{126}, demonstrating how AI systems can arrive at right answers through wrong reasoning. Similarly, in mathematical reasoning tasks, LLMs frequently generate step-by-step solutions that include unnecessary or false intermediate procedures, with research showing that weaker models can even produce training data with higher false positive rates yet still improve reasoning performance \cite{127}—suggesting that the explanations describe neither the model's actual process nor valid reasoning about the world.

Unlike human agents, generative AI systems lack built-in science-sensitive corrective loops—there is no social reality check, no discomfort when their errors are exposed \footnote{RLHF (Reinforcement Learning from Human Feedback) might indeed be considered a form of "discomfort when caught out," as systems receive negative rewards when their responses are not preferred compared to alternatives. That being said, one might make a distinction by arguing that "human preference" is a mixed affective-cognitive attitude and is not functionally equivalent to a cognitive-based "science-sensitive corrective loop."}, no internal standard for coherence beyond statistical optimization. Thus, XAI may not only fail to prevent hallucination but may actively reinforce it, dressing up algorithmic correlations in the language of reason, and doing so in ways that bypass our critical thinking. This pattern manifested dramatically in Google's AI Overviews in 2024, which provided dangerous misinformation (advising users to eat rocks, use glue on pizza, or run with scissors) while offering seemingly logical explanations—describing "health benefits" of rock consumption or explaining how glue could improve pizza "tackiness." The system didn't merely generate false facts; it constructed entire explanatory frameworks to justify these spurious correlations \cite{goodwin_google_ai_overviews_2024}. In short, the uncritical use of explanatory tools across mismatched levels doesn’t just obscure the inner workings of AI—it builds an illusion of transparency that deepens our misunderstanding, and may potentially cause harms.

\subsubsection{Correlation-Causation Confusion in XAI}
\label{subsec:3.2.3}
XAI also suffers from fundamental confusion about the distinction between correlational and causal relationships. This correlation-causation problem manifests regardless of whether XAI tools are explaining model internals or purporting to explain real-world phenomena—creating distinct issues that intersect with the model-versus-world distinction discussed in the following subsection. XAI is confused about the nature of scientific explanation. In scientific contexts, “explanation” refers to specific types of reasoning—a priori, a posteriori, logical, mathematical, and empirical—that aim to establish factual relationships between what is being explained (the explanandum) and what does the explaining (the explanans) \cite{60,128}. For instance, the Deductive-Nomological Model characterizes explanations as deductions in which various phenomena (explanans) follow from nomological principles (laws) and certain initial conditions. It thus presents causal mechanisms underlying those phenomena, not just patterns \cite{129}. When XAI speaks of explaining AI outputs in terms of its internals, it tends to treat the relation between the two as similar to that between a scientific fact and a scientific explanation, whereas AI systems like LLMs and DNNs rely on factors like frequency patterns and probability distributions, rather than induction, deduction, or experimental methods that constitute legitimate scientific reasoning. Their capacity for genuine logical reasoning is disputed \cite{130}. More fundamentally, foundation models might not deduce general causal theories. They excel at training tasks yet can fail to generalize underlying principles to new contexts. For example, models trained on orbital trajectories can fail to apply Newtonian mechanics to new physics tasks, instead relying on task-specific heuristics that do not generalize \cite{131}. They may, however, provide assistive second opinions \cite{132,133}.

This confusion shows up clearly in widely used XAI tools that blur the line between correlation and causation. Take LIME, for instance. If LIME highlights “age” as an important feature in a loan approval decision, all it’s doing is pointing to a potential statistical association. It tells us nothing about whether age actually causes differences in creditworthiness, whether it’s a stand-in for some other variable, or whether the model is merely reflecting historical bias in the data \cite{134,135}. The correlation–causation problem in XAI is more complex than commonly assumed because XAI tools often conflate fundamentally different kinds of relationships. Some are classical statistical correlations—direct causation (X→Y), reverse causation (Y→X), confounding (Z→X,Y), and selection bias (X→Z←Y)—which, despite complicating causal interpretation, still reflect real associations in the data. More troubling are XAI-specific pseudo-correlations, such as suppressor variables that appear important despite having no true correlation with the outcome, training artifacts that arise from quirks in dataset construction rather than real-world patterns, and noise-induced artifacts where adversarial or random inputs are mistakenly treated as meaningful. In fact, the situation is even more complex and potentially misleading than simple statistical association. XAI tools like LIME may highlight variables that are not even statistically associated with the target outcome but are nonetheless crucial for the model’s performance—so-called suppressor variables, where the model learns to use variables that are uninformative with respect to the target in order to denoise other, truly informative variables. Highlighting these instances is frequently more important than identifying genuine correlations \cite{15,73}. For instance, age might show up as "important" in a loan default prediction model not because age predicts default, but because the model uses age to control for age-related effects in other variables that do predict default. The model might subtract age-related variance from truly predictive features, making age appear significant to XAI tools despite having no direct relationship with loan default. This phenomenon reveals that XAI explanations can be misleading by highlighting variables that have no statistical relationship whatsoever with the predicted outcome.  Furthermore, studies have shown that XAI explanations are susceptible to various forms of instability and can be manipulated by adversarial inputs: these "explanations" are often artifacts of random noise rather than meaningful insights into model behavior \cite{137,138,139}. An example discussed earlier shows that XAI failures can have serious consequences; for instance, a low quality explanation can suggested that a criminal justice model depends on a person's race when it instead likely depends on their age, which is correlated with race in one dataset \citep{RudinWaCo2020}. 

A further clarification comes from work in causal discovery and causal inference \cite{140,141,142}, which delineate the boundary XAI often blurs. In these frameworks, an explanation is tied to relations that would remain invariant under interventions—claims about what would change if a variable were actively manipulated—rather than to patterns of co-variation alone. However, causal methods remain limited when applied to XAI: any causal claim depends on substantive assumptions (about confounders, model class, measurement) as well as experimental or quasi-experimental designs; in high-dimensional DL settings these conditions are often nontrivial to meet. The upshot is not that causal techniques “fix” XAI, but that they reframe what counts as an explanation: associational attributions can at best serve as hypothesis-generating hints, whereas stronger explanatory claims are warranted only when supported by evidence of causal structure and robustness across environments. As a result, these so-called “explanations” are not just scientifically invalid; they might actively mislead users into thinking that they have causal insight when they don’t.

% \paragraph{XAI, by default, “explains” the model, not the world}

\noindent\textbf{\textit{3.2.3.1. XAI, by default, “explains” the model, not the world}}

Building on the correlation-causation issues discussed above, there is a distinct and deeper problem: whether XAI explanations pertain to model internals or real-world phenomena. This model-versus-world distinction cuts across both causal and correlational relationships, creating four distinct types of explanatory issues. First, XAI tools may (or may not) correctly identify causal relationships within a model's internal processing. Second, even if do, they may incorrectly present correlational patterns within the model as causal. Third, even if XAI correctly identifies causal relationships in the model's processing, it may wrongly imply that these translate to real-world causation. Fourth, XAI may present correlational model patterns as if they explained real-world phenomena (see Table~\ref{tab:2}). 

\begin{table}[H]
\caption{Model-World Explanatory Matrix in XAI} 
\label{tab:2}
{\fontsize{10}{11}\selectfont
\renewcommand{\arraystretch}{1.7} % Adjust row height
\setlength{\tabcolsep}{10pt} % Adjust column padding
\begin{tabular}{p{0.12\linewidth} p{0.37\linewidth} p{0.37\linewidth}}
\toprule
\multicolumn{1}{c}{\textbf{}}        & \multicolumn{1}{c}{\textbf{Model Explanations}}                                       & \multicolumn{1}{c}{\textbf{World Explanations}}                                              \\ \hline
\textbf{Causal  Relationships}        & Type 1: Correctly identifying causal relationships within model's internal processing & Type 3: Correctly identifying model causation but  incorrectly implying real-world causation \\
\textbf{Correlational Relationships} & Type 2: Incorrectly presenting correlational patterns within model as causal          & Type 4: Presenting correlational model patterns as explanations of real-world phenomena   \\
\bottomrule
\end{tabular}
}
\end{table}

Consider, for example, an issue in gradient-based attributions such as Integrated Gradients (IG). In a neuroradiology protocol-assignment study with BERT, IG heatmaps highlighted tokens whose removal substantially degraded performance. This indicated that those tokens were important to the model’s internal decision rule; yet expert review showed that these attributions often missed clinically decisive concepts and instead emphasized generic or spurious terms. The maps looked like explanations but frequently misrepresented the model’s actual reasoning process, creating a false sense of transparency in a safety-critical setting \cite{143}. This leads to the central point: by default, XAI explains the model, not the world. It tells us how the model processed information—not why the phenomenon it predicted or classified actually occurs. For a mathematical example, if a large language model generates that "$\forall x\, (P(x) \to Q(x)),\ \forall x\, (Q(x) \to R(x)),\ P(a) \vdash R(a)$" it isn’t applying mathematical knowledge (like applying logic rules or model tolerance); it is probably just reproducing a pattern—or generalized form of that pattern \cite{144}—that appears frequently in its training data. If the data distribution changed, the output might too—which is not how real scientific explanation works. This highlights a fundamental shortcoming of large language models: rather than exhibiting genuine logical reasoning, they often operate as advanced pattern recognition systems \cite{145}. When XAI systems try to offer “explanations”, they often construct plausible-sounding stories that seem to explain world phenomena out of what are just model-internal patterns \cite{73}. XAI often leads users to believe that the AI system possesses genuine insight into real-world phenomena when, in reality, what has been revealed is merely which variables the model associates with its outputs—not actual understanding of the underlying world dynamics \cite{137,146}. Presenting such model-internal patterns as world “explanations” is neither professional nor scientific, and degrades XAI to resemble pseudo-science rather than a legitimate scientific field. 

We recognize that the explicit and direct objective of XAI might not be the explanation of world phenomena and values. One might view XAI research as a world-independent, model-centered endeavor aimed at illuminating the relationships between a model’s internal components and its variables. We will explore it further in Section~\ref{subsec:4.4} However, a great deal of XAI research does not exist in a purely theoretical, world-independent vacuum. While causality may not be an explicit goal, the practical application of XAI methods—in critical domains such as healthcare, climate science, and management—implicitly invites causal interpretations. This is frequently reflected in the language of "explanations," "reasons," and "decision factors”  which could naturally lead stakeholders to infer real-world causality.

Furthermore, if XAI is a completely world-independent and only model-centered field, what is the rationale behind its goal of building trust and securing adoption among high-stakes domain experts and investors, such as medical professionals? This goal—of ensuring the experts of real-world relevance fields—indicates something more than just model-centered theoretical interest. XAI products are commonly marketed as solutions for high-stakes, real-world decision-making. The field seeks trust from stakeholders in numerous fields, driven by major motivations such as marketing and deploying AI predictions, and ensuring compliance with regulations like the GDPR’s "right to explanation." This is a clear testament to XAI’s role as a bridge between AI systems and real-world societal use. Therefore, XAI is positioned as a mediating enterprise. Its ultimate function is to support AI's role in the world, which often involves engaging with causal relations in real-world phenomena and values, moving beyond a purely theoretical exercise.

% \begin{figure}[H]
% \centering
% \includegraphics[width=0.9\textwidth]{assets/P4.png}
% \caption{XAI's Correlation-Causation Confusion}
% \label{fig:4}
% \end{figure}

\noindent\textbf{\textit{3.2.3.2. Three Gaps in Deep Learning XAI}}

\begin{figure}
    \centering
    \includegraphics[width=1\linewidth]{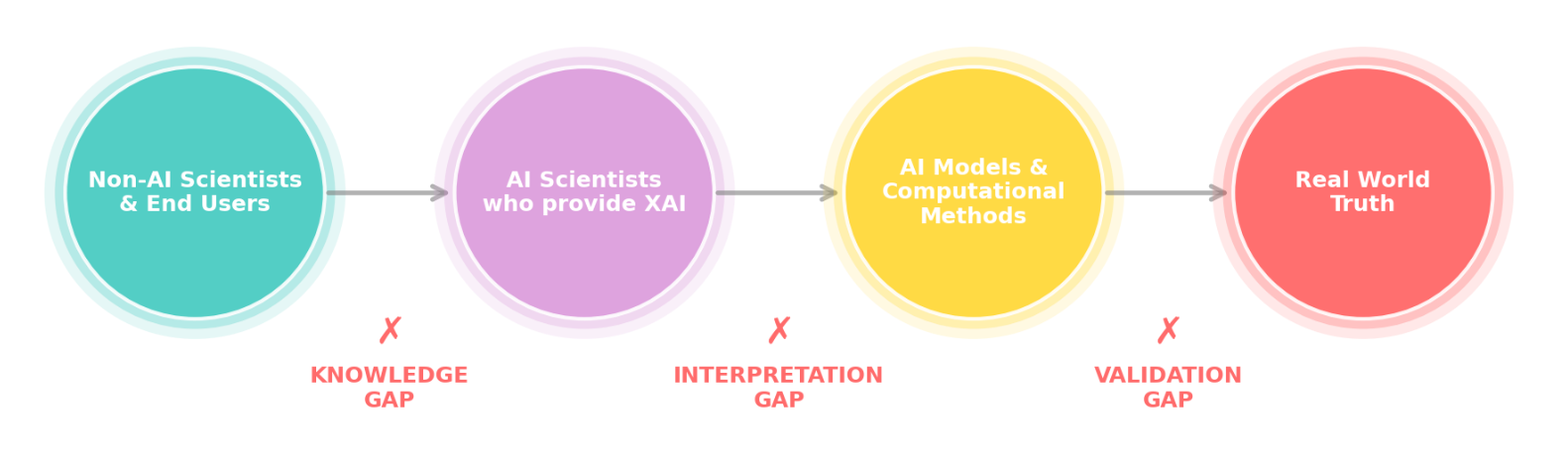}
    \caption{The Three Critical Gaps in XAI}
    \label{fig:gaps}
\end{figure}

This conceptual confusion is more evident in modern XAI methods and the way they treat deep learning (DL) models. Whereas traditional models on tabular data make decisions based on human-interpretable features, DNNs learn complex hierarchical patterns that are difficult to interpret. This gives rise to three key gaps between what the model learns and what actually causes things to happen in the world (see Figure~\ref{fig:gaps}). First, there’s the \textbf{Knowledge Gap}, which refers to the disconnect between the technical expertise of AI scientists and the domain-specific knowledge of the users (e.g., physicians or other domain experts). Whereas the former employ metrics such as SHAP values and feature importance, the latter need explanations in terms of their own frameworks. For instance, while AI scientists might use technical metrics like Grad-CAM or saliency scores to explain CNN-based models \cite{147}, pathologists require explanations grounded in their domain-specific frameworks, such as histomorphological criteria and WHO classifications \cite{148}. That is, by revealing only \textit{where} the model might be looking, but not \textit{why} the model is looking there, the explanation leaves a critical gap that the user might try to fill. This creates a critical disconnect: the simplistic, visually intuitive saliency maps presented to pathologists are often ambiguous, leading them to "fill in the gaps" with their own assumptions and mistakenly attribute human-like reasoning to the AI—a phenomenon known as misleading anthropomorphism \cite{149}. As another example, XAI saliency maps provide no insight to radiologists reading mammograms that have identified breast lesions and are trying to decide whether to biopsy them; a high-quality saliency map simply highlights the lesion that the radiologist has already identified \cite{BarnettEtAl2021}.
These cases exemplify common domain-specific knowledge encoding issues, but the Knowledge Gap reflects a deeper divide; its proper resolution would yield perfectly interpretable AI and eliminate the need for post-hoc XAI entirely \cite{21}.

The second one is the \textbf{Interpretation Gap}, which consists of the disconnect between the “explanations” of a black-box model's behavior provided by XAI methods and the system’s actual internal processes. Since XAI methods employ proxy models with values and variables that differ from those of the original black-box model, these “explanations” may misidentify the model’s true decision-making process. Different XAI methods can yield conflicting “explanations” for the same output \cite{23}, and even an identical XAI method can produce inconsistent explanations when applied to structurally similar inputs with identical predictions \cite{150}. Moreover, black-box models may rely on features or patterns that current XAI proxy models cannot detect or faithfully represent. For example, gradient attribution may highlight specific pixels in medical images, causing radiologists to interpret these regions as diseased tissue, when the model is actually responding to invisible imaging artifacts or metadata patterns that the XAI method fails to capture \cite{151}. Furthermore, this gap compels humans to decide what a given explanation might mean, and unfortunately, there is a human inclination to attach a positive interpretation: we assume that the characteristic we would deem significant is the one the model relied on—an example of the cognitive error called confirmation bias \cite{59}. Third is the \textbf{Validation Gap}, which refers to the disconnect between statistical correlations identified —whether based on relevant or irrelevant variables— by black-box models and causal relationships required for real-world scientific decision-making. What domain scientists may need are causal (or causal-conducive) connections, grounded in scientific knowledge of their domain, rather than mere statistical correlations. For instance, an AI system for predicting clinical outcomes might find correlations between socioeconomic status and treatment response rates. This can mislead clinical decision-making for physical treatments, which require causal explanations involving disease pathophysiology and underlying biological mechanisms \cite{149,152} This conceptual conflation—between genuine scientific explanation (which relies on induction, deduction, and empirical testing within interpretable models whose mechanisms align with domain knowledge) and XAI's statistical pattern recognition through unfaithful or partially faithful proxies of black box models — creates a harmful illusion. It gives developers, researchers, and stakeholders the false sense that they understand how the world works when they have only learned, to some extent, how the black-box model behaves \cite{153,154}. 

Furthermore, this terminological confusion obscures the genuine value of what current XAI methods actually provide: many of these tools function effectively as model debugging aids, or collaborative diagnostic systems that help developers improve model performance through iterative questioning and analysis \cite{155}. These tools, mechanistic diagnostics, and collaborative improvement methods have legitimate scientific value and (as we discuss in Section~\ref{subsec:4.4}) should be recognized as they are, instead of being mislabeled generally and generically as "explanations" \cite{156}.

\subsubsection{Debunking Five Core False Assumptions of XAI}

\begin{figure}[!ht]
    \centering
    \includegraphics[width=1.02\linewidth]{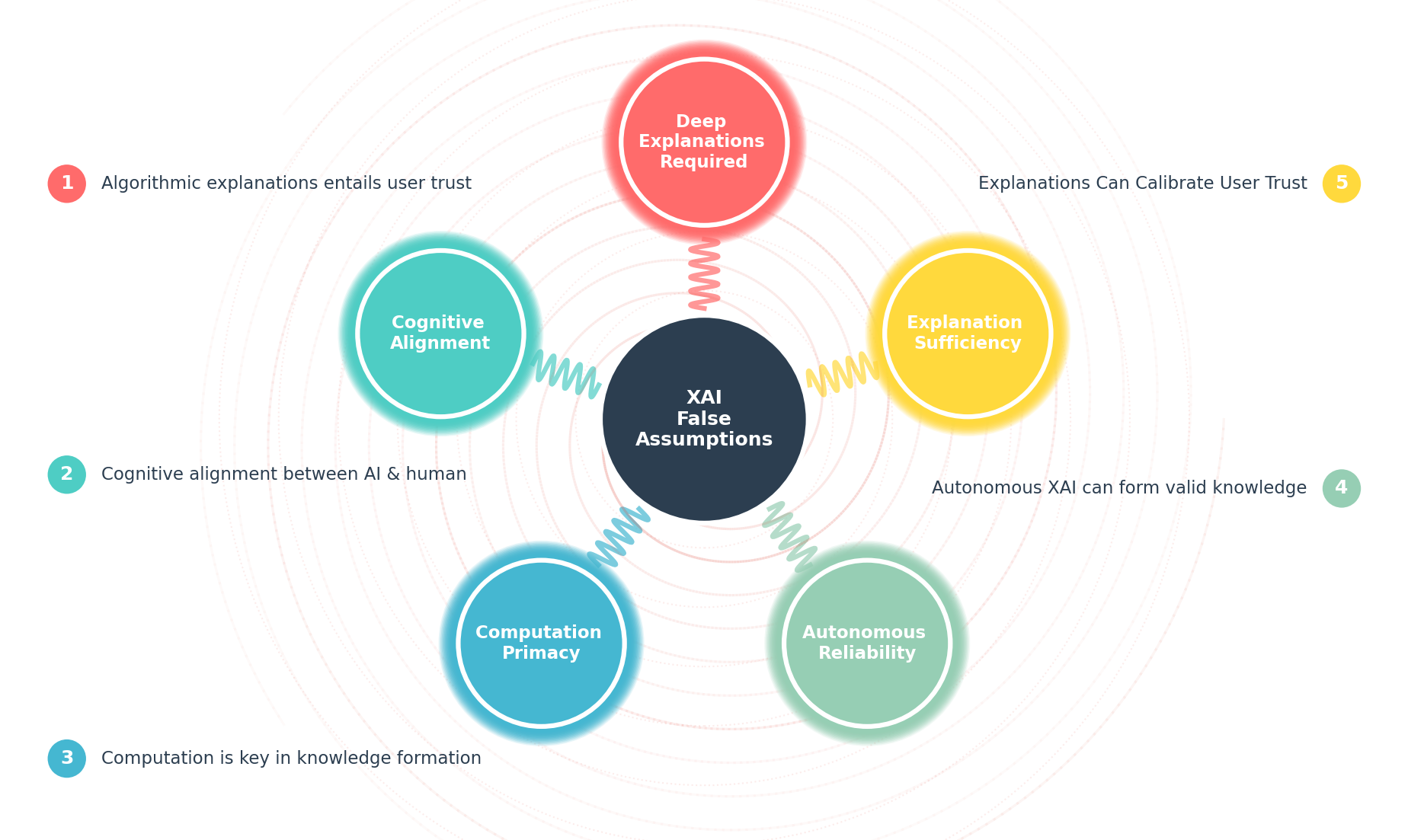}
    \caption{Five False Assumptions of XAI}

      \label{fig:assumptions}
\end{figure}

XAI doesn’t just suffer from a conceptual dilemma and confusions— empirical evidence and etiological analysis reveal that it rests on, at least, five key assumptions about trust, transparency, and how people come to understand things that simply don’t hold up to evidence (see Figure~\ref{fig:assumptions}). As will be elaborated in the following, these assumptions misrepresent how people actually build trust in AI, and this calls the primary motivation behind XAI’s approach into question.

\noindent\textit{\textbf{First assumption}: deep algorithmic “explanations” in XAI are always necessary or beneficial for trust}

The field of XAI has operated on the idea of mitigating and elucidating the black-box, with the core assumption that transparency is a prerequisite for trust \cite{157,158,159}. Yet, this view is contested, and some have argued in defense of black-boxes \cite{160}. Several empirical studies show that such algorithmic transparency may not requisite or necessarily beneficial for trust  \cite{68, 161, 162, 163, 164} (see also Section~\ref{subsec:3.1}). Some empirical studies find that people’s acceptance of AI decisions has nothing to do with whether or not AI systems are black-boxes; rather they pertain to factors such as task difficulty, domain context (particularly health applications), moral implications, and ability to modify algorithms \cite{82, 165}. This is also reinforced by empirical research on user adoption of automated decision-making, where system architecture is not considered, implying that the black-box nature itself may not always be the primary barrier to trust \cite{165}. This is not only true of general users, but also of many professionals \cite{166}. Thus, it seems that trust doesn’t work as XAI assumes.

Although XAI looks neither necessary nor inherently beneficial for trust, it can, in some cases, foster trust contingently. There are many examples where explanations provided by developers—regardless of whether they are false and unfaithful—have increased feelings of trust in users. However, the crucial question is: do unfaithful explanations (by proxy models) lead to justified trust in the original models or misplaced trust? If models that perform well may still rely on superficial heuristics rather than deep structural understanding, then explanations may merely reveal the model's shallow strategies while obscuring its failure to grasp underlying principles, showing that we cannot trust it  \cite{131}.

\noindent\textit{\textbf{Second assumption}: there is a cognitive and methodological alignment between AI and human reasoning.}

Another assumption of XAI is that AI systems make predictions and do reasoning in ways similar to human reasoning, implying a cognitive alignment between AI methods and scientific methodology \cite{167}. However, significant divergence between AI’s information processing methods and human reasoning casts doubt on this assumption \cite{24}. For instance, human generalization typically involves abstraction and concept learning, while AI generalization encompasses out-of-domain generalization, rule-based reasoning, and neurosymbolic abstraction \cite{ilievski2025aligning}. XAI's attempt to interpret AI functions in terms of human-like qualities often anthropomorphizes AI systems \cite{168,169}, though the extent to which this occurs varies across different explanation methods and contexts. Standard scientific methodology involves rule-based hypothesis testing, experiments, and empirical observations to identify underlying causes of the studied phenomena. In contrast, the current AI does not follow these approaches to arrive at decisions but rather works through pattern detection and statistical relationships in large datasets. Thus, there is a significant methodological misalignment between AI processes and typical human reasoning \cite{170,171}.

First, it might be misguided to explain AI processes in terms of human reasoning. Since AI models reason using statistical correlations, it is not clear that those correlations would be understandable to humans at all. The reasoning pathway of the model may be so complex that shallow or incomplete explanations may not suffice -- even if the data generation mechanism itself is perfectly understandable to humans \cite{135, 176, 177}. This methodological incompatibility creates a fundamental dilemma for XAI research. Contemporary AI systems use data-driven, probabilistic reasoning that differs from traditional scientific methodology's causal, hypothesis-driven approaches. This creates two problematic paths for XAI: if it adheres to existing scientific frameworks, it cannot faithfully explain AI systems that operate on different principles. If it attempts to develop entirely new frameworks suited to AI’s unique methodology, it faces the impractical task of revolutionizing scientific methodology itself -- or a task far beyond XAI’s scope and timeline of current research \cite{178,179}. This demands a fundamental reconceptualization of what constitutes a valid scientific explanation in the age of machine learning.

Second, the XAI methods provide incomplete reasoning. \citet{21} shows an example where two different explanations are provided for an image of a dog. In the first case, the XAI algorithm was asked to explain the label of ``husky;'' in the second case, the XAI algorithm was asked to explain the label of ``flute,'' and it provided almost exactly the same explanation in both instances. Even if these explanations were faithful, they were incomplete, leading to the same ``reasoning'' process for different conclusions. 

Third, XAI methods are unfaithful to the underlying models.
Adebayo et al. \cite{176} demonstrated that certain gradient-based methods (particularly Guided BackProp) produce visually similar explanations even when model parameters are systematically randomized or when models are trained on randomly permuted labels, revealing that these methods lack sensitivity to the learned patterns they claim to explain. In other words, no matter what the underlying model is, the XAI algorithms provide the same explanations, which are thus unfaithful to the model. LIME's local linear approximation approach has specific documented limitations. Studies show that what LIME highlights may not necessarily be associational—demonstrating poor faithfulness between the proxy model and the original black-box model it purports to explain \cite{73,172}. Much research demonstrates brittleness issues in LIME and SHAP explanations \cite{173}. Other research shows perverse responsiveness patterns of XAI methods - hypersensitive to some irrelevant perturbations but hyposensitive to some important decisional changes, generating explanations that are simultaneously unpredictable and non-distinguishing \cite{174}. These are not characteristic properties of faithful explanations.

Fourth, XAI is difficult to troubleshoot. Let us consider an example. 
In a neuroradiology protocol assignment, a study using Gradients produced heatmaps that passed an erasure stress-test (removing top-attribution tokens sharply reduced F1), yet expert review showed these attributions frequently missed clinically decisive cues and overweighted generic or procedural terms \cite{143}. What went wrong here? Was the original model not using the correct features and the XAI method identified the problem? Or was the XAI method unfaithfully or incompletely representing the model's reasoning? Or, was the erasure test -- that was meant to verify the explanation -- incomplete, by identifying only a small subset of pixels, instead of a larger collection that is actually driving the predictions? Because the XAI explanations are incomplete and often unfaithful, it is difficult to troubleshoot the system. Not only do we need to troubleshoot the model, but also the XAI. Either one could be misbehaving, and we don't know why.

\noindent\textit{\textbf{Third assumption}: data-driven computation is a major part of human and AI knowledge formation.}

As we will elaborate more in Section~\ref{subsec:4}, XAI assumes that in intelligent agents (including humans and AI systems), knowledge is formed through computation. In AI, it is through algorithmic computation (e.g., data mining and autonomous optimization) that “valid” knowledge is attained (where "valid” knowledge refers to an accurate representation of the actual world, viewed through the lens of the model's intended applications \cite{180}.) On this assumption, trust is secured if information is yielded computationally \cite{52,181}.

However, there is a significant divergence between human and AI knowledge formation because communication (not computation) is key to human knowledge formation. Human agents attain knowledge largely through interactions with each other, such as with teachers, colleagues, or through reading, and not through purely data-driven computation. Human knowledge validation occurs through structured scientific community processes—peer review, expert verification, and systematic validation protocols. Dewey conceptualizes communication as a fundamental mechanism for knowledge formation, serving to transform individual experiences into collectively shared, public knowledge \cite{182}. Moreover, this third assumption faces a critical challenge in an AI-saturated information ecosystem. As AI-generated content proliferates across the internet, the epistemic value of computationally-derived knowledge—including XAI explanations themselves—becomes increasingly questionable. Research on model collapse demonstrates that genuine human interactions with systems become increasingly valuable precisely because LLM-generated content contaminates data crawled from the internet \cite{183}, creating a potentially circular problem where AI systems may be trained on and explain outputs generated by other AI systems. This shows that computation alone cannot guarantee valid knowledge formation; reliable AI deployment should similarly depend on systematic expert verification rather than reliance on opaque autonomous processes or the unfaithful explanations of proxy models. To generate reliable knowledge, generative AI systems must interact with human agents to create shared meanings, rather than just replicating human modes of communication. While autonomous computational processes may engender coherent ``scientific'' knowledge, it might be useless for human society or not measurable by human standards; for instance, it might not align with human values or goals.

\noindent\textit{\textbf{Fourth assumption}: autonomous XAI can form reliable and valid knowledge.}

According to this assumption, AI systems can form reliable and valid knowledge without ongoing feedback and engagement from designers or domain scientists. Leaving AI to function autonomously is to fall into the same black-box problem that necessitated XAI in the first place. Knowledge is fundamentally ``situated in people and in location, and that the social is an essential part of using any knowledge.'' The autonomous formation of knowledge by AI systems risks overlooking these crucial situated and social dimensions that are integral to how knowledge functions in real-world contexts. Given that knowledge and work are closely intertwined in social contexts, this verification cannot be purely technical but must account for the situated nature of how knowledge operates in practice \cite{ackerman2013sharing}. This is also attested to by the scientific practice involving peer review or discussions, as well as by HCI research (e.g., \cite{26}). Thus, for AI-generated knowledge to be reliable, it must be enhanced through systematic verification by human expert domain experts \cite{184}. This suggests that ongoing expert oversight and engagement are necessary in order for an AI system to stay aligned with human goals and values. XAI's proxy models cannot substitute for this expert verification—their explanations may appear intuitively meaningful to non-expert users while remaining unfaithful to what the original black-box model actually computes. Therefore, practical reliance on AI outputs must be grounded in verification protocols rather than in psychological trust generated by post-hoc explanations of uncertain faithfulness. (e.g., \cite{26}). 

\noindent\textit{\textbf{Fifth Assumption}: Explanations Can Calibrate User Trust.}

The fifth false assumption by XAI is that “explanations” guide users toward proper trust calibration. Some experiments in cognitive science, however, show that “explanations” might exacerbate overreliance, despite users having access to ``transparent'' explanations \cite{185}. The study found that ``explanations are interpreted as a general signal of competence—rather than being evaluated individually for their content—and just by their presence can increase the trust in and overreliance on the AI'' \cite{185}. This aligns with decades of automation bias research: passive “explanations” function as cognitive Band-Aids, failing to counteract the human tendency to defer to AI systems, even when they err \cite{186}. The research demonstrated that ``when the AI model was making incorrect predictions, people performed best in conditions that they preferred and trusted the least'' \cite{185}. Also, some clinical evidence supports this concern. In a neuroradiology study, analysis of GPT-4’s free-text rationales showed that detailed, fluent explanations were treated as a cue of competence, prompting deference even when recommendations were wrong. The most concerning flaws were arbitrary errors, where the model confidently recommended non-standard protocols on nonsensical grounds and rarely signaled uncertainty. The very presence of a sensible explanation can increase overreliance, creating conditions for unsafe decisions and potential malpractice \cite{187}. True trust calibration requires interactive safeguards, such as forcing functions that mandate active user verification (e.g., ``Why might this AI recommendation be wrong?'') or friction mechanisms that prevent blind acceptance. Unless these requirements are met, XAI’s deployment and bias could worsen rather than mitigate overreliance \cite{85,188}.

\subsection{XAI’s Formal Logical Limitations}
\label{subsec:3.3}

The symptoms and root problems of XAI raise a critical question about the field's logical limitation and potential for reform.  A critical limitation of XAI is that its post-hoc, after-the-fact explanations are generated after model development rather than interactively during it, resulting in unfaithful and/or incomplete explanations disconnected from the model’s actual behavior.  A deeper formal logical analysis reveals that XAI might face a fundamental theoretical obstacle that strikes at its conceptual foundations: a Russell-type logical paradox inherent in its proxy model architecture.

\subsubsection{The Proxy Model Paradox}

The architectural foundation of XAI rests on a proxy model structure that generates a self-referential logical contradiction analogous to Russell's paradox in set theory. XAI typically employs a two-model architecture where an original black-box model $M_1$ is explained through a proxy model $M_2$ designed to approximate $M_1$'s variable-value mappings. This structure raises the question: does the proxy model itself require explanation?

Formally, let us define the set of incomplete proxy models as
\[
P = \{\, M_2 \mid M_2 \text{ does not completely and faithfully match } M_1 \,\},
\]
where \emph{completeness} refers to sufficient detail about $M_1$'s computational
mechanisms and \emph{faithfulness} refers to accurate representation of $M_1$'s
actual computations. That is, for completeness,
\[
\forall x \in M_1,\ \exists y \in M_2 \text{ such that $y$, in sufficient detail, represents $x$'s computational mechanism},
\]
and for faithfulness,
\[
\text{internal process}(M_2) \approx \text{internal process}(M_1).
\]

This leads to the following two cases:

% ---------- CASE 1 ----------
\paragraph{Case 1: $M_2 \in P$ (Partial Explanation).}
When $M_2 \in P$, the $M_2$--$M_1$ relationship remains opaque, necessitating
a third model $M_3$ to explain their relationship. This creates two
possibilities:

\begin{itemize}
    \item \textbf{$M_3 \notin P$:}  
    $M_3$ completely explains both $M_2$ and $M_1$, making all models
    interpretable and rendering the entire XAI
    architecture unnecessary—contradicting XAI's foundational architecture and
    leading to self-elimination.

    \item \textbf{$M_3 \in P$:}
    $M_3$ requires $M_4$, which requires $M_5$, initiating infinite regress:
    \[
    \forall n \in \mathbb{N},\quad M_n \in P \;\to\; \text{a further model } M_{n+1} \text{ is required},
    \]
    producing the sequence
    \[
    M_1 \xleftarrow{\;\varphi_1\;} M_2 
    \xleftarrow{\;\varphi_2\;} M_3
    \xleftarrow{\;\varphi_3\;} \cdots
    \xleftarrow{\;\varphi_n\;} M_{n+1}
    \xleftarrow{\;\varphi_{n+1}\;} \cdots
    \]
\end{itemize}

% ---------- CASE 2 ----------
\paragraph{Case 2: $M_2 \notin P$ (Complete Explanation).}
When $M_2 \notin P$, the proxy model completely and faithfully explains $M_1$,
making both models interpretable and
rendering the proxy architecture superfluous—again contradicting XAI's
foundational architecture and leading to self-elimination.

% ---------- PARADOX ----------
\paragraph{The paradoxical logical limitation:}
\[
(M_2 \in P \;\rightarrow\; \text{Self-elimination} \;\lor\; \text{Infinite Regress})
\;\;\land\;\;
(M_2 \notin P \;\rightarrow\; \text{Self-elimination}).
\]

Thus, the proxy model appears to face a challenging dilemma. Success requires its own elimination, while failure leads to an infinite regress, indicating that XAI's foundational approach faces the challenge of being inherently self-contradictory.

\begin{figure}[H]
    \centering
    \includegraphics[width=1\linewidth]{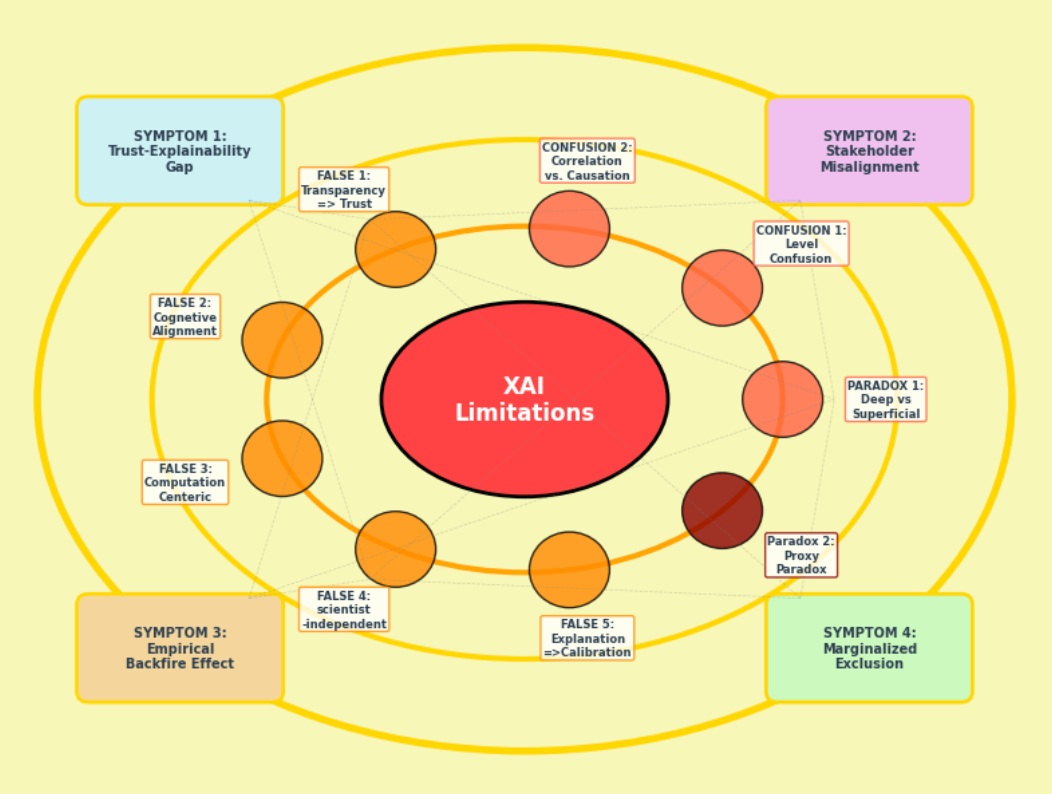}
    \caption{Incurability of XAI with 13 Critical Issues: 4 Surface-Level Symptoms and 9 Root Causes (2 Paradoxes + 2 Conceptual Confusions + 5 False Assumptions)}
    \label{fig:6}
\end{figure}

The empirically observed symptoms, underlying root causes, and formal logical limitation outlined above place XAI in a fundamental predicament (see Figure~\ref{fig:6}). These deep-seated issues raise questions about future attempts to reform the framework, particularly given the risk that efforts to justify a paradoxical foundation may only amplify its misleading potential. 

% \begin{figure}[H]
% \centering
% \includegraphics[width=1.00\textwidth]{assets/F0007.png}
% \caption{Incurability of XAI with 13 Critical Issues: 4 Surface-Level Symptoms and 8 Root Causes (2 Paradoxes + 2 Conceptual Confusions + 5 False Assumptions)}
% \label{fig:6}
% \end{figure}

\subsection{Moving Beyond XAI: From Post-hoc Explanation to Interactive Verification}

Excessive focus on XAI might create the illusion that black-box AI systems can be trusted merely because they provide explanations via proxy models, even when they lack systematic expert verification and scientific community oversight. %For instance, the COMPAS risk assessment tools were autonomously used in courtrooms nationwide because their use was justified through such post-hoc “explanations” \cite{189}. 
%These tools may mask racial bias and incorporated legally irrelevant factors \cite{21}. 
For instance, some large corporations deployed hiring algorithms without human oversight, as “explanations” suggested algorithmic fairness. However, this led to significant discrimination \cite{190}. In other cases, toxic language detectors flag benign content mentioning groups like Muslims due to dataset artifacts—XAI methods would identify these demographic terms as important features, masking rather than exposing spurious correlations  \cite{191}. In healthcare, the “explanations” of sepsis prediction models caused physician overreliance and unsupervised use—later revealed to diagnose sepsis not by learning medical pathophysiology but by detecting hospital workflow patterns \cite{192}. Explainability also has a temporal dimension—systems that appear explainable today may not remain so tomorrow, as use needs evolve and contexts change \cite{corti2024moving}.

These limitations are not restricted to high-stakes domains but also stem from how explainability and interpretability tools are used. As \citet{193} show, explainability tools like SHAP and interpretability tools like InterpretML can be misunderstood, misused, or over-trusted. Even trained professionals may fail to interpret model visualizations correctly or recognize their epistemic limits, treating explanations as evidence of understanding when they merely reproduce model artifacts. This creates a “false interpretability comfort zone,” reinforcing the illusion of control and transparency without enabling genuine verification or error diagnosis; this false sense of accountability could prevent the development of rigorous verification protocols that could actually ensure reliable AI deployment. As a result, most current generated “explanations” lack performance guarantees and are rarely evaluated using proper testing methods \cite{59}. 

Beyond these flaws in single-model systems, XAI's limitations intensify dramatically in agentic AI, where multiple agents interact non-deterministically. Emerging agentic AI architectures expose even more fundamental limitations in the explanatory paradigm. Agentic AI workflows—where multiple reasoning LLM agents are chained together, each invoking tools, plugins, and external interfaces—operate with varying degrees of autonomy, permitting non-deterministic behavior where agents dynamically choose which tools to call, in what sequence, and when to return control. 

This architectural complexity shifts the risk surface from model-level to system-level \cite{194}, where safety and security issues emerge from interactions among agents, tools, and control flows—hazards invisible to isolated model evaluations or traditional XAI post-hoc explanations. Under non-determinism, where the set, order, and frequency of external calls are not pre-specified, XAI cannot reliably capture the true execution path. Post-hoc explanations inevitably oversimplify these dynamic multi-agent interactions, creating an illusion of transparency that masks real hazards including security incidents, data leakage, and cascading failures. Effective risk management for agentic systems demands a shift from explanation to verification: standardized observability frameworks (e.g., OpenTelemetry), comprehensive, tamper-resistant audit trails, runtime policy enforcement, and tooling capable of capturing dynamic agentic behaviors—capabilities that align precisely with an Interactive AI approach \cite{194,195}.

\section{Navigating the Post-XAI Paradigm: Emerging Research Directions}
\label{subsec:4}

\begin{figure}[H]
    \centering
    \includegraphics[width=1\linewidth]{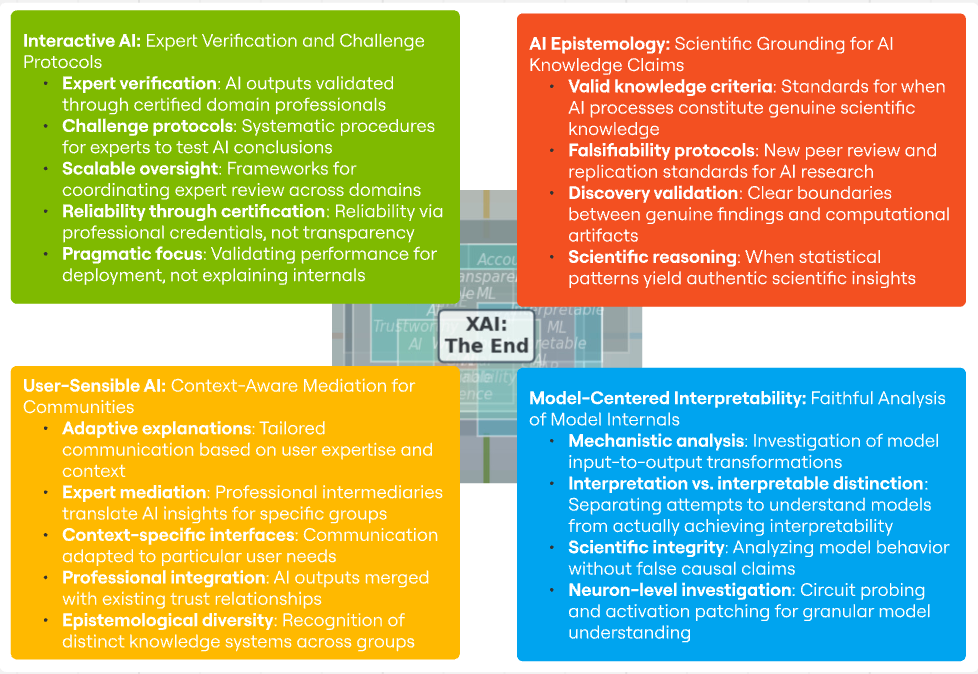}
    \caption{Post-XAI: The Paradigm Shift}
    \label{fig:7}
\end{figure}

Moving to the post-XAI landscape, we propose a four-pronged alternative direction representing a paradigm shift toward reliable and certified and fair AI development. While some relevant ideas underlying these concepts have been discussed in existing literature, we present them here in a novel configuration as components of this paradigm shift. Interactive AI (\ref{subsec:interactive-verification}) solves XAI's post-hoc explanation problem by establishing systematic verification protocols through scientific community interaction, certifying model performance rather than attempting unfaithful explanations. AI Epistemology (\ref{subsec:4.2}) addresses XAI's lack of scientific rigor by establishing methodological foundations for valid AI knowledge generation. User-Sensible AI (\ref{subsec:4.3}) overcomes XAI's impossible universalist goals by creating context-aware systems tailored to specific user communities. Model-Centered Interpretability (\ref{subsec:4.4}) preserves the technical value of model analysis while eliminating XAI's false explanatory claims about real-world phenomena. These alternative directions converge into a new paradigm for reliable and certified AI that sidesteps XAI's inherent contradictions while meeting users' genuine needs for understanding, control, and reliability (see figure ~\ref{fig:7}). 

We distinguish between pragmatic and epistemic levels of AI analysis. At the pragmatic level, we can consider black-boxes as they are and focus on the verification and certification of system performance (see Section~\ref{subsec:interactive-verification})—just as we deal with the fact that humans’ minds are cognitive black-boxes while still certifying their professional competencies. At the epistemic level, we preserve a space for legitimate inquiry into model internals, focusing on genuine scientific discovery rather than the false or unfaithful explanatory claims of XAI (see Sections~\ref{subsec:4.2} and \ref{subsec:4.4}).

% \begin{figure}[H]
% \centering
% \includegraphics[width=1.0\textwidth]{assets/P9.png}
% \caption{Post-XAI: The Paradigm Shift}
% \label{fig:8}
% \end{figure}

\subsection{Interactive AI: Moving from Explanation to Interactive Verification}
\label{subsec:interactive-verification}

IAI, as an alternative to XAI, focuses on the concept of interactions, which have already been recognized as having potential for problem-solving and serving as helpful components \cite{112,196,raees2024explainable}. The purpose of IAI is to incorporate human-AI collaboration into the architecture and functioning of AI systems and to enable rigorous verification and certification through scientific community interaction. Instead of focusing on “explanation” of black-boxes for end-users (even if it were feasible), IAI focuses on real-time interactions, shared agency \cite{197}, and collaborative sensemaking \cite{198,199} for the goal of verifying AI recommendations and outputs, thereby building trust for end-users. This shift from explanation to verification provides more robust guarantees of system performance while avoiding the cognitive overload and judgment impairment that XAI explanations demonstrably cause.

IAI does not focus on “explaining” black-boxes by invoking proxy methods that produce unfaithful explanations. Instead, it seeks to understand and verify its outcomes on two levels: On the epistemic level, IAI involves humans—i.e., developers, designers, and domain experts—in every step of the system’s processing \cite{200}. IAI aligns AI reasoning with domain experts and scientists by grounding the AI in patterns that are more intelligible to humans. This occurs through collective sense-making and shared meaning construction, with humans and AI working as a team, both involved in decision-making. IAI achieves this through interactive training, real-time collaboration interfaces, and continuous feedback loops. This creates a two-way learning process where experts become more familiar with how the AI processes information, while the AI becomes more attuned to expert reasoning patterns and domain-specific thinking. Rather than being passive recipients of AI decisions, domain experts become active participants in the decision-making process, which gives them a sense of control and confidence in the AI's outputs \cite{201,202}. IAI doesn't ignore non-expert end-users; they benefit through a reliance-trust chain (see Table~\ref{tab:1} for the distinction between reliance and trust). Non-experts rely on AI outputs that have been validated by certified domain experts they already trust. This reliance-trust chain allows users without technical expertise to confidently use AI decisions because trusted experts have verified the outputs for them. 

On a pragmatic level, IAI relies on the verification of AI outputs and recommendations by domain experts and scientists. This approach fosters a perception of AI as an assistive partner that contributes to both pragmatic and epistemic goals. It also makes the AI's output, once verified by experts, as reliable as recommendations made solely by humans. IAI emphasizes collaboration and interaction as pivotal for verification and for building reliance upon—though not necessarily trust in—AI systems, at the pragmatic level. The key point is that we need not necessarily understand the AI's internal mechanisms; reliance on outputs verified by domain experts suffices - just as patients rely on prescriptions without understanding drug mechanisms. This mirrors established professional relationships: patients do not understand and cannot directly evaluate or trust the complex medical or pharmacological theories behind their prescriptions. They consider all these prescriptions as black-boxes, but they benefit from them by trusting directly in their certified family physicians and pharmacists. The same principle can work with AI outputs— end-users receive the benefits of verified, high-performance AI systems through established trust relationships with certified human experts in relevant domains, without needing to understand the technical mechanisms themselves. This approach preserves user agency and benefits while eliminating the false promise that end-users must comprehend complex AI internals to use AI systems effectively. This is also supported by several empirical studies showing that successful human-automation collaboration depends more on operators developing a solid grasp of the automation through interaction, rather than just technical transparency  \cite{203,204}.

\subsubsection{Addressing XAI's Limitations Through IAI}

IAI, as a part of the four-pronged paradigm, contributes to addressing fundamental XAI limitations such as the epistemic challenge of distinguishing correlation from causation, the emerging threat of model collapse, and the persistent cognitive alignment problem that has plagued human-AI interaction. The fundamental confusion between correlation and causation (see Section~\ref{subsec:3.2.3}) reinforces why verification-focused approaches are superior to explanatory ones. Rather than trying to make users understand these complex statistical relationships, scientific community verification protocols can systematically validate AI performance. The domain expert's verification can assess whether AI outputs meet domain-specific standards of reliability without necessarily needing to explain the underlying statistical processes to non-specialists. 

IAI's verification-focused approach also gains additional support from research on model collapse, which demonstrates that “the value of data collected about genuine human interactions with systems will be increasingly valuable in the presence of LLM-generated content in data crawled from the Internet” \cite{183}. This evidence supports structured scientific community interaction as essential for maintaining AI system reliability—not through subjective trust relationships but through objective verification protocols that systematically test and validate AI predictions against expert knowledge and real-world outcomes.

Moreover, the cognitive alignment problem that plagues XAI can be addressed in terms of IAI, as collaborative AI systems can adapt to human reasoning patterns  \cite{205}. This alignment could be grounded in mutual theory of mind, where both humans and AI develop reciprocal models of each other's reasoning processes and decision-making patterns \cite{206, 207}. Such alignment occurs through frequent feedback loops that allow behavioral adjustment by both human users and AI systems \cite{208,209}. IAI’s advantages are not confined to reliance-building; it also provides learning opportunities regarding domain-specific problems \cite{210}.

\subsubsection{IAI Implementation: Promising Models for Reliable Certified  AI}

In order to put IAI into practice, organizational and technical changes are required. Here is a potential roadmap of IAI implementation: (1) incorporating real-time human experts’ feedback loops directly into AI system architectures, (2) developing dynamic interaction interfaces that adapt to domain experts’ needs and context, (3) co-designing training data pipelines with domain experts to ensure datasets reflect actual needs and contexts rather than developer assumptions, and (4) developing standardized testing protocols for AI predictions within specific domains, (5) creating expert review panels with domain-specific validation criteria, (6) establishing certification frameworks that document system performance under controlled conditions, (7) implementing systematic audit trails for prediction accuracy across diverse contexts, and (8) developing community-based validation networks where qualified experts can systematically challenge and verify AI outputs.

Building on decades of established human-computer interaction (HCI), human-AI interaction \cite{199} and computer-supported cooperative work (CSCW) research, IAI represents not a leap into the unknown, but a strong connection to proven human-centered design principles adapted for the unique challenges of modern AI systems. IAI incorporates several existing approaches aimed at facilitating human-AI interaction and collaboration, such as \textbf{Participatory AI}, which echoes the rich tradition of participatory design in HCI by ensuring consensus-oriented interventions that meaningfully involve stakeholders in the AI development process \cite{211,212}. \textbf{Dialectical AI}, which tries to foster structured human-AI continuous dialogue \cite{213} and conversational interaction to improve both human understanding and AI performance \cite{214}, \textbf{shared agency} approaches, which divide decision-making tasks and responsibility between humans and AI systems \cite{197}, \textbf{hybrid intelligence} approaches, which recognize the complementary strengths of humans and AI systems \cite{215}, and \textbf{collaborative sense-making} processes, rooted in CSCW traditions of distributed cognition and collective intelligence \cite{216}. 

The IAI paradigm also accommodates several proven approaches that demonstrate superior effectiveness compared to XAI’s explanatory methods, including \textbf{Communicative AI}, which underscores human-AI communications to ensure human contribution to AI knowledge formation \cite{217}; \textbf{Deliberative AI}, which facilitates structured discussions between humans and AI by identifying conflicts, deliberating on evidence, and dynamically updating recommendations \cite{ma2025towards}; \textbf{Adversarial AI}, which aims to expose possible biases by challenging humans and AI systems \cite{218}; \textbf{Evaluative AI}, which provides feedback on human decisions, such as in medical diagnostics and education \cite{24}; \textbf{Scalable Oversight}, which uses debate mechanisms for structured adversarial interactions to verify AI output reliability \cite{219}; and \textbf{Co-Designer AI}, which extends participatory design principles to collaborative AI development \cite{220}. All the aforementioned approaches help to develop responsible AI systems in line with human values and objectives, without a need for fabricated or misleading “explanations” \cite{66,69}. 

\subsubsection{Can Interactive Enhancements Rescue the XAI Framework?}

An objection that comes to mind at this point is why not simply enhance XAI with expert verification processes, such as having domain experts validate AI explanations, instead of completely abandoning the explanatory framework. This way, a well-established field can be preserved while addressing its limitations through scientific rigor.

In response to this objection, we make four points: First, while acknowledging the values and benefits of XAI models, it should be noted that the problems we identified in this framework are not just superficial issues, but fundamental problems, including two paradoxes, conceptual confusions, and flawed assumptions. How can we reform a fundamentally flawed framework through technical changes in minor or subsidiary features? This would be analogous to a situation where a house has a faulty foundation, but we improve its superstructural features such as its walls or stairs. This critique aligns directly with what Ehsan et al. \cite{221,222} terms the ``sociotechnical gap''—the fundamental divide between what XAI systems can support technically and what they must support socially. The paradoxes and confusions we identify are not merely technical limitations but manifestations of this deeper sociotechnical misalignment. The ``reformist'' Human-Centered XAI (HCXAI) movement has emerged precisely to address these limitations of algorithm-centered approaches. However, the interactive enhancements often represent techno-solutionist approaches that focus exclusively on the technical wing (improved algorithms, better interfaces, expert verification) while ignoring critical social components like trust calibration, actionability of explanations, and organizational values. As demonstrated in some case studies, even technically high performance systems (like Kuro with 90\% accuracy) can fail catastrophically when social factors are overlooked—achieving only 10\% user engagement despite substantial investment \cite{221}. This exemplifies the core critique of reformist HCXAI: that purely algorithmic considerations cannot address the explainability needs of human-AI assemblages operating in complex sociotechnical contexts. 

Secondly, supplementing XAI with interactive features could even worsen its already existing complexities and confusions through user engagement because interactive processes may introduce additional layers of interpretation that obscure rather than clarify the underlying issues. Moreover, expert verification of unfaithful “explanations” does not validate those explanations—it merely creates expert-endorsed misinformation about black-box AI system operations. Domain experts verifying a proxy model's output cannot resolve the fundamental faithfulness problem: if the explanation accurately represents the black-box model's actual computations, then the model was never truly a black-box and required no XAI approach; if the explanation differs meaningfully from what the model actually computes, then expert endorsement validates an unfaithful representation. Third, moving beyond current XAI is not about depriving ourselves of its interpretive and pedagogical advantages. What we argue should be reconsidered is the reliance on unfaithful post-hoc “explanations,” particularly when the stakes are too high, as in healthcare and criminal justice settings. In its current form, XAI often undermines user trust, which suggests the need for a shift toward a paradigm that can enhance the reliability of AI systems. Finally, the reformist HCXAI scholars’ proposal to shift from “filling gaps” to “gap understanding” may indeed facilitate a more systematic mapping of sociotechnical tensions \cite{221}. However, it seems unlikely to resolve the fundamental issues rooted in the framework's inherent paradox. That being said, we do not dismiss efforts to reform existing XAI frameworks; rather, after diagnosing their foundational flaws, it may be unwise to build a house on sand.

\subsection{AI Epistemology: Grounding AI in Scientific Rigor}
\label{subsec:4.2}

The foundational issue is whether AI systems, and particularly LLMs and deep learning models, can be said to employ recognizable scientific reasoning procedures or are going through entirely new types of information processing requiring new epistemological categories \cite{226}. Traditional scientific epistemology—induction, deduction, and controlled experimentation—are the products of centuries-long human cognitive development and social warranting processes that potentially cannot be directly exported to AI systems whose outputs are a function of statistical pattern recognition and probabilistic inference \cite{227}. Complicating this further, the reproducibility of AI outputs can vary considerably between different AI architectures and training methodologies. The issue is not whether artificial intelligence reasoning is similar to human scientific reasoning, but whether artificial intelligence processes can be formalized into methodologically sound structures permitting systematic verification of knowledge and cumulative scientific progress \cite{108,228}.

Current approaches to AI testing focus on predictive accuracy and benchmarking performance while mostly overlooking the epistemological foundations needed to ground AI as a credible scientific tool \cite{229}. Pattern recognition, probabilistic inference, and statistical correlation—today's dominant modes of operation for AI—all require rigorous methodological frameworks that specify under what conditions these activities yield genuine scientific knowledge rather than mere computational outputs \cite{230}. At the epistemic level, we seek to understand when and how these processes might constitute valid forms of reasoning, while at the pragmatic level, IAI's verification protocols can certify system performance without requiring such understanding.

Developing AI epistemology will require addressing a series of fundamental questions that current research has generally overlooked. First, what is the character of evidence in AI systems and how should evidence of this kind be tested against external reality as opposed to training distributions? Second, how can AI-made hypotheses be subjected to critical testing in the manner that characterizes good scientific investigation? Third, what is the proper role for human scientific communities to play in verifying AI-generated knowledge claims—both in IAI's pragmatic verification protocols and in epistemic-level validation of scientific discoveries—and how might auditing procedures be structured to avoid the anthropocentric assumptions that might blind us to genuinely novel kinds of machine reasoning?

The epistemological framework for AI will also have to establish solid lines of demarcation between genuine scientific discovery and computational byproduct, so that insights generated from AI are properly validated before they are adopted as scientific evidence. This includes the formulation of new forms of peer review, replication protocols, and falsifiability standards specifically designed for AI research but that maintain the stringency of systematic procedure that marks sound scientific inquiry \cite{231}. The payoffs of this epistemological project extend beyond the academic community: with AI systems increasingly shaping scientific research across fields \cite{232}, the absence of robust epistemological foundations creates risks for maintaining scientific rigor and public trust in AI-assisted research.

Contrary to XAI’s assurances of deceptive “explanation,” AI epistemology research takes seriously the complex epistemological questions posed by AI systems and proceeds toward principled solutions that can benefit both AI and scientific knowledge in general. Recent advances in causal inference and discovery \cite{233,234} and mechanistic interpretability provide promising directions for building more robust AI epistemological foundations \cite{235}Yet this is not even enough to begin building an AI epistemology, because these tools primarily address how AI systems process information internally, while leaving unanswered the more fundamental questions of whether and how AI-generated claims constitute genuine knowledge at all.

\subsection{User-Sensible AI: Making AI Understandable Without Universal Explanations}
\label{subsec:4.3}

A universal explanation for AI appears an impractical (if even feasible) goal. Effective transparency requires tailoring explanations to different users' age, cognition, and expertise. Adaptive AI systems that provide tailored explanations for different users and contexts overcome XAI’s universalist limitations through two key features \cite{236}. To begin, this approach shifts away from the paradoxical effort of explaining complex algorithms to all users toward the more achievable goal of creating appropriate expert-mediated interfaces that allow different user communities to benefit from verified AI outputs through context-appropriate professional guidance  \cite{89,237}. Second, this concept is context-awareness in character and applicable to specific situations, adapting to users’ requirements and circumstances \cite{238}, rather than imposing on universal “explanation” systems \cite{239}. Rather than pursuing XAI’s inconsistent goal, \textbf{user-sensible AI} \cite{240}, user-tailored or user-understandable AI \cite{241} manifests in user-specific paths that adjust their communication style, interface, and interaction to where their users are \cite{muralidharan2024ai}. 

Every branch of sensible AI creates different expert-mediated pathways appropriate to specific user communities and professional contexts. Pediatric- and Child-friendly AI systems work through child development specialists who translate verified AI insights into age-appropriate guidance \cite{242}, requiring radically different professional mediation frameworks from healthcare AI systems that operate through certified medical professionals who provide patient-centered communication of verified diagnostic insights \cite{243}. 

These distinctions reflect differences in human cognitive development, cultural knowledge systems, professional expertise, and communication preferences that cannot be addressed with universal explanatory solutions. For example, \textbf{Pediatric-friendly AI} operates on profoundly dissimilar epistemological foundations than \textbf{Caring AI} in healthcare environments—the former is based on developmental psychology and theory of narrative cognition \cite{244}, while the latter has its basis in medical communication studies and patient advocacy models \cite{245}. These are not permutations of the same methodology but typically other methodologies that appreciate the profound differences in how different user communities process information, develop trust, and integrate technology into existing practice. The user-sensible or understandable AI acknowledges that heterogeneous communities require profoundly different types of expert-mediated AI interaction—not technical “explanations” or direct user understanding, but appropriate professional guidance that integrates verified AI outputs with existing trust relationships and professional practice standards. By focusing on user-specific clarity rather than universally explanatory ones, this model offers a science-grounded, evidence-based counter to the verifiably false XAI presumptions while advocating for the actual purpose of beneficial AI implementation among diverse human groups.

\subsection{Model-Centered Interpretability: Reframing Model Analysis for Developers}
\label{subsec:4.4}

We need to distinguish between interpretation attempt and interpretable AI, and between explanation attempt and XAI (see Table~\ref{tab:1}) for definitions), because these distinctions separate processes from outcomes and goals from methods. Distinguishing "interpretation attempt" from "interpretable AI" recognizes that trying to understand a model's internals (the attempt) doesn't guarantee success—interpretable AI is the achieved property, not merely the pursuit of it (see Table~\ref{tab:1}). Similarly, separating "explanation attempt" from "XAI" acknowledges that XAI is just one specific technical approach (using unfaithful proxy models) among potentially many ways to pursue explanations. Without these distinctions, we risk conflating failure with success, confusing a particular methodology (XAI's proxies) with the broader goal of genuine understanding, and mistaking activity for achievement. These conceptual boundaries are crucial for evaluating whether we've actually made AI systems understandable or merely created an illusion of transparency.

That being said, Model-Centered interpretability as an interpretation attempt, allows developers to observe what their models are accomplishing without falsely claiming to represent real-world causal processes. This epistemic-level research complements IAI's pragmatic verification protocols by providing the scientific foundation for understanding when and why certain model behaviors occur, even as we accept black-box deployment through expert certification processes. Model-Centered Interpretability preserves this technical utility (e.g., how neural networks and variables are causally related in a model \cite{246}) while eliminating XAI's false hope and unfaithful explanatory promises. The core issue with XAI lies not in the analytical tools themselves, which remain important and valuable for model analysis and exploratory work in development settings, but in their misapplication as credible justifications for decisions made by end-users and stakeholders in real-life contexts.  While XAI unfaithfully promises to “explain” why AI systems make particular decisions about real-world phenomena, Model-Centered Interpretability honestly investigates how models transform inputs into outputs through their learned representations and computational pathways \cite{247,248,249,250}. This approach provides genuine scientific value for developers, researchers, and engineers who need to understand model behavior for debugging, optimization, and safety evaluation purposes—For instance, in the use-case of correcting model mistakes, detecting harmful queries and contents in LLMs, and steering models’ behaviors in vision and language models \cite{251,252,253}.

Redefining model analysis as Model-Centered Interpretability rather than “explanation” is not merely semantic precision—it is an essential commitment to scientific integrity in terms of what current techniques can and cannot accomplish. This reframing does not dispense with the productive technical work presently mistakenly labeled "explainable AI" but does eliminate the epistemic misunderstandings that have bedeviled the field \cite{254,255,256}.  XAI methods like LIME, SHAP, and gradient-based approaches might reveal statistical associations without attaining external validity \cite{137,176}. Model-Centered Interpretability positions these tools as mechanistic analysis tools for reverse-engineering representations learned, debugging training dynamics, and understanding computational behavior — without falsely claiming to explain external phenomena or domain-specific causal dependencies. They associate certain aspects of LLMs' behavior with the behavior of specific nodes in the LLMs. However, they provide only partial explanations for model behavior. Mechanistic analysis methods may someday help build the understanding needed to create interpretable LLMs; i.e., LLMs constrained so that their reasoning process is inherently understandable. 

It also offers conceptual space for the development of real scientific methods of AI knowledge verification in the form of the distinct separation of understanding models and knowing the world \cite{26,257}. This reframing preserves the substantial value that XAI methods provide for internal model development work while clearly distinguishing these legitimate use cases from inappropriate claims about external explanatory power—placing them in their proper position among the general scope of AI research methodologies.

Model-Centered Interpretability would accommodate all kinds of methods analyzing machine learning models to assist developers in model behavior understanding and performance improvement, consisting of several subcategories, such as: \textbf{Intrinsic Interpretability} builds transparent models by design like decision trees \cite{258} and more recently interpretable neural networks, also known as concept bottleneck models (CBMs) \cite{252,253,259}. \textbf{Model-Specific Interpretability} applies architecture-tailored techniques like attention visualization in transformers \cite{260}, prototypical deep learning \cite{261,261b,262}, deep k nearest neighbors\cite{263}, interpretable entity representations \cite{50,264,265}. \textbf{Mechanistic Interpretability} pursues neuron-level understanding through automated explainability techniques \cite{247,266}, activation patching and circuit probing \cite{267}. \textbf{Causal Interpretability} examines internal model decision processes through causal inference techniques rather than claiming to provide necessary external causal explanations \cite{268,269,270,271}, etc.

Model-Centered Interpretability research has been useful for model failure detection and alleviation \cite{272}, train dynamics understanding \cite{273}, and the development of more reliable AI systems for all analytical approaches \cite{273}. Moreover, the safety implications of successful Model-Centered Interpretability research should not be underestimated, particularly as AI systems become more powerful and are deployed in high-stakes settings \cite{267}. It also enables effective inference time techniques that can be used for model behavior steering, guiding models towards safe generation and away from harmful outputs in real-time. This area of investigation has true scientific value precisely because it abandons the misleading explanatory goals of XAI in favor of rigorous empirical investigation into model internals \cite{219}. Furthermore, it is important that the interpretations are reliable and faithful, which is an important problem that is being investigated in the recent work \cite{247,253}.

\subsection{Limitations and Risk Mitigation}

While our proposed four-pronged paradigm offers promising pathways beyond XAI's fundamental contradictions, each component may face significant challenges that must be acknowledged. Interactive AI's reliance on expert verification may introduce disciplinary biases, create deployment bottlenecks requiring extensive coordination, and risk capture by powerful interests through professional gatekeeping—potentially creating new digital divides between those with and without access to certified expert guidance. More fundamentally, research on scalable oversight reveals how cognitive biases shape expert judgment even on contentious issues like Covid-19 or climate change, where value pluralism creates legitimate disagreement \cite{219}. AI Epistemology faces the danger of becoming excessively abstract and philosophical, while establishing consensus across diverse research communities (computer science, statistics, philosophy, domain sciences) may prove difficult given fundamental disagreements about knowledge validity. User-sensible AI confronts implementation challenges in training expert mediators who possess both domain knowledge and AI systems expertise. Model-Centered Interpretability, while valuable for technical analysis, may inadvertently obscure the sociotechnical context of AI deployment—neglecting crucial considerations of data provenance and downstream impacts—while facing an inherent tension between depth of mechanistic analysis and practical utility for system improvement.

Effective mitigation strategies must address these interconnected challenges through coordinated approaches: establishing diverse, transparent expert panels with accessible pathways for community participation; balancing theoretical rigor with practical applicability through iterative frameworks; developing scalable training programs for expert mediators across domains; and ensuring interpretability research maintains connections to deployment concerns while communicating findings to non-specialist stakeholders. These limitations underscore the need for thoughtful, adaptive implementation that learns from XAI's failures while avoiding new forms of opacity, exclusion, or impracticality.

\section{Concluding Remarks}
\label{subsec:5}

This article has systematically demonstrated that XAI faces significant challenges and limitations. While we acknowledge its valuable contributions to debugging models and enhancing model performance, these fundamental concerns suggest that XAI, in its current form, faces difficulties serving as an effective framework for ongoing research on trustworthy AI. We have identified XAI's underlying presuppositions as problematic, and its actual applications as insufficient for building user trust. The limitations are not only some implementation issues but to-be-expected consequences of XAI’s weak conceptual foundations. XAI's theoretical and epistemic concerns are even more fundamental: XAI’s conflation of correlation with causation, its focus at the wrong levels of “explanation”, and reliance on unfaithful post-hoc narratives raise serious epistemic questions culminating in a fundamental tension—XAI struggles to provide explanations that simultaneously are accurate and explicable. Thus, the very central promise of XAI faces inherent challenges. Moreover, the implications go far beyond theoretical or methodological considerations in academia. As AI technology extends to health care, education, and law enforcement, ongoing reliance on XAI's limitations threatens potential harm to public health and democratic government. The accountability offered by XAI is, in some high-stakes contexts, worse than useless. It risks creating illusions of transparency while actually preserving the opacity and bias it claims to combat.

We support a four-pronged paradigm shift toward reliable and certified AI development. The proposed lines of inquiry—IAI, AI Epistemology, User-Sensible AI, and Model-Centered Interpretability—IAI provides a unifying umbrella for post-XAI development focused on prioritizing human welfare over technical convenience. These four post-XAI research paths work synergistically to address different aspects of XAI's limitations: IAI prevents opacity through collaboration, AI Epistemology provides scientific rigor, User-Sensible AI ensures user-appropriate understanding, and Model-Centered Interpretability offers honest technical analysis. Together, they constitute a comprehensive paradigm for reliable and certified AI that addresses XAI's conceptual contradictions while meeting legitimate needs for transparency, understanding, and accountability.

\section*{Acknowledgement}
This research is supported in part by the following foundations/institutes: \textbf{J.J.} and \textbf{S.A.} acknowledge support from the National Science Foundation under grant number 2125858 and UT-Good Systems. \textbf{M.Gh.} acknowledges supported in part by a National Science Foundation CAREER Award (2339381), and an AI2050 Award Early Career Fellowship (G-25-68042), a Gordon \& Betty Moore Foundation Award, and a Sloan Research Fellow Award (FG-2025-24277). \textbf{A.F.} acknowledges support from the National Science Foundation under Cooperative Agreement 2421782 and the Simons Foundation grant MPS-AI-00010515 awarded to NSF-Simons AI Institute for Cosmic Origins (CosmicAI, https://www.cosmicai.org/). \textbf{D.A.M.} acknowledges support from the Kempner Institute for the Study of Natural and Artificial Intelligence at Harvard University; the Aramont Fellowship Fund; and the National Science Foundation through the AI Institute for Societal Decision Making (IIS-2229881). \textbf{F.I.D.} \& \textbf{H.G.} were supported in part by CHANSE and NORFACE through the MICRO project, funded by ESRC/UKRI (grant ref. UKRI572). \textbf{A.K.} acknowledges support by NSERC Discovery Grants. \textbf{F.J.B.} acknowledges support from AWS (AI/ML for Identifying Social Determinants of Health). \textbf{B.L.Y.}  acknowledges support by the National Research Foundation, Singapore and Infocomm Media Development Authority under its Trust Tech Funding Initiative (Award No. DTC-RGC-09). Any opinions, findings and conclusions or recommendations expressed in this material are those of the author(s) and do not reflect the views of the foundations/institutes. S.A. is also grateful to Cynthia Rudin, peter stone, and Jason D'Cruz for their invaluable comments and support. Furthermore, in accordance with MLA (Modern Language Association) guidelines, we note the use of AI-powered tools, such as Anthropic's and OpenAI's applications, for assistance in editing and brainstorming. 

This paper was developed through a multi-round collaborative review process. Authorship was granted based on the fulfillment of the following criteria: (1) making substantial contributions to the conception, design, validation, or interpretation of the work, and endorsing its findings; (2) participating in drafting the manuscript or critically revising it for important intellectual content within the established review timeline; and (3) providing final approval of the version to be published and consenting to its submission within the designated review period. Accordingly, authorship required substantive and comprehensive engagement with the full manuscript, including the provision of feedback within the established timeline, as well as final approval of the submitted version. Contributors who participated in the review process but did not fulfill all three criteria – for instance, because their feedback did not address the manuscript as a whole, was not submitted within the established timeline, or was not accompanied by final approval – are gratefully acknowledged. These criteria and procedures were applied consistently and uniformly to all invited contributors.

\section*{Competing interests}
 M.R.K.M. is a co-founder and the chief scientific advisor of Nexilico, Inc., a start-up developing AI-driven microbiome engineering technologies. The other
authors declare no competing interests.

{\fontsize{8}{10}\selectfont
\bibliography{main}

@article{1,
  title={Reviewing the need for explainable artificial intelligence (xAI)},
  author={Gerlings, Julie and Shollo, Arisa and Constantiou, Ioanna},
  journal={arXiv preprint arXiv:2012.01007},
  year={2020}
}

@article{2,
  author={Palikhe, A. and Yu, Zhenyu and Wang, Zichong and Zhang, Wenbin},
  title={Towards Transparent AI: A Survey on Explainable Large Language Models},
  journal={arXiv preprint arXiv:2506.21812},
  year={2025}
}

@article{3,
  title={Explainable and interpretable multimodal large language models: A comprehensive survey},
  author={Dang, Yunkai and Huang, Kaichen and Huo, Jiahao and Yan, Yibo and Huang, Sirui and Liu, Dongrui and Gao, Mengxi and Zhang, Jie and Qian, Chen and Wang, Kun and others},
  journal={arXiv preprint arXiv:2412.02104},
  year={2024}
}

@inproceedings{4,
  title={Symbolic vs sub-symbolic ai methods: Friends or enemies?},
  author={Ilkou, Eleni and Koutraki, Maria},
  booktitle={CIKM (Workshops)},
  volume={2699},
  year={2020}
}

@article{5,
  author={Gunning, David and Vorm, Eric and Wang, Yunyan and Turek, Matt},
  title={DARPA’s explainable AI (XAI) program: A retrospective},
  journal={Applied AI Letters},
  volume={2},
  number={4},
  pages={e61},
  year={2021},
  doi={10.1002/ail2.61}
}

@article{6,
  title={{XAI} is in trouble},
  author={Weber, Rosina O and Johs, Adam J and Goel, Prateek and Silva, Jo{\~a}o Marques},
  journal={AI Magazine},
  volume={45},
  number={3},
  pages={300--316},
  year={2024},
  publisher={Wiley Online Library}
}

@article{7,
  title={DARPA’s explainable artificial intelligence (XAI) program},
  author={Gunning, David and Aha, David},
  journal={AI magazine},
  volume={40},
  number={2},
  pages={44--58},
  year={2019}
}

@inproceedings{8,
  title={" Why should i trust you?" Explaining the predictions of any classifier},
  author={Ribeiro, Marco Tulio and Singh, Sameer and Guestrin, Carlos},
  booktitle={Proceedings of the 22nd ACM SIGKDD international conference on knowledge discovery and data mining},
  pages={1135--1144},
  year={2016}
}

@inproceedings{9,
  title={Grad-cam: Visual explanations from deep networks via gradient-based localization},
  author={Selvaraju, Ramprasaath R and Cogswell, Michael and Das, Abhishek and Vedantam, Ramakrishna and Parikh, Devi and Batra, Dhruv},
  booktitle={Proceedings of the IEEE international conference on computer vision},
  pages={618--626},
  year={2017}
}

@book{10,
  title={Rule based expert systems: the mycin experiments of the stanford heuristic programming project (the Addison-Wesley series in artificial intelligence)},
  author={Buchanan, Bruce G and Shortliffe, Edward H},
  year={1984},
  publisher={Addison-Wesley Longman Publishing Co., Inc.}
}

@inproceedings{11,
  title={Exact rule learning via boolean compressed sensing},
  author={Malioutov, Dmitry and Varshney, Kush},
  booktitle={International conference on machine learning},
  pages={765--773},
  year={2013},
  organization={PMLR}
}

@book{12,
  title={The first electronic computer: The Atanasoff story},
  author={Burks, Alice R and Burks, Arthur Walter},
  year={1988},
  publisher={University of Michigan Press}
}

@inproceedings{13,
  title={Case-based explanation of non-case-based learning methods},
  author={Caruana, Rich and Kangarloo, Hooshang and Dionisio, John David and Sinha, Usha and Johnson, David},
  booktitle={Proceedings of the AMIA Symposium},
  pages={212},
  year={1999}
}

@article{14,
  author={Baehrens, D. and Fiddike, T. and Harmeling, S. and Kawanabe, M. and Hansen, K. and Müller, K.-R.},
  title={How to Explain Individual Classification Decisions},
  journal={Journal of Machine Learning Research},
  volume={11},
  year={2009},
  url={files/12518/Baehrens et al. - 2009 - How to Explain Individual Classification Decisions.pdf}
}

@article{15,
  author={Haufe, Stefan and Meinecke, Frank and G{\"o}rgen, Kai and D{\"a}hne, Sven and Haynes, John-Dylan and Blankertz, Benjamin and Bie{\ss}mann, Felix},
  title={On the interpretation of weight vectors of linear models in multivariate neuroimaging},
  journal={Neuroimage},
  volume={87},
  pages={96--110},
  year={2014},
  doi={10.1016/j.neuroimage.2013.10.067}
}

@misc{16,
  author={Gunning, D.},
  title={Explainable artificial intelligence (xai)},
  year={2017},
  journal={Defense Advanced Research Projects Agency (DARPA), Web},
  note={2(2), 1}
}

@article{17,
  author={Gunning, D. and Aha, D.},
  title={DARPA’s Explainable Artificial Intelligence (XAI) Program},
  journal={AI Mag},
  volume={40},
  number={2},
  pages={44--58},
  year={2019},
  doi={10.1609/aimag.v40i2.2850}
}

@article{18,
  author={Gunning, D. and Vorm, E. and Wang, J. Y. and Turek, M.},
  title={DARPA’s explainable AI (XAI) program: A retrospective},
  journal={Applied AI Letters},
  volume={2},
  number={4},
  pages={e61},
  year={2021},
  doi={10.1002/ail2.61}
}

@misc{19,
  title={European Union Regulations on Algorithmic Decision Making and a ‘Right to Explanation’},
  year={2025},
  month={Oct},
  doi={10.1609/aimag.v38i3.2741}
}

@report{20,
  author={H. J., R. and Hamon, I. S.},
  title={Robustness and explainability of artificial intelligence},
  institution={Publications Office of the European Union},
  volume={207},
  number={40},
  year={2020}
}

@article{21,
  author={Rudin, Cynthia},
  title={Stop explaining black box machine learning models for high stakes decisions and use interpretable models instead},
  journal={Nature Machine Intelligence},
  volume={1},
  number={5},
  pages={206--215},
  year={2019},
  doi={10.1038/s42256-019-0048-x}
}

@article{RudinEtAlSurvey2022,
author = {Cynthia Rudin and Chaofan Chen and Zhi Chen and Haiyang Huang and Lesia Semenova and Chudi Zhong},
title = {{Interpretable machine learning: Fundamental principles and 10 grand challenges}},
volume = {16},
journal = {Statistics Surveys},
number = {none},
publisher = {Amer. Statist. Assoc., the Bernoulli Soc., the Inst. Math. Statist., and the Statist. Soc. Canada},
pages = {1 -- 85},
year = {2022},
doi = {10.1214/21-SS133},
URL = {https://doi.org/10.1214/21-SS133}
}

@article{22,
  author={Swamy, V. and Frej, J. and Käser, T.},
  title={Viewpoint: The Future of Human-Centric Explainable Artificial Intelligence (XAI) is not Post-Hoc Explanations},
  journal={Journal of Artificial Intelligence Research},
  volume={84},
  year={2025},
  doi={10.1613/jair.1.17970}
}

@article{23,
  author={Lipton, Zachary C},
  title={The Mythos of Model Interpretability},
  journal={arXiv},
  year={2017},
  doi={10.48550/arXiv.1606.03490}
}

@article{24,
  author={Miller, T.},
  title={Explainable AI is Dead, Long Live Explainable AI! Hypothesis-driven decision support},
  journal={arXiv},
  doi={10.48550/arXiv.2302.12389}
}

@article{25,
  author={Hoffman, R. R. et al.},
  title={Evaluating machine-generated explanations: a ‘Scorecard’ method for XAI measurement science},
  journal={Front Comput Sci},
  year={2023}
}

@article{26,
  author={Miller, T.},
  title={Explanation in artificial intelligence: Insights from the social sciences},
  journal={Artif Intell},
  volume={267},
  pages={1--38},
  year={2019},
  doi={10.1016/j.artint.2018.07.007}
}

@incollection{27,
  author={Ehsan, U. and Riedl, M. O.},
  title={Human-Centered Explainable AI: Towards a Reflective Sociotechnical Approach},
  booktitle={Springer International Publishing},
  year={2020},
  pages={449--466},
  doi={10.1007/978-3-030-60117-1_33}
}

@article{28,
  author={Hoffman, R. R. and Mueller, S. T. and Klein, G. and Litman, J.},
  title={Measures for explainable AI: Explanation goodness, user satisfaction, mental models, curiosity, trust, and human-AI performance},
  journal={Frontiers in Computer Science},
  volume={5},
  doi={10.3389/fcomp.2023.1096257}
}

@article{29,
  author={Linardatos, P. and Papastefanopoulos, V. and Kotsiantis, S.},
  title={Explainable AI: A Review of Machine Learning Interpretability Methods},
  journal={Entropy},
  volume={23},
  number={1},
  pages={18},
  year={2021},
  doi={10.3390/e23010018}
}

@article{30,
  author={Ali, S. et al.},
  title={Explainable Artificial Intelligence (XAI): What we know and what is left to attain Trustworthy Artificial Intelligence},
  journal={Information Fusion},
  volume={99},
  pages={101805},
  year={2023},
  doi={10.1016/j.inffus.2023.101805}
}

@inproceedings{31,
  author={Zhong, J. and Negre, E.},
  title={AI: To interpret or to explain?},
  booktitle={INFORSID},
  year={2021},
  url={https://www.semanticscholar.org/paper/AI%3A-To-interpret-or-to-explain-Zhong-Negre/aea5940e91ccf37676b8c3a3273356752b9b680b}
}

@article{32,
  author={Angelov, P. P. and Soares, E. A. and Jiang, R. and Arnold, N. I. and Atkinson, P. M.},
  title={Explainable artificial intelligence: an analytical review},
  journal={WIREs Data Mining and Knowledge Discovery},
  volume={11},
  number={5},
  pages={e1424},
  year={2021},
  doi={10.1002/widm.1424}
}

@incollection{33,
  author={Hansen, L. K. and Rieger, L.},
  title={Interpretability in Intelligent Systems – A New Concept?},
  booktitle={Explainable AI: Interpreting, Explaining and Visualizing Deep Learning},
  publisher={Springer},
  pages={41--49},
  doi={10.1007/978-3-030-28954-6_3}
}

@misc{34,
  author={Hare, A.},
  title={Explainable vs Interpretable AI: An Intuitive Example},
  howpublished={Medium},
  year={2020}
}

@article{35,
  author={De, T. and Giri, P. and Mevawala, A. and Nemani, R. and Deo, A.},
  title={Explainable AI: A Hybrid Approach to Generate Human-Interpretable Explanation for Deep Learning Prediction},
  journal={Procedia Comput Sci},
  volume={168},
  pages={40--48},
  year={2020},
  doi={10.1016/j.procs.2020.02.255}
}

@article{36,
  author={Fazzinga, B. and Flesca, S. and Furfaro, F. and Pontieri, L.},
  title={Process Mining meets argumentation: Explainable interpretations of low-level event logs via abstract argumentation},
  journal={Inf Syst},
  volume={107},
  pages={101987},
  year={2022},
  doi={10.1016/j.is.2022.101987}
}

@article{37,
  author={Laato, S. and Tiainen, M. and Islam, A. K. M. N. and Mäntymäki, M.},
  title={How to explain AI systems to end users: a systematic literature review and research agenda},
  journal={Internet Research},
  volume={32},
  number={7},
  pages={1--31},
  year={2022},
  doi={10.1108/INTR-08-2021-0600}
}

@inproceedings{38,
  author={Gwak, J.-Y. et al.},
  title={Debugging Malware Classification Models Based on Event Logs with Explainable AI},
  booktitle={2023 IEEE International Conference on Data Mining Workshops (ICDMW)},
  pages={939--948},
  year={2023},
  doi={10.1109/ICDMW60847.2023.00125}
}

@article{39,
  author={Fresz, B. and Dubovitskaya, E. and Brajovic, D. and Huber, M. F. and Horz, C.},
  title={How Should AI Decisions Be Explained? Requirements for Explanations from the Perspective of European Law},
  journal={Proceedings of the AAAI/ACM Conference on AI, Ethics, and Society},
  volume={7},
  number={1},
  pages={438--450},
  year={2024},
  doi={10.1609/aies.v7i1.31648}
}

@article{40,
  author={Adadi, A. and Berrada, M.},
  title={Peeking Inside the Black-Box: A Survey on Explainable Artificial Intelligence (XAI)},
  journal={IEEE Access},
  volume={6},
  pages={52138--52160},
  year={2018},
  doi={10.1109/ACCESS.2018.2870052}
}

@inproceedings{41,
  author={Gilpin, L. H. and Bau, D. and Yuan, B. Z. and Bajwa, A. and Specter, M. and Kagal, L.},
  title={Explaining Explanations: An Overview of Interpretability of Machine Learning},
  booktitle={2018 IEEE 5th International Conference on Data Science and Advanced Analytics (DSAA)},
  pages={80--89},
  year={2018},
  doi={10.1109/DSAA.2018.00018}
}

@article{42,
  author={Hoffman, R. R. and Mueller, S. T. and Klein, G. and Litman, J.},
  title={Metrics for Explainable AI: Challenges and Prospects},
  journal={arXiv},
  year={2019},
  doi={10.48550/arXiv.1812.04608}
}

@incollection{43,
  author={Ehsan, U. and Riedl, M. O.},
  title={Human-Centered Explainable AI: Towards a Reflective Sociotechnical Approach},
  booktitle={HCI International 2020 - Late Breaking Papers: Multimodality and Intelligence},
  publisher={Springer},
  pages={449--466},
  year={2020},
  doi={10.1007/978-3-030-60117-1_33}
}

@article{44,
  author={Lipton, Z. C.},
  title={The Mythos of Model Interpretability: In machine learning, the concept of interpretability is both important and slippery},
  journal={Queue},
  volume={16},
  number={3},
  pages={31--57},
  year={2018},
  doi={10.1145/3236386.3241340}
}

@article{45,
  author={Mueller, S. T. and Hoffman, R. R. and Clancey, W. and Emrey, A. and Klein, G.},
  title={Explanation in Human-AI Systems: A Literature Meta-Review, Synopsis of Key Ideas and Publications, and Bibliography for Explainable AI},
  journal={arXiv},
  year={2019},
  doi={10.48550/arXiv.1902.01876}
}

@article{46,
  title={Explainable Artificial Intelligence (XAI): Concepts, taxonomies, opportunities and challenges toward responsible AI},
  author={Arrieta, Alejandro Barredo and D{\'\i}az-Rodr{\'\i}guez, Natalia and Del Ser, Javier and Bennetot, Adrien and Tabik, Siham and Barbado, Alberto and Garc{\'\i}a, Salvador and Gil-L{\'o}pez, Sergio and Molina, Daniel and Benjamins, Richard and others},
  journal={Information fusion},
  volume={58},
  pages={82--115},
  year={2020},
  publisher={Elsevier}
}

@inproceedings{47,
  author    = {Clinciu, M. A. and Hastie, H. F.},
  title     = {A Survey of Explainable AI Terminology},
  booktitle = {Proceedings of the 1st Workshop on Interactive Natural Language Technology for Explainable Artificial Intelligence},
  publisher = {Association for Computational Linguistics},
  pages     = {8--13},
  doi       = {10.18653/v1/W19-8403}
}

@article{48,
  author  = {Moreno-S{\'a}nchez, P. A.},
  title   = {Improvement of a prediction model for heart failure survival through explainable artificial intelligence},
  journal = {Frontiers in Cardiovascular Medicine},
  volume  = {10},
  year    = {2023},
  doi     = {10.3389/fcvm.2023.1219586}
}

@article{49,
  author  = {Baron, S.},
  title   = {Explainable AI and Causal Understanding: Counterfactual Approaches Considered},
  journal = {Minds and Machines},
  volume  = {33},
  number  = {2},
  pages   = {347--377},
  year    = {2023},
  doi     = {10.1007/s11023-023-09637-x}
}

@misc{50,
  author       = {Garcia-Olano, D. and Onoe, Y. and Ghosh, J. and Wallace, B.},
  title        = {Intermediate Entity-based Sparse Interpretable Representation Learning},
  year         = {2022},
  howpublished = {Online},
  url          = {https://www.researchgate.net/publication/366027059_Intermediate_Entity-based_Sparse_Interpretable_Representation_Learning}
}

@article{51,
  title={Unmasking Clever Hans predictors and assessing what machines really learn},
  author={Lapuschkin, Sebastian and W{\"a}ldchen, Stephan and Binder, Alexander and Montavon, Gr{\'e}goire and Samek, Wojciech and M{\"u}ller, Klaus-Robert},
  journal={Nature communications},
  volume={10},
  number={1},
  pages={1096},
  year={2019},
  publisher={Nature Publishing Group UK London}
}

@article{52,
  author  = {Minh, D. and Wang, H. X. and Li, Y. F. and Nguyen, T. N.},
  title   = {Explainable artificial intelligence: a comprehensive review},
  journal = {Artificial Intelligence Review},
  volume  = {55},
  number  = {5},
  pages   = {3503--3568},
  doi     = {10.1007/s10462-021-10088-y}
}

@inproceedings{53,
  author    = {Schoeffer, J. M. D.-A. and Kuehl, N.},
  title     = {Explanations, Fairness, and Appropriate Reliance in Human-AI Decision-Making},
  booktitle = {Proceedings of the 2024 CHI Conference on Human Factors in Computing Systems},
  year      = {2024}
}

@article{54,
  author  = {Leichtmann, B. and Humer, C. and Hinterreiter, A. and Streit, M. and Mara, M.},
  title   = {Effects of Explainable Artificial Intelligence on trust and human behavior in a high-risk decision task},
  journal = {Computers in Human Behavior},
  volume  = {139},
  pages   = {107539},
  doi     = {10.1016/j.chb.2022.107539}
}

@inproceedings{55,
  author    = {Weitz, K. and Schiller, D. and Schlagowski, R. and Huber, T. and Andr{\'e}, E.},
  title     = {{`Do you trust me?': Increasing User-Trust by Integrating Virtual Agents in Explainable AI Interaction Design}},
  booktitle = {Proceedings of the 19th ACM International Conference on Intelligent Virtual Agents (IVA '19)},
  publisher = {Association for Computing Machinery},
  pages     = {7--9},
  doi       = {10.1145/3308532.3329441}
}

@inproceedings{56,
  author    = {Ferrario, A. and Loi, M.},
  title     = {How Explainability Contributes to Trust in AI},
  booktitle = {Proceedings of the 2022 ACM Conference on Fairness, Accountability, and Transparency (FAccT '22)},
  publisher = {Association for Computing Machinery},
  pages     = {1457--1466},
  doi       = {10.1145/3531146.3533202}
}

@article{57,
  author  = {Hoffman, R. R. and Mueller, S. T. and Klein, G. and Litman, J.},
  title   = {Measures for explainable AI: Explanation goodness, user satisfaction, mental models, curiosity, trust, and human-AI performance},
  journal = {Frontiers in Computer Science},
  volume  = {5},
  year    = {2023},
  doi     = {10.3389/fcomp.2023.1096257}
}

@inproceedings{58,
  author    = {Poursabzi-Sangdeh, F. and Goldstein, D. G. and Hofman, J. M. and Wortman Vaughan, J. W. and Wallach, H.},
  title     = {Manipulating and Measuring Model Interpretability},
  booktitle = {CHI '21},
  publisher = {Association for Computing Machinery},
  year      = {2021},
  pages     = {1--52},
  doi       = {10.1145/3411764.3445315}
}

@article{59,
  author  = {Ghassemi, M. and Oakden-Rayner, L. and Beam, A. L.},
  title   = {The false hope of current approaches to explainable artificial intelligence in health care},
  journal = {The Lancet Digital Health},
  volume  = {3},
  number  = {1},
  pages   = {745--750},
  year    = {2021}
}

@incollection{60,
  author    = {Woodward, J. and Ross, L.},
  title     = {Scientific Explanation},
  booktitle = {The Stanford Encyclopedia of Philosophy, Summer 2021},
  editor    = {Zalta, E. N.},
  publisher = {Metaphysics Research Lab, Stanford University},
  year      = {2021},
  howpublished = {Online},
  url       = {https://plato.stanford.edu/archives/sum2021/entries/scientific-explanation/}
}

@article{63,
  author  = {S. D. F. S. S. T. S. V. E. A. K. N. B. J. W. and M. W. S. Jabbour},
  title   = {Measuring the Impact of AI in the Diagnosis of Hospitalized Patients},
  journal = {JAMA},
  volume  = {330},
  number  = {23},
  year    = {2023}
}

@article{64,
  author  = {Fazelpour, S. and Danks, D.},
  title   = {Algorithmic bias: Senses, sources, solutions},
  journal = {Philosophy Compass},
  volume  = {16},
  number  = {8},
  pages   = {e12760},
  year    = {2021},
  doi     = {10.1111/phc3.12760}
}

@misc{65,
  author       = {Passi, S. and Vorvoreanu, M.},
  title        = {Overreliance on AI: Literature Review},
  year         = {2022},
  howpublished = {Online},
  url          = {https://www.microsoft.com/en-us/research/publication/overreliance-on-ai-literature-review/}
}

@article{66,
  author  = {Afroogh, S. A. A. M. E. K. M. A. H.},
  title   = {Trust in AI: progress, challenges, and future directions},
  journal = {Humanities and Social Sciences Communications},
  volume  = {11},
  number  = {1},
  pages   = {1--30},
  year    = {2024}
}

@inproceedings{67,
  author    = {Kim, S. S. and Wortman Vaughan, Jennifer and Liao, Q. Vera and Lombrozo, Tania and Russakovsky, Olga},
  title     = {Fostering appropriate reliance on large language models: The role of explanations, sources, and inconsistencies},
  booktitle = {Proceedings of the 2025 CHI Conference on Human Factors in Computing Systems},
  year      = {2025}
}

@inproceedings{68,
  title={To trust or to think: cognitive forcing functions can reduce overreliance on {AI} in {AI}-assisted decision-making},
  author={Bu{\c{c}}inca, Zana and Malaya, Maja Barbara and Gajos, Krzysztof Z},
  booktitle={Proceedings of the ACM Conference on Human-Computer Interaction},
  volume={5},
  pages={1--21},
  year={2021},
  publisher={ACM New York, NY, USA}
}

@article{69,
  author  = {Le, T. and Miller, T. and Sonenberg, L. and Singh, R.},
  title   = {Towards the New {XAI}: A Hypothesis-Driven Approach to Decision Support Using Evidence},
  journal = {arXiv},
  doi     = {10.48550/arXiv.2402.01292}
}

@article{70,
  author  = {Yadollahi, E. et al.},
  title   = {Expectations, Explanations, and Embodiment: Attempts at Robot Failure Recovery},
  journal = {arXiv},
  year    = {2025},
  doi     = {10.48550/arXiv.2504.07266}
}

@inproceedings{71,
  author    = {Ferguson, A. N. and Franklin, M. and Lagnado, D.},
  title     = {Explanations that backfire: Explainable artificial intelligence can cause information overload},
  booktitle = {Proceedings of the Annual Meeting of the Cognitive Science Society},
  volume    = {44},
  number    = {44},
  year      = {2022},
  howpublished = {Online},
  url       = {https://escholarship.org/uc/item/3d97g0n3}
}

@article{72,
  author  = {Hoffman, R. R. and Mueller, S. T. and Klein, G. and Litman, J.},
  title   = {Measures for explainable {AI}: Explanation goodness, user satisfaction, mental models, curiosity, trust, and human-{AI} performance},
  journal = {Frontiers in Computer Science},
  volume  = {5},
  year    = {2023},
  doi     = {10.3389/fcomp.2023.1096257}
}

@inproceedings{73,
  author    = {Wilming, R. and Kieslich, L. and Clark, B. and Haufe, S.},
  title     = {Theoretical Behavior of {XAI} Methods in the Presence of Suppressor Variables},
  booktitle = {International Conference on Machine Learning},
  publisher = {PMLR},
  year      = {2023},
  pages     = {37091--37107},
  url       = {https://proceedings.mlr.press/v202/wilming23a.html}
}

@article{74,
  author  = {Turpin, M. and Michael, J. and Perez, E. and Bowman, S. R.},
  title   = {Language Models Don’t Always Say What They Think: Unfaithful Explanations in Chain-of-Thought Prompting},
  journal = {arXiv},
  year    = {2023},
  doi     = {10.48550/arXiv.2305.04388}
}

@inproceedings{75,
  author    = {Jacovi, A. and Goldberg, Y.},
  title     = {Towards Faithfully Interpretable NLP Systems: How Should We Define and Evaluate Faithfulness?},
  booktitle = {Proceedings of the 58th Annual Meeting of the Association for Computational Linguistics (ACL 2020)},
  editor    = {Jurafsky, D. and Chai, J. and Schluter, N. and Tetreault, J.},
  publisher = {Association for Computational Linguistics},
  year      = {2020},
  pages     = {4198--4205},
  doi       = {10.18653/v1/2020.acl-main.386}
}

@article{76,
  author  = {Lyu, Q. and Apidianaki, M. and Callison-Burch, C.},
  title   = {Towards Faithful Model Explanation in NLP: A Survey},
  journal = {Computational Linguistics},
  volume  = {50},
  number  = {2},
  pages   = {657--723},
  year    = {2024},
  doi     = {10.1162/coli_a_00511}
}

@misc{77,
  author       = {Mueller, S. T. and Hoffman, R. R. and Clancey, W. and Emrey, A. and Klein, G.},
  title        = {DARPA XAI Literature Review: Explanation in Human-AI Systems. A Literature Meta-Review, Synopsis of Key Ideas and Publications, and Bibliography for Explainable AI. Prepared by Task Area 2 Institute for Human and Machine Cognition},
  year         = {2019},
  note         = {Macrocognition}
}

@article{78,
  author  = {Rastogi, N. and Pant, S. and Dhanuka, D. and Saxena, A. and Mairal, P.},
  title   = {Too Much to Trust? Measuring the Security and Cognitive Impacts of Explainability in AI-Driven SOCs},
  journal = {arXiv},
  year    = {2025},
  doi     = {10.48550/arXiv.2503.02065}
}

@article{79,
  author  = {Muir, B. M.},
  title   = {Trust in automation: Part I. Theoretical issues in the study of trust and human intervention in automated systems},
  journal = {Ergonomics},
  volume  = {37},
  number  = {11},
  pages   = {1905--1922},
  year    = {1994},
  doi     = {10.1080/00140139408964957}
}

@article{80,
  author  = {Abbass, H. A.},
  title   = {Social Integration of Artificial Intelligence: Functions, Automation Allocation Logic and Human-Autonomy Trust},
  journal = {Cognitive Computation},
  volume  = {11},
  number  = {2},
  pages   = {159--171},
  year    = {2019},
  doi     = {10.1007/s12559-018-9619-0}
}

@misc{81,
  author       = {Do, H. et al.},
  title        = {Facilitating Human-LLM Collaboration through Factuality Scores and Source Attributions},
  year         = {2024},
  howpublished = {Online},
  url          = {files/12615/Do et al. - 2024 - Facilitating Human-LLM Collaboration through Factu.pdf}
}

@article{82,
  author  = {Dietvorst, B. and Simmons, J. and Massey, C.},
  title   = {Algorithm Aversion: People Erroneously Avoid Algorithms After Seeing Them Err},
  journal = {Journal of Experimental Psychology: General},
  volume  = {144},
  doi     = {10.1037/xge0000033}
}

@article{83,
  author  = {Jones-Jang, S. M. and Park, Y. J.},
  title   = {How do people react to AI failure? Automation bias, algorithmic aversion, and perceived controllability},
  journal = {Journal of Computer-Mediated Communication},
  volume  = {28},
  number  = {1},
  pages   = {zmac029},
  year    = {2022},
  doi     = {10.1093/jcmc/zmac029}
}

@article{84,
  author  = {Rastogi, C. and Zhang, Y. and Wei, D. and Varshney, K. R. and Dhurandhar, A. and Tomsett, R.},
  title   = {Deciding Fast and Slow: The Role of Cognitive Biases in AI-assisted Decision-making},
  journal = {Proceedings of the ACM on Human-Computer Interaction},
  volume  = {6},
  number  = {CSCW1},
  pages   = {83:1--83:22},
  doi     = {10.1145/3512930}
}

@article{85,
  author  = {Bu{\c c}inca, Z. and Malaya, M. B. and Gajos, K. Z.},
  title   = {To Trust or to Think},
  journal = {Proceedings of the ACM on Human-Computer Interaction},
  volume  = {5},
  number  = {CSCW1},
  pages   = {1--21},
  year    = {2021},
  doi     = {10.1145/3449287}
}

@article{86,
  author  = {Ashktorab, Z. et al.},
  title   = {Who’s Sorry Now: User Preferences Among Rote, Empathic, and Explanatory Apologies from LLM Chatbots},
  journal = {arXiv},
  year    = {2025},
  doi     = {10.48550/arXiv.2507.02745}
}

@article{87,
  author  = {Narayanan, M. E. C. J. H. B. K. S. G. and F. D.-V.},
  title   = {How do humans understand explanations from machine learning systems. An evaluation of the human-interpretability of explanation},
  journal = {arXiv},
  year    = {2018},
  note    = {arXiv:1802.00682v1 [cs.AI] 2 Feb 2018}
}

@article{88,
  author  = {Nguyen, T. A. C. and Zhu, J.},
  title   = {How human-centered explainable AI interface are designed and evaluated: A systematic survey},
  journal = {arXiv preprint arXiv:2403.14496},
  year    = {2024}
}

@article{89,
  author  = {Haque, A. B. A. N. I. and P. M.},
  title   = {Explainable Artificial Intelligence (XAI) from a user perspective: A synthesis of prior literature and problematizing avenues for future research},
  journal = {Technological Forecasting and Social Change},
  year    = {2023}
}

@inproceedings{94,
  author    = {Mhasawade, V. and Rahman, S. and Haskell-Craig, Z. and Chunara, R.},
  title     = {Understanding Disparities in Post Hoc Machine Learning Explanation},
  booktitle = {FAccT '24: The 2024 ACM Conference on Fairness, Accountability, and Transparency},
  publisher = {ACM},
  year      = {2024},
  pages     = {2374--2388},
  doi       = {10.1145/3630106.3659043}
}

@article{95,
  author  = {Turpin, M. and Michael, J. and Perez, E. and Bowman, S. R.},
  title   = {Language Models Don’t Always Say What They Think: Unfaithful Explanations in Chain-of-Thought Prompting},
  journal = {arXiv},
  doi     = {10.48550/arXiv.2305.04388}
}

@incollection{96,
  author    = {de Kleijn, Roy},
  title     = {Artificial Intelligence Versus Biological Intelligence: A Historical Overview},
  booktitle = {Law and Artificial Intelligence},
  editor    = {Custers, B. and Fosch-Villaronga, E.},
  series    = {Information Technology and Law Series},
  volume    = {35},
  publisher = {T.M.C. Asser Press},
  address   = {The Hague},
  year      = {2022},
  doi       = {10.1007/978-94-6265-523-2_2}
}

@incollection{97,
  author    = {Chalmers, D.},
  title     = {The Hard Problem of Consciousness},
  booktitle = {The Blackwell Companion to Consciousness},
  publisher = {John Wiley \& Sons, Ltd},
  year      = {2017},
  pages     = {32--42},
  howpublished = {Online},
  url       = {https://onlinelibrary.wiley.com/doi/abs/10.1002/9781119132363.ch3}
}

@article{98,
  author  = {Jain, S. and Wallace, B. C.},
  title   = {Attention is not Explanation},
  journal = {arXiv},
  doi     = {10.48550/arXiv.1902.10186}
}

@article{99,
  author  = {Bogert, E. and Schecter, A. and Watson, R. T.},
  title   = {Humans rely more on algorithms than social influence as a task becomes more difficult},
  journal = {Scientific Reports},
  volume  = {11},
  number  = {1},
  pages   = {8028},
  year    = {2021},
  doi     = {10.1038/s41598-021-87480-9}
}

@article{100,
  author  = {Promberger, M. and Baron, J.},
  title   = {Do patients trust computers?},
  journal = {Journal of Behavioral Decision Making},
  volume  = {19},
  number  = {5},
  pages   = {455--468},
  year    = {2006},
  doi     = {10.1002/bdm.542}
}

@article{101,
  author  = {Ismatullaev, U. V. U. and Kim, S.-H.},
  title   = {Review of the Factors Affecting Acceptance of AI-Infused Systems},
  journal = {Human Factors},
  volume  = {66},
  number  = {1},
  pages   = {126--144},
  year    = {2024},
  doi     = {10.1177/00187208211064707}
}

@article{102,
  author  = {Dietvorst, B. J. and Simmons, J. P. and Massey, C.},
  title   = {Overcoming Algorithm Aversion: People Will Use Imperfect Algorithms If They Can (Even Slightly) Modify Them},
  journal = {Management Science},
  volume  = {64},
  number  = {3},
  pages   = {1155--1170},
  year    = {2018},
  doi     = {10.1287/mnsc.2016.2643}
}

@article{103,
  author  = {Jung, N. and Wranke, C. and Hamburger, K. and Knauff, M.},
  title   = {How emotions affect logical reasoning: evidence from experiments with mood-manipulated participants, spider phobics, and people with exam anxiety},
  journal = {Frontiers in Psychology},
  volume  = {5},
  year    = {2014},
  doi     = {10.3389/fpsyg.2014.00570}
}

@article{104,
  author  = {Okon-Singer, H. and Hendler, T. and Pessoa, L. and Shackman, A. J.},
  title   = {The neurobiology of emotion--cognition interactions: fundamental questions and strategies for future research},
  journal = {Frontiers in Human Neuroscience},
  volume  = {9},
  pages   = {58},
  year    = {2015},
  doi     = {10.3389/fnhum.2015.00058}
}

@article{105,
  author  = {Wang, Y. and Zhao, J. and Ones, D. S. and He, L. and Xu, X.},
  title   = {Evaluating the ability of large language models to emulate personality},
  journal = {Scientific Reports},
  volume  = {15},
  number  = {1},
  pages   = {519},
  year    = {2025},
  doi     = {10.1038/s41598-024-84109-5}
}

@article{106,
  author  = {Lo, J.-H. and Huang, H.-P. and Lo, J.-S.},
  title   = {LLM-based robot personality simulation and cognitive system},
  journal = {Scientific Reports},
  volume  = {15},
  number  = {1},
  pages   = {16993},
  year    = {2025},
  doi     = {10.1038/s41598-025-01528-8}
}

@inproceedings{107,
  author    = {O'Shea, J. and Crockett, Keeley and Khan, Wasiq and Kindynis, Philippos and Antoniades, Athos and Boultadakis, G.},
  title     = {Intelligent deception detection through machine based interviewing},
  booktitle = {International Joint Conference on Neural Networks (IJCNN)},
  year      = {2018}
}

@article{108,
  author  = {Barez, F. et al.},
  title   = {Chain-of-thought is not explainability},
  journal = {arXiv},
  year    = {2025}
}

@article{109,
  author  = {Ott, M. and Choi, Y. and Cardie, C. and Hancock, J. T.},
  title   = {Finding Deceptive Opinion Spam by Any Stretch of the Imagination},
  journal = {arXiv},
  year    = {2011},
  doi     = {10.48550/arXiv.1107.4557}
}

@inproceedings{110,
  author    = {Rashkin, H. and Choi, E. and Jang, J. Y. and Volkova, S. and Choi, Y.},
  title     = {Truth of Varying Shades: Analyzing Language in Fake News and Political Fact-Checking},
  booktitle = {Proceedings of EMNLP 2017},
  editor    = {Palmer, M. and Hwa, R. and Riedel, S.},
  publisher = {Association for Computational Linguistics},
  year      = {2017},
  pages     = {2931--2937},
  doi       = {10.18653/v1/D17-1317}
}

@article{111,
  author  = {Jiang, S. W. C.},
  title   = {Linguistic Signals under Misinformation and Fact-Checking: Evidence from User Comments on Social Media},
  journal = {Proceedings of the ACM on Human-Computer Interaction},
  volume  = {2},
  number  = {CSCW},
  year    = {2018}
}

@article{112,
  author  = {Liao, Q. V. and Varshney, K. R.},
  title   = {Human-Centered Explainable AI (XAI): From Algorithms to User Experiences},
  journal = {arXiv},
  year    = {2022},
  doi     = {10.48550/arXiv.2110.10790}
}

@inproceedings{113,
  author    = {Eiband, M. and Buschek, D. and Kremer, A. and Hussmann, H.},
  title     = {The Impact of Placebic Explanations on Trust in Intelligent Systems},
  booktitle = {CHI EA '19},
  publisher = {Association for Computing Machinery},
  year      = {2019},
  pages     = {1--6},
  doi       = {10.1145/3290607.3312787}
}

@book{114,
  author    = {Konar, A.},
  title     = {Artificial Intelligence and Soft Computing: Behavioral and Cognitive Modeling of the Human Brain},
  publisher = {CRC Press},
  doi       = {10.1201/9781315219738}
}

@article{116,
  author  = {Sakai, T. and Nagai, T.},
  title   = {Explainable autonomous robots: a survey and perspective},
  volume  = {36},
  number  = {5},
  pages   = {219--238},
  doi     = {10.1080/01691864.2022.2029720}
}

@article{117,
  author  = {Parasuraman, R. and Riley, V.},
  title   = {Humans and Automation: Use, Misuse, Disuse, Abuse},
  journal = {Human Factors},
  volume  = {39},
  number  = {2},
  pages   = {230--253},
  year    = {1997},
  doi     = {10.1518/001872097778543886}
}

@inproceedings{118,
  author    = {Lima, G. and Grgi{\'c}-Hla{\v c}a, N. and Jeong, J. K. and Cha, M.},
  title     = {The Conflict Between Explainable and Accountable Decision-Making Algorithms},
  booktitle = {FAccT '22},
  publisher = {Association for Computing Machinery},
  year      = {2022},
  pages     = {2103--2113},
  doi       = {10.1145/3531146.3534628}
}

@article{119,
  author  = {Logg, J. M. and Minson, J. A. and Moore, D. A.},
  title   = {Algorithm appreciation: People prefer algorithmic to human judgment},
  journal = {Organizational Behavior and Human Decision Processes},
  volume  = {151},
  pages   = {90--103},
  year    = {2019},
  doi     = {10.1016/j.obhdp.2018.12.005}
}

@article{122,
  author  = {Afroogh, S.},
  title   = {The Mirage of Motivation Reason Internalism},
  journal = {Journal of Value Inquiry},
  volume  = {58},
  number  = {1},
  pages   = {111--129},
  year    = {2024},
  doi     = {10.1007/s10790-021-09871-5}
}

@book{123,
  author    = {Doris, J. M.},
  title     = {Talking to Our Selves: Reflection, Ignorance, and Agency},
  publisher = {Oxford University Press}
}

@book{124,
  author    = {Hirstein},
  title     = {Confabulation: Views from neuroscience, psychiatry, psychology and philosophy},
  edition   = {1},
  publisher = {Oxford University Press}
}

@article{125,
  author  = {Kalai, A. T. and Nachum, Ofir and Vempala, Santosh S. and Zhang, Edwin},
  title   = {Why language models hallucinate},
  journal = {arXiv preprint arXiv:2509.04664},
  year    = {2025}
}

@misc{126,
  author       = {Gabriel, S. and Lyu, L. and Siderius, J. and Ghassemi, M. and Andreas, J. and Ozdaglar, A.},
  title        = {MisinfoEval: Generative AI in the Era of `Alternative Facts'},
  howpublished = {Online},
  url          = {files/12688/Gabriel et al. - MisinfoEval Generative AI in the Era of “Alternat.pdf}
}

@article{127,
  author  = {Bansal, H. and Hosseini, A. and Agarwal, R. and Tran, V. Q. and Kazemi, M.},
  title   = {Smaller, Weaker, Yet Better: Training LLM Reasoners via Compute-Optimal Sampling},
  journal = {arXiv},
  year    = {2024},
  doi     = {10.48550/arXiv.2408.16737}
}

@article{128,
  author  = {Hempel, C. G. and Oppenheim, P.},
  title   = {Studies in the Logic of Explanation},
  journal = {Philosophy of Science},
  volume  = {15},
  number  = {2},
  pages   = {135--175},
  year    = {1948},
  doi     = {10.1086/286983}
}

@book{129,
  author    = {Salmon, W. C.},
  title     = {Four Decades of Scientific Explanation},
  publisher = {University of Pittsburgh Press},
  year      = {2006},
  howpublished = {Online},
  url       = {https://www.google.com/books?id=JhoFR_F4G4EC}
}

@article{130,
  author  = {Marcus, G.},
  title   = {The Next Decade in {AI}: Four Steps Towards Robust Artificial Intelligence},
  journal = {arXiv},
  year    = {2020},
  doi     = {10.48550/arXiv.2002.06177}
}

@article{131,
  author  = {Vafa, K. and Chang, P. G. and Rambachan, A. and Mullainathan, S.},
  title   = {What Has a Foundation Model Found? Using Inductive Bias to Probe for World Models},
  journal = {arXiv},
  year    = {2025},
  doi     = {10.48550/arXiv.2507.06952}
}

@inproceedings{132,
  author    = {Abdelaal, M. M. A. and Sena, H. A. and Farouq, M. W. and Salem, A. B. M.},
  title     = {Using pattern recognition approach for providing second opinion of breast cancer diagnosis},
  booktitle = {2010 The 7th International Conference on Informatics and Systems (INFOS)},
  year      = {2010},
  pages     = {1--7},
  howpublished = {Online},
  url       = {https://ieeexplore.ieee.org/abstract/document/5461804}
}

@misc{133,
  author  = {Dai, T. and Singh, S.},
  title   = {Using AI as Gatekeeper or Second Opinion: Designing Patient Pathways for AI-Augmented Healthcare},
  doi     = {10.13140/RG.2.2.21491.85286}
}

@inproceedings{134,
  author    = {Ribeiro, M. T. and Singh, S. and Guestrin, C.},
  title     = {{`Why Should I Trust You?': Explaining the Predictions of Any Classifier}},
  booktitle = {KDD '16},
  publisher = {Association for Computing Machinery},
  year      = {2016},
  pages     = {1135--1144},
  doi       = {10.1145/2939672.2939778}
}

@inproceedings{135,
  author    = {Slack, D. and Hilgard, S. and Jia, E. and Singh, S. and Lakkaraju, H.},
  title     = {Fooling LIME and SHAP: Adversarial Attacks on Post hoc Explanation Methods},
  booktitle = {AIES '20},
  publisher = {Association for Computing Machinery},
  year      = {2020},
  pages     = {180--186},
  doi       = {10.1145/3375627.3375830}
}

@article{137,
  author  = {Ghorbani, A. and Abid, A. and Zou, J.},
  title   = {Interpretation of Neural Networks Is Fragile},
  journal = {Proceedings of the AAAI Conference on Artificial Intelligence},
  volume  = {33},
  number  = {01},
  pages   = {3681--3688},
  year    = {2019},
  doi     = {10.1609/aaai.v33i01.33013681}
}

@inproceedings{138,
  author    = {Slack, D. and Hilgard, S. and Jia, E. and Singh, S. and Lakkaraju, H.},
  title     = {Fooling LIME and SHAP: Adversarial attacks on post hoc explanation methods},
  booktitle = {AIES 2020 - Proceedings of the AAAI/ACM Conference on AI, Ethics, and Society},
  publisher = {Association for Computing Machinery, Inc},
  year      = {2020},
  month     = {Feb},
  pages     = {180--186},
  doi       = {10.1145/3375627.3375830}
}

@article{139,
  author  = {Dombrowski, P. and Kessel, P. and others},
  title   = {Explanations can be manipulated and geometry is to blame},
  journal = {Advances in Neural Information Processing Systems},
  volume  = {23},
  year    = {2019}
}

@book{140,
  author    = {Pearl, J.},
  title     = {Causality},
  publisher = {Cambridge University Press},
  year      = {2009}
}

@book{141,
  author    = {Imbens, G. W. and Rubin, D. B.},
  title     = {Causal inference in statistics, social, and biomedical sciences},
  publisher = {Cambridge University Press},
  year      = {2015}
}

@book{142,
  author    = {Hern{\'a}n, M. A. and Robins, J. M.},
  title     = {Causal Inference: What If},
  publisher = {Taylor \& Francis},
  year      = {2020}
}

@article{143,
  author  = {Talebi, S. and Tong, E. and Li, A. and Yamin, G. and Zaharchuk, G. and Mofrad, M. R. K.},
  title   = {Exploring the performance and explainability of fine-tuned BERT models for neuroradiology protocol assignment},
  journal = {BMC Medical Informatics and Decision Making},
  volume  = {24},
  number  = {1},
  pages   = {40},
  year    = {2024},
  doi     = {10.1186/s12911-024-02444-z}
}

@article{144,
  author  = {Ruis, L. et al.},
  title   = {Procedural knowledge in pretraining drives reasoning in large language models},
  journal = {arXiv preprint arXiv:2411.12580},
  year    = {2024}
}

@article{145,
  author  = {Mirzadeh, I. K. A. H. S. O. T. S. B. and Farajtabar, M.},
  title   = {Gsm-symbolic: Understanding the limitations of mathematical reasoning in large language models},
  journal = {arXiv preprint arXiv:2410.05229},
  year    = {2024}
}

@misc{146,
  author       = {Adebayo, J. and Gilmer, J. and Muelly, M. and Goodfellow, I. and Hardt, M. and Kim, B.},
  title        = {Sanity Checks for Saliency Maps},
  year         = {2018},
  howpublished = {Online},
  url          = {https://proceedings.neurips.cc/paper/2018/hash/294a8ed24b1ad22ec2e7efea049b8737-Abstract.html}
}

@inproceedings{147,
  author    = {Selvaraju, R. R. and Cogswell, M. and Das, A. and Vedantam, R. and Parikh, D. and Batra, D.},
  title     = {Grad-CAM: Visual Explanations From Deep Networks via Gradient-Based Localization},
  booktitle = {Proceedings of the IEEE International Conference on Computer Vision},
  year      = {2017},
  pages     = {618--626},
  howpublished = {Online},
  url       = {https://openaccess.thecvf.com/content_iccv_2017/html/Selvaraju_Grad-CAM_Visual_Explanations_ICCV_2017_paper.html}
}

@article{148,
  author  = {Stenzinger, A. et al.},
  title   = {Artificial intelligence and pathology: From principles to practice and future applications in histomorphology and molecular profiling},
  journal = {Seminars in Cancer Biology},
  volume  = {84},
  pages   = {129--143},
  year    = {2022},
  doi     = {10.1016/j.semcancer.2021.02.011}
}

@article{149,
  author  = {Evans, T. et al.},
  title   = {The explainability paradox: Challenges for xAI in digital pathology},
  journal = {Future Generation Computer Systems},
  year    = {2022}
}

@misc{150,
  author       = {Hajiramezanali, E. and Maleki, S. and Tseng, A. and Bentaieb, A. and Scalia, G. and Biancalani, T.},
  title        = {On the Consistency of GNN Explainability Methods},
  howpublished = {Online},
  url          = {files/12770/Hajiramezanali et al. - On the Consistency of GNN Explainability Methods.pdf}
}

@article{151,
  author  = {Zech, J. R. and Badgeley, M. A. and Liu, M. and Costa, A. B. and Titano, J. J. and Oermann, E. K.},
  title   = {Variable generalization performance of a deep learning model to detect pneumonia in chest radiographs: A cross-sectional study},
  journal = {PLoS Medicine},
  volume  = {15},
  number  = {11},
  year    = {2018},
  month   = {Nov},
  doi     = {10.1371/journal.pmed.1002683}
}

@article{152,
  author  = {Sanchez, P. and Voisey, J. P. and Xia, T. and Watson, H. I. and O'Neil, A. Q. and Tsaftaris, S. A.},
  title   = {Causal machine learning for healthcare and precision medicine},
  journal = {Royal Society Open Science},
  volume  = {9},
  number  = {8},
  pages   = {220638},
  year    = {2022},
  doi     = {10.1098/rsos.220638}
}

@article{153,
  author  = {Miller, T.},
  title   = {Explanation in artificial intelligence: Insights from the social sciences},
  journal = {Artificial Intelligence},
  volume  = {267},
  pages   = {1--38},
  year    = {2019},
  doi     = {10.1016/j.artint.2018.07.007}
}

@article{154,
  author  = {Doshi-Velez, F. and Kim, B.},
  title   = {Towards A Rigorous Science of Interpretable Machine Learning},
  journal = {arXiv},
  year    = {2017},
  doi     = {10.48550/arXiv.1702.08608}
}

@inproceedings{155,
  author    = {Bhatt, U. et al.},
  title     = {Explainable machine learning in deployment},
  booktitle = {FAT* 2020 - Proceedings of the 2020 Conference on Fairness, Accountability, and Transparency},
  publisher = {Association for Computing Machinery, Inc},
  year      = {2020},
  month     = {Jan},
  pages     = {648--657},
  doi       = {10.1145/3351095.3375624}
}

@inproceedings{156,
  author    = {R{\"a}uker, T. and Ho, A. and Casper, S. and Hadfield-Menell, D.},
  title     = {Toward Transparent AI: A Survey on Interpreting the Inner Structures of Deep Neural Networks},
  booktitle = {2023 IEEE Conference on Secure and Trustworthy Machine Learning (SaTML)},
  year      = {2023},
  pages     = {464--483},
  doi       = {10.1109/SaTML54575.2023.00039}
}

@article{157,
  author  = {Naiseh, M. and Al-Thani, D. and Jiang, N. and Ali, R.},
  title   = {Explainable recommendation: when design meets trust calibration},
  volume  = {24},
  number  = {5},
  pages   = {1857--1884},
  doi     = {10.1007/s11280-021-00916-0}
}

@article{158,
  author  = {de Bruijn, H. and Warnier, M. and Janssen, M.},
  title   = {The perils and pitfalls of explainable AI: Strategies for explaining algorithmic decision-making},
  volume  = {39},
  number  = {2},
  pages   = {101666},
  doi     = {10.1016/j.giq.2021.101666}
}

@incollection{159,
  author    = {Samek, W. and M{\"u}ller, K.-R.},
  title     = {Towards Explainable Artificial Intelligence},
  booktitle = {Explainable AI: Interpreting, Explaining and Visualizing Deep Learning},
  editor    = {Samek, W. and Montavon, G. and Vedaldi, A. and Hansen, L. K. and M{\"u}ller, K.-R.},
  publisher = {Springer International Publishing},
  pages     = {5--22},
  doi       = {10.1007/978-3-030-28954-6_1}
}

@article{160,
  author  = {Holm, E. A.},
  title   = {In defense of the black box},
  journal = {Science},
  volume  = {364},
  number  = {6435},
  pages   = {26--27},
  year    = {2019}
}

@inproceedings{161,
  author    = {Lim, B. Y. and Dey, A. K. and Avrahami, D.},
  title     = {Why and why not explanations improve the intelligibility of context-aware intelligent systems},
  booktitle = {Proceedings of the SIGCHI Conference on Human Factors in Computing Systems},
  year      = {2009}
}

@inproceedings{162,
  author    = {Ehsan, U. M. R.},
  title     = {Explainable AI reloaded: Challenging the XAI status quo in the era of large language models},
  booktitle = {Proceedings of the Halfway to the Future Symposium},
  year      = {2024}
}

@inproceedings{163,
  author    = {Ehsan, U. M. R.},
  title     = {Explainable AI reloaded: Challenging the XAI status quo in the era of large language models},
  booktitle = {Proceedings of the Halfway to the Future Symposium},
  year      = {2024}
}

@inproceedings{164,
  author    = {Wang, X. and Wang, M. Y.},
  title     = {Are explanations helpful? A comparative study of the effects of explanations in AI-assisted decision-making},
  booktitle = {Proceedings of the 26th International Conference on Intelligent User Interfaces},
  year      = {2021}
}

@article{165,
  author  = {Bogert, E. and Schecter, A. and Watson, R. T.},
  title   = {Humans rely more on algorithms than social influence as a task becomes more difficult},
  journal = {Scientific Reports},
  volume  = {11},
  number  = {1},
  pages   = {8028},
  year    = {2021},
  doi     = {10.1038/s41598-021-87480-9}
}

@article{166,
  author  = {Hoffman, R. R. and Mueller, S. T. and Klein, G. and Litman, J.},
  title   = {Measures for explainable AI: Explanation goodness, user satisfaction, mental models, curiosity, trust, and human-AI performance},
  journal = {Frontiers in Computer Science},
  volume  = {5},
  year    = {2023},
  doi     = {10.3389/fcomp.2023.1096257}
}

@misc{167,
  author       = {Rothermel, M. and Daftarian, S. S. and Koosha, T. A. and Mahani, M.-A. N. and Jamalabadi, H.},
  title        = {On the cognitive alignment between humans and machines},
  howpublished = {Online},
  url          = {https://openreview.net/forum?id=SkLYMkxE6n#discussion}
}

@article{168,
  author  = {Mathew, D. E. and Ebem, D. U. and Ikegwu, A. C. and Ukeoma, P. E. and Dibiaezue, N. F.},
  title   = {Recent Emerging Techniques in Explainable Artificial Intelligence to Enhance the Interpretable and Understanding of AI Models for Human},
  journal = {Neural Processing Letters},
  volume  = {57},
  number  = {1},
  pages   = {16},
  year    = {2025},
  doi     = {10.1007/s11063-025-11732-2}
}

@article{169,
  author  = {Mueller, S. T.},
  title   = {Cognitive Anthropomorphism of AI: How Humans and Computers Classify Images},
  journal = {Ergonomics in Design},
  volume  = {28},
  number  = {3},
  pages   = {12--19},
  year    = {2020},
  doi     = {10.1177/1064804620920870}
}

@article{170,
  author  = {Salles, A. and Evers, K. and Farisco, M.},
  title   = {Anthropomorphism in AI},
  journal = {AJOB Neuroscience},
  volume  = {11},
  number  = {2},
  pages   = {88--95},
  year    = {2020},
  doi     = {10.1080/21507740.2020.1740350}
}

@article{171,
  author  = {Li, J. J. and Tong, X.},
  title   = {Statistical hypothesis testing versus machine learning binary classification: Distinctions and guidelines},
  journal = {Patterns},
  volume  = {1},
  number  = {7},
  year    = {2020}
}

@inproceedings{172,
  author    = {Slack, D. and Hilgard, S. and Jia, E. and Singh, S. and Lakkaraju, H.},
  title     = {Fooling LIME and SHAP: Adversarial attacks on post hoc explanation methods},
  booktitle = {AIES 2020 - Proceedings of the AAAI/ACM Conference on AI, Ethics, and Society},
  publisher = {Association for Computing Machinery, Inc},
  year      = {2020},
  month     = {Feb},
  pages     = {180--186},
  doi       = {10.1145/3375627.3375830}
}

@article{173,
  author  = {Alvarez-Melis, D. and Jaakkola, T. S.},
  title   = {On the robustness of interpretability methods},
  journal = {arXiv},
  year    = {2018}
}

@article{174,
  author  = {Gu, J. and Tresp, V.},
  title   = {Saliency methods for explaining adversarial attacks},
  journal = {arXiv preprint arXiv:1908.08413},
  year    = {2019}
}

@misc{176,
  author       = {Adebayo, J. and Gilmer, J. and Muelly, M. and Goodfellow, I. and Hardt, M. and Kim, B.},
  title        = {Sanity Checks for Saliency Maps},
  year         = {2018},
  howpublished = {Curran Associates, Inc.},
  url          = {https://proceedings.neurips.cc/paper/2018/hash/294a8ed24b1ad22ec2e7efea049b8737-Abstract.html}
}

@article{177,
  author  = {Jain, S. and Wallace, B. C.},
  title   = {Attention is not Explanation},
  journal = {arXiv},
  year    = {2019},
  url     = {http://arxiv.org/abs/1902.10186}
}

@article{178,
  author  = {Marcus, G.},
  title   = {Deep Learning: A Critical Appraisal},
  journal = {arXiv},
  year    = {2018},
  doi     = {10.48550/arXiv.1801.00631}
}

@article{179,
  author  = {Mitchell, M.},
  title   = {Why {AI} is Harder Than We Think},
  journal = {arXiv},
  year    = {2021},
  doi     = {10.48550/arXiv.2104.12871}
}

@article{180,
  author  = {Roy, C. J. and Oberkampf, W. L.},
  title   = {A comprehensive framework for verification, validation, and uncertainty quantification in scientific computing},
  journal = {Computer Methods in Applied Mechanics and Engineering},
  volume  = {200},
  number  = {25},
  pages   = {2131--2144},
  year    = {2011},
  doi     = {10.1016/j.cma.2011.03.016}
}

@book{181,
  author    = {Hanne, T. and Dornberger, R.},
  title     = {Computational Intelligence in Logistics and Supply Chain Management},
  series    = {International Series in Operations Research \& Management Science},
  volume    = {244},
  publisher = {Springer International Publishing},
  doi       = {10.1007/978-3-319-40722-7}
}

@book{182,
  author    = {Dewey, John},
  title     = {Democracy and Education},
  note      = {Democracy And Education: Dewey, John: 9780684836317},
  url       = {https://www.amazon.com/Democracy-Education-John-Dewey/dp/0684836319}
}

@article{183,
  author  = {Shumailov, I. and Shumaylov, Z. and Zhao, Y. and Papernot, N. and Anderson, R. and Gal, Y.},
  title   = {{AI} models collapse when trained on recursively generated data},
  journal = {Nature},
  volume  = {631},
  number  = {8022},
  pages   = {755--759},
  year    = {2024},
  doi     = {10.1038/s41586-024-07566-y}
}

@book{184,
  author    = {Vygotsky, L. S.},
  title     = {Mind in Society: Development of Higher Psychological Processes},
  publisher = {Harvard University Press},
  doi       = {10.2307/j.ctvjf9vz4}
}

@article{185,
  author  = {Bu{\c c}inca, Z. and Malaya, M. B. and Gajos, K. Z.},
  title   = {To Trust or to Think: Cognitive Forcing Functions Can Reduce Overreliance on {AI} in {AI}-assisted Decision-making},
  journal = {Proceedings of the ACM on Human-Computer Interaction},
  volume  = {5},
  number  = {CSCW1},
  pages   = {188:1--188:21},
  year    = {2021},
  doi     = {10.1145/3449287}
}

@article{186,
  author  = {Parasuraman, R. and Manzey, D. H.},
  title   = {Complacency and Bias in Human Use of Automation: An Attentional Integration},
  journal = {Human Factors},
  volume  = {52},
  number  = {3},
  pages   = {381--410},
  year    = {2010},
  doi     = {10.1177/0018720810376055}
}

@article{187,
  author  = {Talebi, S. and Tong, E. and Mofrad, M. R. K.},
  title   = {Beyond the Hype: Assessing the Performance, Trustworthiness, and Clinical Suitability of {GPT3.5}},
  journal = {arXiv},
  year    = {2023},
  doi     = {10.48550/arXiv.2306.15887}
}

@article{188,
  author  = {de Brito Duarte, R. and Correia, F. and Arriaga, P. and Paiva, A.},
  title   = {{AI} Trust: Can Explainable {AI} Enhance Warranted Trust?},
  journal = {Human Behavior and Emerging Technologies},
  volume  = {2023},
  number  = {1},
  pages   = {4637678},
  year    = {2023},
  doi     = {10.1155/2023/4637678}
}

@inproceedings{190,
  author    = {Raghavan, M. and Barocas, S. and Kleinberg, J. and Levy, K.},
  title     = {Mitigating bias in algorithmic hiring: evaluating claims and practices},
  booktitle = {FAT* '20},
  publisher = {Association for Computing Machinery},
  year      = {2020},
  pages     = {469--481},
  doi       = {10.1145/3351095.3372828}
}

@article{191,
  author  = {Hartvigsen, T. and Gabriel, S. and Palangi, H. and Sap, M. and Ray, D. and Kamar, E.},
  title   = {{ToxiGen}: A Large-Scale Machine-Generated Dataset for Adversarial and Implicit Hate Speech Detection},
  journal = {arXiv},
  year    = {2022},
  doi     = {10.48550/arXiv.2203.09509}
}

@article{192,
  author  = {Coupland, H. and S. N. and K. A. et al.},
  title   = {Exploring the potential and limitations of deep learning and explainable {AI} for longitudinal life course analysis},
  journal = {BMC Public Health},
  volume  = {25},
  number  = {1520},
  year    = {2025}
}

@inproceedings{193,
  author    = {Kaur, H. and Nori, H. and Jenkins, S. and Caruana, R. and Wallach, H. and Wortman Vaughan, J.},
  title     = {Interpreting Interpretability: Understanding Data Scientists' Use of Interpretability Tools for Machine Learning},
  booktitle = {CHI '20},
  publisher = {Association for Computing Machinery},
  year      = {2020},
  pages     = {1--14},
  doi       = {10.1145/3313831.3376219}
}

@misc{194,
  author       = {Shah, R. et al.},
  title        = {An Approach to Technical {AGI} Safety and Security},
  howpublished = {Online},
  url          = {files/12845/Shah et al. - An Approach to Technical AGI Safety and Security.pdf}
}

@misc{195,
  author       = {{IBM AI Ethics Board}},
  title        = {{AI} agents: Opportunities, risks, and mitigations},
  year         = {2025}
}

@article{196,
  author  = {Amershi, S. and Cakmak, Maya and Knox, William Bradley and Kulesza, Todd},
  title   = {Power to the people: The role of humans in interactive machine learning},
  journal = {AI Magazine},
  volume  = {35},
  number  = {4},
  year    = {2014}
}

@article{197,
  author  = {Wu, M. and Liu, W. and Wang, Y. and Yao, M.},
  title   = {Negotiating the Shared Agency between Humans \& {AI} in the Recommender System},
  journal = {arXiv},
  year    = {2024},
  doi     = {10.48550/arXiv.2403.15919}
}

@article{198,
  author  = {Shneiderman, B.},
  title   = {Human-Centered Artificial Intelligence: Reliable, Safe \& Trustworthy},
  volume  = {36},
  number  = {6},
  pages   = {495--504},
  doi     = {10.1080/10447318.2020.1741118}
}

@inproceedings{199,
  author    = {Amershi, S. et al.},
  title     = {Guidelines for Human-{AI} Interaction},
  booktitle = {CHI '19},
  publisher = {Association for Computing Machinery},
  year      = {2019},
  pages     = {1--13},
  doi       = {10.1145/3290605.3300233}
}

@article{200,
  author  = {Holzinger, A. and Biemann, C. and Pattichis, C. S. and Kell, D. B.},
  title   = {What do we need to build explainable {AI} systems for the medical domain?},
  journal = {arXiv},
  year    = {2017},
  doi     = {10.48550/arXiv.1712.09923}
}

@article{201,
  author  = {Tsiakas, K. D. C. T. H. M. D. R. S. R. H. T. and Barakova, E. I.},
  title   = {Negotiating Learning Goals with Your Future Learning-Self},
  journal = {Technologies (Basel)},
  volume  = {10},
  number  = {2},
  year    = {2022}
}

@article{202,
  author  = {Dietvorst, B. J. and Simmons, J. P. and Massey, C.},
  title   = {Overcoming Algorithm Aversion: People Will Use Imperfect Algorithms If They Can (Even Slightly) Modify Them},
  journal = {Management Science},
  volume  = {64},
  number  = {3},
  pages   = {1155--1170},
  year    = {2018},
  doi     = {10.1287/mnsc.2016.2643}
}

@article{203,
  author  = {Hoff, K. A. and Bashir, M.},
  title   = {Trust in Automation: Integrating Empirical Evidence on Factors That Influence Trust},
  journal = {Human Factors: The Journal of the Human Factors and Ergonomics Society},
  volume  = {57},
  number  = {3},
  pages   = {407--434},
  year    = {2015},
  doi     = {10.1177/0018720814547570}
}

@misc{204,
  author = {Parasuraman, R.},
  title  = {Humans and Automation: Use, Misuse, Disuse, Abuse},
  year   = {1997}
}

@inproceedings{205,
  author    = {Lim, B. Y. and Cahaly, J. P. and Sng, C. Y. F. and Chew, A.},
  title     = {Diagrammatization and Abduction to Improve {AI} Interpretability With Domain-Aligned Explanations for Medical Diagnosis},
  booktitle = {CHI 2025: CHI Conference on Human Factors in Computing Systems},
  publisher = {ACM},
  year      = {2025},
  pages     = {1--25},
  doi       = {10.1145/3706598.3714058}
}

@article{206,
  author  = {Wang, Q. and Goel, Ashok K.},
  title   = {Mutual theory of mind for human-{AI} communication},
  journal = {arXiv preprint arXiv:2210.03842},
  year    = {2022}
}

@inproceedings{207,
  author    = {Ashktorab, Z. et al.},
  title     = {Bridging the Gap: Unifying {HCI} \& {ML} Perspectives on Mutual Theory of Mind},
  booktitle = {International Joint Conference on Artificial Intelligence},
  year      = {2025},
  howpublished = {Online},
  url       = {https://research.ibm.com/publications/bridging-the-gap-unifying-hci-and-ml-perspectives-on-mutual-theory-of-mind}
}

@article{208,
  author  = {Rahwan, I. et al.},
  title   = {Machine behaviour},
  journal = {Nature},
  volume  = {568},
  number  = {7753},
  pages   = {477--486},
  year    = {2019},
  doi     = {10.1038/s41586-019-1138-y}
}

@article{209,
  author  = {Dafoe, A. et al.},
  title   = {Open Problems in Cooperative {AI}},
  journal = {arXiv},
  year    = {2020},
  doi     = {10.48550/arXiv.2012.08630}
}

@inproceedings{210,
  author    = {Holstein, K. and Wortman Vaughan, J. and Daum{\'e}, H. and Dudik, M. and Wallach, H.},
  title     = {Improving Fairness in Machine Learning Systems: What Do Industry Practitioners Need?},
  booktitle = {CHI '19},
  publisher = {Association for Computing Machinery},
  year      = {2019},
  pages     = {1--16},
  doi       = {10.1145/3290605.3300830}
}

@article{211,
  author  = {Kyng, M.},
  title   = {Designing for cooperation: cooperating in design},
  journal = {Communications of the ACM},
  volume  = {34},
  number  = {12},
  pages   = {65--73},
  year    = {1991},
  doi     = {10.1145/125319.125323}
}

@misc{212,
  author       = {Suvarna, A. et al.},
  title        = {ModelCitizens: Representing Community Voices in Online Safety},
  howpublished = {Online},
  url          = {files/12871/Suvarna et al. - ModelCitizens Representing Community Voices in On.pdf}
}

@misc{213,
  author       = {Dollinger, M. and Lodge, J.},
  title        = {Co-creation strategies for learning analytics},
  year         = {2018},
  howpublished = {Online},
  url          = {files/12298/Dollinger and Lodge - 2018 - Co-creation strategies for learning analytics.pdf}
}

@article{214,
  author  = {Mageed, I. A.},
  title   = {The Unfolding Dialectic-A Comparative Analysis of Human and Artificial Intelligence, its Open Challenges, and Future Prospects},
  journal = {Preprints},
  year    = {2025},
  doi     = {10.20944/preprints202506.1896.v1}
}

@article{215,
  author  = {Dellermann, D. and Ebel, P. and S{\"o}llner, M. and Leimeister, J. M.},
  title   = {Hybrid Intelligence},
  journal = {Business \& Information Systems Engineering},
  volume  = {61},
  number  = {5},
  pages   = {637--643},
  year    = {2019},
  doi     = {10.1007/s12599-019-00595-2}
}

@article{216,
  author  = {Coeckelbergh, M.},
  title   = {Narrative responsibility and artificial intelligence},
  journal = {AI \& Society},
  volume  = {38},
  number  = {6},
  pages   = {2437--2450},
  year    = {2023},
  doi     = {10.1007/s00146-021-01375-x}
}

@article{217,
  author  = {Bolin, G.},
  title   = {Communicative {AI} and Techno-Semiotic Mediatization: Understanding the Communicative Role of the Machine},
  volume  = {7},
  number  = {1},
  doi     = {10.30658/hmc.7.4}
}

@article{218,
  author  = {Walter, M. J. and Barrett, A. and Walker, D. J. and Tam, K.},
  title   = {Adversarial {AI} Testcases for Maritime Autonomous Systems},
  number  = {1},
  doi     = {10.5772/acrt.15}
}

@article{219,
  author  = {Rahman, S. S. I. A. S. G. L. J. S. J. L. M. R. P. et al.},
  title   = {{AI} Debate Aids Assessment of Controversial Claims},
  journal = {arXive},
  year    = {2025}
}

@misc{220,
  title        = {Reconsidering Artificial Intelligence as Co-Designer},
  howpublished = {Online},
  url          = {https://openaccess.wgtn.ac.nz/articles/conference_contribution/Reconsidering_Artificial_Intelligence_as_Co-Designer/24123639?file=42320835}
}

@article{221,
  title={Charting the sociotechnical gap in explainable AI: A framework to address the gap in XAI},
  author={Ehsan, Upol and Saha, Koustuv and De Choudhury, Munmun and Riedl, Mark O},
  journal={Proceedings of the ACM on human-computer interaction},
  volume={7},
  number={CSCW1},
  pages={1--32},
  year={2023},
  publisher={ACM New York, NY, USA}
}

@article{222,
  author  = {Ehsan, U. Q. V. L. S. P. M. O. R. and H. D. III},
  title   = {Seamful {XAI}: Operationalizing seamful design in explainable {AI}},
  journal = {Proceedings of the ACM on Human-Computer Interaction},
  volume  = {8},
  number  = {CSCW1},
  year    = {2024}
}

@article{226,
  author  = {Shanahan, M.},
  title   = {Talking About Large Language Models},
  journal = {arXiv},
  year    = {2023},
  doi     = {10.48550/arXiv.2212.03551}
}

@book{227,
  author    = {Kitcher, P.},
  title     = {The Advancement of Science: Science without Legend, Objectivity without Illusions},
  address   = {Oxford, New York},
  publisher = {Oxford University Press},
  year      = {1995},
  howpublished = {Online},
  url       = {files/11771/the-advancement-of-science-9780195096538.html}
}

@article{228,
  author  = {Zednik, C. and Boelsen, H.},
  title   = {Scientific Exploration and Explainable Artificial Intelligence},
  volume  = {32},
  number  = {1},
  pages   = {219--239},
  doi     = {10.1007/s11023-021-09583-6}
}

@inproceedings{229,
  author    = {Rogers, A.},
  title     = {Changing the World by Changing the Data},
  booktitle = {Proceedings of the 59th Annual Meeting of the Association for Computational Linguistics and the 11th International Joint Conference},
  year      = {2021},
  doi       = {10.18653/v1/2021.acl-long.170}
}

@article{230,
  author  = {Spirtes, P. and Zhang, K.},
  title   = {Causal discovery and inference: concepts and recent methodological advances},
  journal = {Applied Informatics (Berlin)},
  volume  = {3},
  number  = {1},
  pages   = {3},
  year    = {2016},
  doi     = {10.1186/s40535-016-0018-x}
}

@article{231,
  author  = {Hutson, M.},
  title   = {Artificial intelligence faces reproducibility crisis},
  journal = {Science},
  volume  = {359},
  number  = {6377},
  pages   = {725--726},
  year    = {2018},
  doi     = {10.1126/science.359.6377.725}
}

@misc{232,
  author       = {Gottweis, J. and F. and V. N. R. L.},
  title        = {Accelerating scientific breakthroughs with an {AI} co-scientist},
  howpublished = {Google Research},
  year         = {2025}
}

@article{233,
  author  = {Spirtes, P. and Zhang, K.},
  title   = {Causal discovery and inference: concepts and recent methodological advances},
  journal = {Applied Informatics (Berlin)},
  volume  = {3},
  number  = {1},
  pages   = {3},
  year    = {2016},
  doi     = {10.1186/s40535-016-0018-x}
}

@article{234,
  author  = {Vowels, M. J. and Camgoz, N. C. and Bowden, R.},
  title   = {D'ya Like {DAG}s? A Survey on Structure Learning and Causal Discovery},
  journal = {ACM Computing Surveys},
  volume  = {55},
  number  = {4},
  pages   = {1--36},
  year    = {2023},
  doi     = {10.1145/3527154}
}

@article{235,
  author  = {Sch{\"o}lkopf, B. et al.},
  title   = {Toward Causal Representation Learning},
  journal = {Proceedings of the IEEE},
  volume  = {109},
  number  = {5},
  pages   = {612--634},
  year    = {2021},
  doi     = {10.1109/JPROC.2021.3058954}
}

@article{236,
  author  = {Varshney, K. R. and Ashktorab, Z. and Bouneffouf, D. and Riemer, M. and Weisz, J. D.},
  title   = {Scopes of Alignment},
  journal = {arXiv},
  year    = {2025},
  doi     = {10.48550/arXiv.2501.12405}
}

@inproceedings{237,
  author    = {Rebanal, J. and Combitsis, Jordan and Tang, Yuqi and Chen, Xiang'Anthony},
  title     = {{Xalgo}: a design probe of explaining algorithms' internal states via question-answering},
  booktitle = {Proceedings of the 26th International Conference on Intelligent User Interfaces},
  year      = {2021}
}

@inproceedings{238,
  author    = {Guerdan, L. and Raymond, Alex and Gunes, Hatice},
  title     = {Toward affective {XAI}: facial affect analysis for understanding explainable human-{AI} interactions},
  booktitle = {Proceedings of the IEEE/CVF International Conference on Computer Vision},
  pages     = {3796--3805},
  year      = {2021}
}

@inproceedings{239,
  author    = {Jones, B. and Xu, Y. and Li, Q. and Scherer, S.},
  title     = {Designing a Proactive Context-Aware {AI} Chatbot for People's Long-Term Goals},
  booktitle = {CHI EA '24},
  publisher = {Association for Computing Machinery},
  year      = {2024},
  pages     = {1--7},
  doi       = {10.1145/3613905.3650912}
}

@inproceedings{240,
  author    = {Kaur, H. and Adar, E. and Gilbert, E. and Lampe, C.},
  title     = {Sensible {AI}: Re-imagining Interpretability and Explainability using Sensemaking Theory},
  booktitle = {{FAccT} '22: 2022 {ACM} Conference on Fairness, Accountability, and Transparency},
  publisher = {ACM},
  year      = {2022},
  pages     = {702--714},
  doi       = {10.1145/3531146.3533135}
}

@article{241,
  author  = {Martens, H.},
  title   = {A Greener, Safer, and More Understandable {AI} for Natural Science and Technology},
  journal = {Journal of Chemometrics},
  volume  = {39},
  number  = {2},
  pages   = {e3643},
  year    = {2025},
  doi     = {10.1002/cem.3643}
}

@article{242,
  author  = {Kurian, N.},
  title   = {Once Upon an {AI}: Six Scaffolds for Child-{AI} Interaction Design, Inspired by Disney},
  journal = {arXiv},
  year    = {2025}
}

@article{243,
  author  = {Wei, H. and Watson, J.},
  title   = {Preserving Professional Human Caring in Nursing in the Era of Artificial Intelligence},
  journal = {Advances in Nursing Science},
  year    = {2025},
  pages   = {10.1097/ANS.0000000000000573},
  doi     = {10.1097/ANS.0000000000000573}
}

@article{244,
  author  = {Kurian, N.},
  title   = {Developmentally aligned {AI}: a framework for translating the science of child development into {AI} design},
  journal = {{AI}, Brain and Child},
  volume  = {1},
  number  = {1},
  pages   = {9},
  year    = {2025},
  doi     = {10.1007/s44436-025-00009-z}
}

@article{245,
  author  = {De Togni, G. and Erikainen, S. and Chan, S. and Cunningham-Burley, S.},
  title   = {What makes {AI} `intelligent' and `caring'? Exploring affect and relationality across three sites of intelligence and care},
  journal = {Social Science \& Medicine},
  volume  = {277},
  pages   = {113874},
  year    = {2021},
  doi     = {10.1016/j.socscimed.2021.113874}
}

@article{246,
  author  = {Mueller, A. J. B. M. L. S. M. K. P. N. P. C. R. et al.},
  title   = {The quest for the right mediator: A history, survey, and theoretical grounding of causal interpretability},
  year    = {2024}
}

@article{247,
  author  = {Oikarinen, T. and Yan, G. and Weng, T.-W.},
  title   = {Evaluating Neuron Explanations: A Unified Framework with Sanity Checks},
  journal = {arXiv},
  year    = {2025},
  doi     = {10.48550/arXiv.2506.05774}
}

@article{248,
  author  = {Bai, N. and Iyer, R. A. and Oikarinen, T. and Kulkarni, A. and Weng, T.-W.},
  title   = {Interpreting Neurons in Deep Vision Networks with Language Models},
  journal = {arXiv},
  year    = {2025},
  doi     = {10.48550/arXiv.2403.13771}
}

@misc{249,
  author       = {Srinivas, A. A. and Oikarinen, T. and Srivastava, D. and Weng, W.-H. and Weng, T.-W.},
  title        = {{SAND}: Enhancing Open-Set Neuron Descriptions through Spatial Awareness},
  howpublished = {Online},
  url          = {files/13142/Srinivas et al. - SAND Enhancing Open-Set Neuron Descriptions throu.pdf}
}

@article{250,
  author  = {Wu, T.-Y. and Lin, Y.-X. and Weng, T.-W.},
  title   = {{AND}: Audio Network Dissection for Interpreting Deep Acoustic Models},
  journal = {arXiv},
  year    = {2024},
  doi     = {10.48550/arXiv.2406.16990}
}

@article{251,
  author  = {Oikarinen, T. and Das, S. and Nguyen, L. M. and Weng, T.-W.},
  title   = {Label-Free Concept Bottleneck Models},
  journal = {arXiv},
  year    = {2023},
  doi     = {10.48550/arXiv.2304.06129}
}

@article{252,
  author  = {Srivastava, D. and Yan, G. and Weng, T.-W.},
  title   = {{VLG}{-}{CBM}: Training Concept Bottleneck Models with Vision-Language Guidance},
  journal = {arXiv},
  year    = {2025},
  doi     = {10.48550/arXiv.2408.01432}
}

@article{253,
  author  = {Sun, C.-E. and Oikarinen, T. and Ustun, B. and Weng, T.-W.},
  title   = {Concept Bottleneck Large Language Models},
  journal = {arXiv},
  year    = {2025},
  doi     = {10.48550/arXiv.2412.07992}
}

@book{254,
  author    = {Clark, A.},
  title     = {Surfing Uncertainty: Prediction, Action, and the Embodied Mind},
  address   = {Oxford, New York},
  publisher = {Oxford University Press},
  year      = {2015}
}

@article{255,
  author  = {Lipton, Z. C.},
  title   = {The Mythos of Model Interpretability: In machine learning, the concept of interpretability is both important and slippery},
  journal = {Queue},
  volume  = {16},
  number  = {3},
  pages   = {31--57},
  year    = {2018},
  doi     = {10.1145/3236386.3241340}
}

@article{256,
  author  = {Murdoch, W. J. and Singh, C. and Kumbier, K. and Abbasi-Asl, R. and Yu, B.},
  title   = {Definitions, methods, and applications in interpretable machine learning},
  journal = {Proceedings of the National Academy of Sciences},
  volume  = {116},
  number  = {44},
  pages   = {22071--22080},
  year    = {2019},
  doi     = {10.1073/pnas.1900654116}
}

@misc{257,
  author       = {Kim, B.},
  title        = {Beyond interpretability: developing a language to shape our relationships with {AI}},
  howpublished = {Medium},
  year         = {2022}
}

@misc{258,
  author  = {Drousiotis, E. and Joyce, D. and Varsi, A. and Spirakis, P. and Maskell, S.},
  title   = {Intrinsically Interpretable Decision Trees for Healthcare Applications},
  year    = {2024},
  doi     = {10.21203/rs.3.rs-4608203/v1}
}

@article{259,
  author  = {Koh, P. W. et al.},
  title   = {Concept Bottleneck Models},
  journal = {arXiv},
  year    = {2020},
  doi     = {10.48550/arXiv.2007.04612}
}

@inproceedings{260,
  author    = {Rigotti, M. and Miksovic, C. and Giurgiu, I. and Gschwind, T. and Scotton, P.},
  title     = {Attention-based Interpretability with Concept Transformers},
  booktitle = {International Conference on Learning Representations},
  year      = {2021},
  howpublished = {Online},
  url       = {https://openreview.net/forum?id=kAa9eDS0RdO}
}

@inproceedings{261,
title={Deep Learning for Case-based Reasoning through Prototypes: A Neural Network that Explains its Predictions},
booktitle={Proceedings of {AAAI} Conference on Artificial Intelligence}, 
year={2018},
author={Oscar Li and Hao Liu and Chaofan Chen and Cynthia Rudin}
}

@inproceedings{261b,
title={This Looks Like That: Deep Learning for Interpretable Image Recognition},
author={Chaofan Chen and Oscar Li and Alina Barnett and Jonathan Su and Cynthia Rudin},
booktitle={Proceedings of Neural Information Processing Systems {(NeurIPS)}},
year={2019}
}

@article{262,
  author  = {Gee, A. H. and Garcia-Olano, D. and G. and G.},
  title   = {Explaining Deep Classification of Time-Series Data with Learned Prototypes},
  journal = {CEUR Workshop Proceedings},
  volume  = {2019},
  number  = {2429}
}

@article{263,
  author  = {Papernot, N. and McDaniel, P.},
  title   = {Deep k-Nearest Neighbors: Towards Confident, Interpretable and Robust Deep Learning},
  journal = {arXiv},
  year    = {2018},
  doi     = {10.48550/arXiv.1803.04765}
}

@inproceedings{264,
  author    = {Onoe, Y. and Durrett, G.},
  title     = {Interpretable Entity Representations through Large-Scale Typing},
  booktitle = {Findings 2020},
  editor    = {Cohn, T. and He, Y. and Liu, Y.},
  publisher = {Association for Computational Linguistics},
  year      = {2020},
  pages     = {612--624},
  doi       = {10.18653/v1/2020.findings-emnlp.54}
}

@misc{265,
  author       = {Garcia-Olano, Diego},
  title        = {In-process diagnostic methods for entity representation learning on sequential data at scale},
  howpublished = {UT Electronic Theses and Dissertations},
  year         = {2022}
}

@article{266,
  author  = {Oikarinen, T. and Weng, T.-W.},
  title   = {{CLIP}-Dissect: Automatic Description of Neuron Representations in Deep Vision Networks},
  journal = {arXiv},
  year    = {2023},
  doi     = {10.48550/arXiv.2204.10965}
}

@article{267,
  author  = {Bereska, L. and Gavves, E.},
  title   = {Mechanistic interpretability for {AI} safety--a review},
  journal = {arXiv},
  year    = {2024}
}

@article{268,
  author  = {Moraffah, R. and Karami, M. and Guo, R. and Raglin, A. and Liu, H.},
  title   = {Causal Interpretability for Machine Learning -- Problems, Methods and Evaluation},
  journal = {SIGKDD Explorations Newsletter},
  volume  = {22},
  number  = {1},
  pages   = {18--33},
  year    = {2020},
  doi     = {10.1145/3400051.3400058}
}

@article{269,
  author  = {Bargagli Stoffi, F. J. and C. and Gnecco, G.},
  title   = {Simple models in complex worlds: occam's razor and statistical learning theory},
  journal = {Minds and Machines},
  volume  = {32},
  number  = {1},
  year    = {2022}
}

@article{270,
  author  = {Bargagli-Stoffi, F. J. and W. and Gnecco, G.},
  title   = {Heterogeneous causal effects with imperfect compliance: A Bayesian machine learning approach},
  journal = {The Annals of Applied Statistics},
  volume  = {16},
  number  = {3},
  year    = {2022}
}

@article{271,
  author  = {Bargagli-Stoffi, F. J. and K. L. and Dominici, F.},
  title   = {Causal rule ensemble: Interpretable discovery and inference of heterogeneous treatment effects},
  year    = {2020}
}

@misc{272,
  author       = {Wang, F. and Adebayo, J. and Tan, S. and Garcia-Olano, D. and Kokhlikyan, N.},
  title        = {Error Discovery By Clustering Influence Embeddings},
  howpublished = {Online},
  url          = {files/13250/Wang et al. - Error Discovery By Clustering Influence Embeddings.pdf}
}

@article{273,
  author  = {Swayamdipta, S. et al.},
  title   = {Dataset Cartography: Mapping and Diagnosing Datasets with Training Dynamics},
  journal = {arXiv},
  year    = {2020},
  doi     = {10.48550/arXiv.2009.10795}
}

@inproceedings{panigutti2022understanding,
  title={Understanding the impact of explanations on advice-taking: a user study for AI-based clinical Decision Support Systems},
  author={Panigutti, Cecilia and Beretta, Andrea and Giannotti, Fosca and Pedreschi, Dino},
  booktitle={Proceedings of the 2022 CHI Conference on Human Factors in Computing Systems},
  pages={1--9},
  year={2022}
}

@article{jacobs2021machine,
  title={How machine-learning recommendations influence clinician treatment selections: the example of antidepressant selection},
  author={Jacobs, Maia and Pradier, Melanie F and McCoy Jr, Thomas H and Perlis, Roy H and Doshi-Velez, Finale and Gajos, Krzysztof Z},
  journal={Translational psychiatry},
  volume={11},
  number={1},
  pages={108},
  year={2021},
  publisher={Nature Publishing Group UK London}
}

@online{goodwin_google_ai_overviews_2024,
  author       = {Goodwin, Danny},
  title        = {Google {AI} Overviews under fire for giving dangerous and wrong answers},
  year         = {2024},
  month        = may,
  day          = {24},
  url          = {https://searchengineland.com/google-ai-overview-fails-442575},
  note         = {Accessed: 2025-12-18},
  organization = {Search Engine Land}
}

@article{ackerman2013sharing,
  title={Sharing knowledge and expertise: The CSCW view of knowledge management},
  author={Ackerman, Mark S and Dachtera, Juri and Pipek, Volkmar and Wulf, Volker},
  journal={Computer Supported Cooperative Work (CSCW)},
  volume={22},
  number={4},
  pages={531--573},
  year={2013},
  publisher={Springer}
}

@inproceedings{corti2024moving,
  title={``It Is a Moving Process": Understanding the Evolution of Explainability Needs of Clinicians in Pulmonary Medicine},
  author={Corti, Lorenzo and Oltmans, Rembrandt and Jung, Jiwon and Balayn, Agathe and Wijsenbeek, Marlies and Yang, Jie},
  booktitle={Proceedings of the 2024 CHI Conference on Human Factors in Computing Systems},
  pages={1--21},
  year={2024}
}

@article{raees2024explainable,
  title={From explainable to interactive AI: A literature review on current trends in human-AI interaction},
  author={Raees, Muhammad and Meijerink, Inge and Lykourentzou, Ioanna and Khan, Vassilis-Javed and Papangelis, Konstantinos},
  journal={International Journal of Human-Computer Studies},
  volume={189},
  pages={103301},
  year={2024},
  publisher={Elsevier}
}

@inproceedings{ma2025towards,
  title={Towards human-ai deliberation: Design and evaluation of llm-empowered deliberative ai for ai-assisted decision-making},
  author={Ma, Shuai and Chen, Qiaoyi and Wang, Xinru and Zheng, Chengbo and Peng, Zhenhui and Yin, Ming and Ma, Xiaojuan},
  booktitle={Proceedings of the 2025 CHI Conference on Human Factors in Computing Systems},
  pages={1--23},
  year={2025}
}

@article{ilievski2025aligning,
  title={Aligning generalization between humans and machines},
  author={Ilievski, Filip and Hammer, Barbara and van Harmelen, Frank and Paassen, Benjamin and Saralajew, Sascha and Schmid, Ute and Biehl, Michael and Bolognesi, Marianna and Dong, Xin Luna and Gashteovski, Kiril and others},
  journal={Nature Machine Intelligence},
  pages={1--12},
  year={2025},
  publisher={Nature Publishing Group UK London}
}

@InProceedings{vashistha2025i,
  title = 	 {I-trustworthy Models. A framework for trustworthiness evaluation of probabilistic classifiers},
  author =       {Vashistha, Ritwik and Farahi, Arya},
  booktitle = 	 {Proceedings of The 28th International Conference on Artificial Intelligence and Statistics},
  pages = 	 {4726--4734},
  year = 	 {2025},
  volume = 	 {258},
  series = 	 {Proceedings of Machine Learning Research},
  month = 	 {03--05 May},
  publisher =    {PMLR},
  url = 	 {https://proceedings.mlr.press/v258/vashistha25a.html},
}

@inproceedings{vashistha2024u,
  title={U-trustworthy models. reliability, competence, and confidence in decision-making},
  author={Vashistha, Ritwik and Farahi, Arya},
  booktitle={Proceedings of the AAAI Conference on Artificial Intelligence},
  volume={38},
  number={18},
  pages={19956--19964},
  year={2024}
}

@inproceedings{deck2024critical,
  title={A critical survey on fairness benefits of explainable AI},
  author={Deck, Luca and Schoeffer, Jakob and De-Arteaga, Maria and K{\"u}hl, Niklas},
  booktitle={Proceedings of the 2024 ACM Conference on Fairness, Accountability, and Transparency},
  pages={1579--1595},
  year={2024}
}

@article{muralidharan2024ai,
  title={AI and the need for justification (to the patient)},
  author={Muralidharan, Anantharaman and Savulescu, Julian and Schaefer, G Owen},
  journal={Ethics and Information Technology},
  volume={26},
  number={1},
  pages={16},
  year={2024},
  publisher={Springer}
}

@article{RudinWaCo2020,
journal = {Harvard Data Science Review},
number = {1},
note = {https://hdsr.mitpress.mit.edu/pub/7z10o269},
title = {The Age of Secrecy and Unfairness in Recidivism Prediction},
url = {https://hdsr.mitpress.mit.edu/pub/7z10o269},
volume = {2},
author = {Rudin, Cynthia and Wang, Caroline and Coker, Beau},
date = {2020-01-31},
year = {2020},
month = {1},
day = {31},
}

@article{BarnettEtAl2021,
  author    = {Alina Jade Barnett and
               Fides Regina Schwartz and
               Chaofan Tao and
               Chaofan Chen and
               Yinhao Ren and
               Joseph Y. Lo and
               Cynthia Rudin},
  title     = {{IAIA-BL:} {A} Case-based Interpretable Deep Learning Model for Classification
               of Mass Lesions in Digital Mammography},
  journal   = {Nature Machine Intelligence},
  volume = {3},
  pages = {1061--1070},
  year      = {2021}
}
}

\end{document}